\documentclass[a4paper,11pt]{article}
\usepackage{jheppub}
\usepackage{braket}
\usepackage{comment}
\usepackage{appendix}
\usepackage{mathtools}
\usepackage{caption}
\usepackage{subcaption}
\usepackage{bbold}
\usepackage{float}
\usepackage[utf8]{inputenc}
\usepackage{tikz-cd}
\usepackage[T1]{fontenc} 
\usepackage{scalerel}
\usepackage{amsmath}
\usepackage{amsthm}

\usepackage{mathrsfs}
\usepackage{physics}
\usepackage[compat=1.1.0]{tikz-feynman}
\usepackage{tensor}
\usepackage[]{graphicx}
\usepackage[export]{adjustbox}
\usepackage{slashed}
\usepackage{stackrel,amssymb}
\usepackage{tikz}
\usetikzlibrary{math}
\usetikzlibrary{calc,intersections}
\usetikzlibrary{arrows.meta,shapes.misc}
\usetikzlibrary{arrows.meta}
\tikzset{>={Latex[scale=1.1]}}
\usetikzlibrary{arrows.meta,
	bending,
	decorations.markings, decorations.text}

\makeatletter
\newcommand*{\letterdef@}{}
\newcommand*{\letterdef}[3]{%
\def\letterdef@##1{\expandafter\newcommand\csname #1\endcsname{#2{##1}}}%
	\@tfor\@tempa :=#3\do{\expandafter\letterdef@\expandafter{\@tempa}}}
\makeatother
\letterdef{c#1} {\mathcal}{ABCDEFGHIJKLMNOPQRSTUVWXYZ} 
\letterdef{rm#1}{\mathrm} {dDeimM} 
\author[a]{Carlos Barredo Mart\'inez}
\affiliation[a]{Abdus Salam Centre for Theoretical Physics, Imperial College London,\\
6 Prince Consort Road, London, SW7 2AZ, UK.}
\emailAdd{carlos.barredo21@imperial.ac.uk}
\abstract{We consider three families of semiclassical string solutions on AdS$_5\cross S^5/\mathbb{Z}_L$ orbifold backgrounds which correspond to `twisted' single-trace operators in the SU$(2)$, SU$(3)$ and SL$(2)$ sectors, respectively, of the dual 4d $\mathcal{N}=2$ quiver gauge theory. The leading quantum correction to the classical energy of each solution is computed, and is matched, despite the lack of maximal supersymmetry, to the corresponding results from finite-size corrections to the twisted Bethe ansatz in the continuum $J\to\infty$ limit and the associated Landau-Lifshitz low-energy effective theory. The $\mathbb{Z}_L$ quotient allows the string to have fractional winding branches $\mu=\widetilde{m}+\frac{n}{L}$ along orbifolded directions. As such, stability is a property of the full winding branch $\mu$, rather than of the individual twisted sectors $n$. Restricting to the purely fractional branch ($\widetilde m=0$), we obtain new stable configurations which were absent in the analogous semiclassical string solutions on AdS$_5\cross S^5$. The energies of these stable states connect smoothly to those of states in relevant BPS sectors as $\mu\to0$.}
\begin{document}
\begin{flushright}Imperial$-$TP$-$2026$-$CBM$-$01\end{flushright}
\title{On quantum corrections to semiclassical strings on $\mathrm{AdS}_5\cross S^5/\mathbb{Z}_{L}$ orbifold backgrounds}
\maketitle
\section{Introduction}
By now, it is well established that the AdS/CFT correspondence \cite{Maldacena:1997re, Witten:1998qj} matches the energies $E$ of non-interacting (solitonic) quantum closed strings on AdS$_5\cross S^5$ to the conformal dimensions $\Delta$ of single-trace operators of 4d $\mathcal{N}=4$ SU$(N)$ super Yang-Mills (SYM) in the planar (large $N$) limit with fixed 't Hooft coupling $\lambda\equiv g_{\rm{YM}}^2N$ (see, for example, \cite{Beisert:2003tq, Zarembo:2004hp, Tseytlin:2010jv}). It is natural to expect that both $E$ and $\Delta$ should be functions of $\lambda$ (the effective string tension being $T=\frac{\sqrt{\lambda}}{2\pi}$), the conserved quantum numbers $Q$ of the solution, and/or different winding numbers $m$, with AdS/CFT predicting the equality $E(\lambda,Q,m)=\Delta(\lambda,Q,m)$ for \textit{any} value of the arguments. For a specific subsector of non-BPS `semiclassical' states with $Q, \sqrt{\lambda}\to\infty$ and fixed $\frac{Q}{\sqrt{\lambda}}\gg 1$, it was found that both perturbative string theory and perturbative gauge theory predict that the respective large $Q$ expansions of $E$ and $\Delta$ should be regular and have similar structures \cite{Frolov_2003, Arutyunov:2003uj, Arutyunov:2003za, Beisert:2003xu, Beisert:2003ea, Engquist:2003rn}.\footnote{These expansions are obtained in `opposite' regimes. On the string side, one first expands in large $T$ while keeping $\frac {Q}{\sqrt{\lambda}}$ large but fixed, while on the gauge theory side one expands in $\lambda$ in the usual perturbation theory sense, and then expands the $k^{th}$ loop anomalous dimension at large $Q$.} 

On the string side, quantum corrections to the energy of a given semiclassical string solution may be computed by quantising its fluctuations in a suitable regularisation scheme (e.g. \cite{Frolov:2003tu, Frolov:2004bh, Park:2005ji}). The large-charge expansion of such corrections may then be compared against the continuum $(Q\to\infty)$ finite-size expansion of the solutions of the corresponding spin-chain Bethe equations \cite{Arutyunov:2004yx, Beisert:2005mq, Kazakov:2004qf}, which determine the anomalous dimensions perturbatively.\footnote{The terminology `finite-size corrections' is inherited from \cite{Beisert:2005mq}, but should not be confused with wrapping corrections to the thermodynamic Bethe ansatz, which start contributing at $\order{\lambda^Q}$.} They may also be compared against the associated Landau-Lifshitz (LL) effective theory after assuming a particular regularisation scheme \cite{Kruczenski:2003gt, Kruczenski:2004kw, Hernandez:2004uw, Stefanski:2004cw, Kruczenski:2004cn}. The LL model describes the low-energy dynamics of the relevant integrable spin-chain, and its action may be derived both in the continuum coherent state limit of the one-loop gauge theory dilatation operator, as well as from the `fast-string' limit $\left(\frac{\lambda}{Q^2}\ll1\right)$ of the worldsheet sigma model. It therefore provides a common effective description which acts as a bridge between the string theory and the spin-chain description.

For type IIB theory on AdS$_5\cross S^5$, agreement was established through second order in the effective `fast-string' coupling, that is, to $\order{\frac{\lambda^2}{Q^4}}$, amongst all three approaches, with the leading $1/Q$ corrections from one-loop string worldsheet calculations agreeing with the corresponding finite-size corrections to the Bethe ansatz and LL results for several semiclassical solutions \cite{Beisert:2005mq, Minahan:2005mx}. Discrepancies at higher orders are attributed to the structure of the dilatation operator on the gauge theory side \cite{Serban:2004jf, Beisert:2004hm}, as well as due to a non-commutativity of the weak/strong coupling and $Q\to\infty$ limits \cite{Beisert:2005cw}.

While the aforementioned program has been extended to different classes of backgrounds \cite{McLoughlin:2008he, Astolfi:2008ji, Hernandez:2014eta}, including integrable deformations of AdS$_5\cross S^5$ \cite{Frolov:2005ty, Borsato:2022drc}, comparatively little attention has been devoted to an arguably simpler deformation of AdS$_5\cross S^5$ preserving integrability and $\mathcal{N}=2$ supersymmetry, which consists on taking its $\mathbb{Z}_L$ orbifold. These models have gained interest in recent years, in particular, due to advancements in supersymmetric localisation (see e.g. \cite{Billo:2022fnb, Beccaria:2023qnu, Korchemsky:2025eyc} and references therein) allowing to compute certain observables at arbitrary values of $\lambda$. It is therefore of interest to extend the semiclassical string program to this class of backgrounds. 
\subsection{Background}
\begin{figure}[htbp]
\centering
\begin{tikzpicture}[
    node/.style={circle, draw, minimum size=0.7cm, thick},
    scale=1
]
\node[node] (1) at (90:2) {$1$};
\node[node] (2) at (30:2) {$2$};
\node[node] (3) at (-30:2) {$3$};
\node[node] (b1) at (-90:2) {};
\node[node] (b2) at (-150:2) {};
\node[node] (L) at (150:2) {$L$};

\draw[thick] (L) to[bend left=20] (1);
\draw[thick] (1) to[bend left=20] (2);
\draw[thick] (2) to[bend left=20] (3);
\draw[thick] (3) to[bend left=20] (b1);
\draw[thick] (b1) to[bend left=20] (b2);
\path (b2) to[bend left=20]
    coordinate[pos=0.2] (d1)
    coordinate[pos=0.5] (d2)
    coordinate[pos=0.8] (d3)
(L);
\foreach \p in {d1,d2,d3}
    \fill (\p) circle (0.05);
\end{tikzpicture}
\caption{Diagrammatic representation of the quiver theory. Nodes represent $\mathcal{N}=2$ vector multiplets in the adjoint representation of $\mathrm{SU}(N)_i$. Each line connecting adjacent nodes gives rise to an $\mathcal{N}=2$ hypermultiplet in the bifundamental representation of SU$(N)_i\cross\mathrm{SU}(N)_{i+1}$. For $L=2$, an additional $\mathrm{SU}(2)_L$ global symmetry arises since the two hypermultiplets connecting the two nodes may be rotated into each other.}
\label{fig:quiverdiagram}
\end{figure}
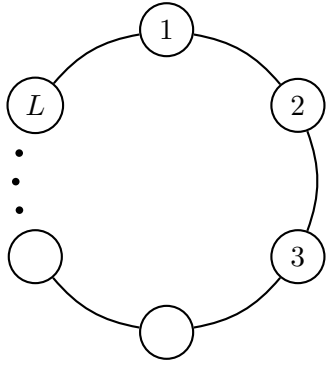

On the field theory side, the $\mathbb{Z}_L$ quotient acts on 4d $\mathcal{N}=4$ SU$(LN)$ SYM theory, resulting in a 4d $\mathcal{N}=2$ circular quiver gauge theory with gauge group $\mathrm{SU}(N)^L$ and matter content encoded in the diagram of Figure \ref{fig:quiverdiagram} \cite{Kachru1998, Lawrence:1998ja, Bershadsky:1998cb, Klebanov1999, Gukov:1998kk, Klebanov:1998hh, Oz1998}. We consider the theory at the orbifold point, where the couplings at each node are equal (i.e. $g_{\mathrm{YM}_i}=g$).
The $\mathbb{Z}_L$ generator, $\xi$, acts on the gauge indices of the $\mathcal{N}=4$ theory via the unitary `twist matrix'
\begin{equation}
\gamma=\mathrm{diag}\left(\mathbb{1}_N,\xi\cdot\mathbb{1}_N,\xi^2\cdot\mathbb{1}_N,\dots\,,\xi^{L-1}\cdot\mathbb{1}_N\right)\,,\qquad \xi\equiv e^{i\frac{2\pi}{L}}\,,
\end{equation}
breaking the gauge group $\mathrm{SU}(LN)\to \mathrm{SU}(N)^L$. It also acts on the R-symmetry indices via the element $R(\xi)\in\mathrm{SU}(4)_R$. Using the subgroup embedding SU$(4)_R\supset \mathrm{SU}(2)_L\cross\mathrm{SU}(2)_R\cross\mathrm{U}(1)_r$, one may choose $R(\xi)$ to be an element of order $L$ in $\mathrm{SU}(2)_L$. This choice preserves $\mathrm{SU}(2)_R\cross\mathrm{U}(1)_r$ R-symmetry and thus $\mathcal{N}=2$ supersymmetry. The surviving fields $\Phi$ must therefore obey the projection
\begin{equation}\label{eq:fieldtheoryprojection}
  \Phi=\frac{1}{L}\sum_{l=0}^{L-1}(\xi^{h_{\Phi}}\gamma)^l\Phi_{\mathcal{N}=4}\gamma^{-l}\,,\qquad h_{\Phi}\in\mathbb{Z}_{L}\,,
\end{equation}
where $h_{\Phi}$ is the $\mathbb{Z}_L$ charge of the given $\mathcal{N}=4$ field, $\Phi_{\mathcal{N}=4}$, which depends on the SU$(4)_R$ charges associated to it \cite{Gaberdiel:2022iot}.\footnote{The $\mathcal{N}=4$ gauge field $A_{\mu}$ is an SU$(4)$ singlet, hence $h_{A}=0$. Only $L$ diagonal $N\cross N$ components of the $LN\cross LN$ matrix survive, giving rise to the $\mathrm{SU}(N)^L$ gauge group.} For the three complex scalars $\{\mathcal{X},\mathcal{Y},\mathcal{Z}\}$ of the $\mathcal{N}=4$ theory, we have that $(h_{\mathcal{X}},h_{\mathcal{Y}},h_{\mathcal{Z}})=(1,-1,0)$. Applying \eqref{eq:fieldtheoryprojection} yields the $\mathcal{N}=2$ scalars
\begin{equation}
X=\begin{pmatrix}
0 & X_{12} & \dots&\\
 & 0 & X_{23}&\;\;\\
& &\ddots & \\
X_{L1} & \dots& &0
\end{pmatrix}\,,\qquad Y=\begin{pmatrix}
0 & \dots & &Y_{1L}\\
Y_{21} & 0 & &\vdots\\
& Y_{32}&\ddots & \\
 & & &0
\end{pmatrix}\,,\qquad Z=\begin{pmatrix}
Z_{11} & 0 & \dots&0\\
0 & Z_{22}& &\vdots\\
\vdots& &\ddots & \\
0 & \dots& &Z_{LL}
\end{pmatrix}\,,
\end{equation}
which obey
\begin{equation}\label{eq:scalarZL}
    \gamma^{\dagger}X\gamma=\xi X\,,\qquad \gamma^{\dagger}Y\gamma=\xi^{-1}Y\,,\qquad \gamma^{\dagger}Z\gamma=Z\,.
\end{equation}
On the string side, the orbifold group acts on the embedding coordinates of $S^{5}\subset\mathbb{C}^{3}$ as
\begin{equation}
\label{orbifoldaction}
\Gamma_{L}:\quad (X,Y,Z)\to (\xi X,\, \xi^{-1}Y, \,Z)\,,
\end{equation}
featuring a non-trivial locus of fixed points parametrised by a circle $S^{1}\subset S^{5}$. In order to form a modular-invariant theory, one must include both the untwisted sector (formed by $\mathbb{Z}_{L}$-invariant string states), as well as the $(L-1)$ twisted sectors spanned by strings which close up to the action of $\Gamma_L^n$, $n\in\{1,2,\dots, L-1\}$ \cite{Dixon:1985jw, Dixon:1986jc} ($n=0$ denotes the untwisted sector). The latter have non-trivial dynamics with respect to the `parent' theory, and, in particular, light twisted sector states are described by an effective six-dimensional theory on $\mathrm{AdS}_{5}\times S^{1}\subset \mathrm{AdS}_5\cross S^5/\mathbb{Z}_L$ \cite{Gukov:1998kk, Billo:2022fnb, Skrzypek:2023fkr, Martinez:2025jjq}. 
\subsection{Results}
The first step in extending the semiclassical string program, is to therefore construct classical solutions on AdS$_5\cross S^5/\mathbb{Z}_L$ and compute the leading quantum correction to their classical energies. We focus on solutions dual to operators in three bosonic subsectors of the quiver theory that are closed under the one-loop dilatation operator, namely the SU$(2)$, SU$(3)$ and SL$(2)$ sectors. With reference to \eqref{orbifoldaction}, if the string does not wrap an orbifolded direction, the semiclassical solution is `insensitive' to the $\mathbb{Z}_L$ quotient and is therefore the same as in AdS$_5\cross S^5$ background. Within the class of extended semiclassical solutions considered here, sensitivity to the twisted boundary conditions is achieved by winding an orbifolded plane. The allowed winding is of the form $\mu=\tilde{m}+\frac{n}{L}$, with $\tilde{m}\in\mathbb{Z}$, along the given direction. A twisted sector $n$ therefore only fixes the fractional part of $\mu$. In particular, stability is a property of the full winding branch $\mu$, rather than of $n$ alone. As such, different winding branches within the same twisted sector may have different stability properties. Throughout this paper, the explicit string, LL and Bethe-ansatz comparisons below are performed on the purely fractional branch $\mu=\frac{n}{L}$ with $\widetilde m=0$. Results often follow by the effective replacement of the relevant integer winding in the analogous $\mathrm{AdS}_5\times S^5$ solutions by $\mu$. The availability of fractional values of $\mu$ gives rise to novel stable winding branches in the orbifold theory.

Classical spinning string solutions on $S^{5}/\mathbb{Z}_{L}$ with non-trivial angular momenta $J=\sum_i J_i$ admit regular large $J$ expansions of the form \cite{Tseytlin:2002ny,Gubser:2002tv,Frolov:2002av}
\begin{equation}
\label{eq:energyexpansion}
E=J\bigg[1+\frac{\lambda}{J^{2}}\bigg(c_{1}+\frac{d_{1}}{J}+...\bigg)+\bigg(\frac{\lambda}{J^{2}}\bigg)^{2}\bigg(c_{2}+\frac{d_{2}}{J}+...\bigg)+...\bigg]\,,
\end{equation}
where the coefficients $c_{n}\equiv c_{n}(\frac{J_{i}}{J},L)$, $d_{n}\equiv d_{n}(\frac{J_{i}}{J},L)$ now depend on the orbifold parameter $L$, and should be finite in the standard semiclassical limit with fixed $\frac{\lambda}{J^{2}}\ll 1$ and $L$. The quantum corrections we compute correspond to the coefficients $d_1$ in the above expansion.

On the other hand, operators in the $n^{th}$ twisted sector of the quiver theory have the general form
\begin{equation}
    \mathcal O_{n,\Psi}
    =
    \sum_W
    \Psi_W\,
    \mathrm{Tr}\!\left[\gamma^n W\right],
    \qquad
    W=\Phi_1\Phi_2\cdots\Phi_{J},
\label{eq:twistedoperatorgeneral}
\end{equation}
where $\Phi_i$ is an arbitrary $\mathcal{N}=2$ field and the sum runs over `words' $W$ of length $J$ and fixed global charges, with $\Psi_W$ denoting the corresponding spin-chain wavefunction. For a scalar word containing $(J_1,J_2,J_3)$ copies of $(X,Y,Z)$, respectively, closure of the gauge index
requires its total orbifold charge to vanish, implying \cite{Ideguchi2004} (cf. \eqref{eq:scalarZL})
\begin{equation}\label{eq:angmomentaprojection}
J_1-J_2 = 0\mod L\,,
\end{equation}
that is, that the sequence of `letters' defines a closed path along the circular quiver.

In the planar limit, at one-loop, the local interaction between neighboring letters is inherited from the $\mathcal{N}=4$ theory, with the insertion $\gamma^n$ appearing through twisted boundary conditions and a twisted momentum constraint. The mixing problem in each of the subsectors of interest is therefore described by a finite twisted spin-chain \cite{Minahan:2002ve,Beisert:2003tq,Beisert:2005he,Astolfi:2006is}, whose Hamiltonian is the one-loop dilatation operator in the given sector.\footnote{The $\mathcal{N}=2$ dilatation operator may be obtained from the known $\mathcal{N}=4$ result by applying the orbifold projection. Restricting to the SU$(2)$ sector, the result was derived explicitly at one and two loops in \cite{Astolfi:2006is}.} The continuum limit $J\to\infty$ then allows for direct comparison to semiclassical strings and the LL description. Despite the lack of maximal supersymmetry, we assume that \eqref{eq:energyexpansion} should be matched (to order $\lambda^2/J^4$) by the corresponding large $J$ expansion of the full scaling dimension of the dual quiver theory operators, which is of the form
\begin{equation}
\label{eq:conformaldimexpansion}
    \Delta=J\left[1+\frac{\lambda}{J^2}\left(a_1+\frac{b_1}{J}+\dots\right)+\left(\frac{\lambda}{J^2}\right)^2\left(a_2+\frac{b_2}{J}+\dots\right)+\dots\right]\,,
\end{equation}
where, again, the coefficients $a_{n}\equiv a_{n}(\frac{J_{i}}{J},L)$ and $b_{n}\equiv b_{n}(\frac{J_{i}}{J},L)$ depend on the $\mathbb{Z}_L$ parameter. Computing the coefficients $b_1$ both from the corresponding finite-size corrections to the orbifolded Bethe ansatz, as well as from the LL effective theory, we indeed find $d_1=b_1$ in each of the sectors.\footnote{For the SU$(3)$ sector, only the `non-anomalous' finite-size correction is computed, which matches the zero-mode part of $d_1$ in the expansion of the energy of the $J_1=J_2$, $J_3\neq 0$ solution.} This provides an example of agreement between all three approaches at one-loop in models with less than maximal supersymmetry. In the SU$(2)$ sector, this agreement also extends to the stability boundary.

The three families of semiclassical solutions studied in this paper are associated with the following twisted spin-chain eigenstates:
\begin{equation}
\label{eq:stateopmap}
\begin{split}
&\mathrm{Solutions\; \eqref{classicalsoln}}
\;\leftrightarrow\;
\mathrm{Tr}\!\left[\gamma^n X^{J/2}Y^{J/2}+\mathrm{perms.}\right]
\;\rightarrow\;
\textrm{twisted}\;\mathrm{SU}(2)\;\mathrm{XXX}_{1/2}\;\textrm{spin-chain}\,,\\
&\mathrm{Solutions\; \eqref{eq:3spinclassicalsoln}}
\;\leftrightarrow\;
\mathrm{Tr}\!\left[\gamma^n X^{J/2}Y^{J/2}Z^{J_3}+\mathrm{perms.}\right]
\;\rightarrow\;
\textrm{twisted}\;\mathrm{SU}(3)\;\mathrm{XXX}\;\textrm{nested spin-chain}\,,\\
&\mathrm{Solutions\; \eqref{eq:orbifoldSJsolution}}
\;\leftrightarrow\;
\mathrm{Tr}\!\left[\gamma^n D^{s_1}X\cdots D^{s_J}X\right]
\;\rightarrow\;
\textrm{twisted}\;\mathrm{SL}(2)\;\mathrm{XXX}_{-1/2}\;\textrm{spin-chain}\,.
\end{split}
\end{equation}
where $\sum _is_i=S$ with $i\in[1,J]$ and perms. denotes all distinct orderings of `letters' in a given `word'.\footnote{$DX=n^{\mu}D_{\mu}X=n^{\mu}(\partial_{\mu}X-ig_{\rm{YM}}[A_{\mu},X])$ is a covariant derivative along a light-cone direction with fixed null vector $n^{\mu}$.} In the compact SU$(2)$ and SU$(3)$ sectors each spin-chain site has a finite-dimensional local state space, spanned by $(X,Y)$ and $(X,Y,Z)$, respectively. By contrast, an SL$(2)$ site carries the infinite tower $\{D^{s_i} X\}_{s_i}$. In addition, the Bethe ansatz description requires a choice of an ordered basis of fields and a vacuum. For the SU$(2)$ and SL$(2)$ cases, the choice $(X,Y,Z)$ results in the `vacuum' $\mathrm{Tr}[\gamma^nX^J]$, which is part of the physical spectrum of the quiver theory if $n=0$ and $J=0\,\mathrm{mod}\,L$ \cite{lePlat:2025eod}. For $n\neq 0$, the corresponding `vacua' should therefore be understood as virtual reference states on top of which one can build physical spin-chain excitations (magnons), rather than as being part of the physical spectrum. For the SU$(3)$ sector, the choice of vacuum depends on the relative scaling of $J$ and $J_3$. For $\frac{J}{2}\geq J_3$, the appropriate bifundamental `vacuum' is $\mathrm{Tr}[\gamma^nX^{J+J_3}]$, existing only for $n=0$ and $J+J_3=0\,\mathrm{mod}\,L$. Conversely, for $J_3\geq \frac{J}{2}$, one may choose the adjoint vacuum $\mathrm{Tr}[\gamma^nZ^{J+J_3}]$, which exists for all operator lengths (cf. \eqref{eq:scalarZL}). A summary of the results obtained in each of the sectors of semiclassical string states considered may be found in Table \ref{tab:sector_summary}.

\begin{table}[t]
    \centering
    \renewcommand{\arraystretch}{1.5}
    \setlength{\tabcolsep}{4pt}
    \resizebox{\textwidth}{!}{%
    \begin{tabular}{|p{1.2cm}|p{3.5cm}|p{5cm}|p{4cm}|p{3.8cm}|p{3.8cm}|}
    \hline
    \textbf{Sector}
    &
    \textbf{Physical charge constraints}
    &
    \textbf{Stability}
    &
    \textbf{Bethe `vacuum'}
    &
    \textbf{Twisted momentum constraints}
    &
    \textbf{Matching status}
    \\
    \hline\hline
    $\mathrm{SU}(2)$ & $J_1=J_2=\frac{J}{2}$, $J_3=0$, together with projection condition \eqref{eq:angmomentaprojection}. & $|\mu|<\frac12\to$  stable,\newline
    $|\mu|=\frac12\to $ marginal,\newline
    $|\mu|>\frac12\to$ unstable. &
    $\Tr[\gamma^n X^J]\to$ only part of the physical spectrum for $n=0$ and $J=0\mod\,L$, else virtual reference state.
    &
    $e^{iP}=\xi^n$, with $P=2\pi\mu\,\mod 2\pi$.
    &
    Complete agreement at one-loop between all three approaches.\\
    \hline
    $\mathrm{SU}(3)$ & $J_1=J_2=\frac{J}{2}$, $J_3\neq0$, together with projection condition \eqref{eq:angmomentaprojection}. & {\raggedright
$|\mu|<\frac{1}{2}\to$ stable $\forall\, q\in(0,1]$,\newline
$|\mu|=\frac{1}{2}\to$ marginal,\newline
$|\mu|>\frac{1}{2}\to$\par
\vspace{-2pt}\hspace{12pt}
$\begin{cases}
    \mathrm{stable\;for}\;q<1-\frac{1}{4\mu^2},\\
    \mathrm{marginal\;for\;}q=1-\frac{1}{4\mu^2},\\
    \mathrm{unstable\;for\;}q>1-\frac{1}{4\mu^2}\,.
\end{cases}$
}
    &$\Tr[\gamma^n X^{J+J_3}]$ if $\frac{J}{2}\geq J_3$.\newline
    $\Tr[\gamma^n Z^{J+J_3}]$ if $J_3\geq \frac{J}{2}$.\newline
    The latter is $\mathbb{Z}_L$ neutral and is a physical vacuum for arbitrary $J+J_3$.
    &
    $e^{iP}=\xi^n$ if $\frac{J}{2}\geq J_3$, and $e^{iP}=1$ if $J_3\geq \frac{J}{2}$.
    &
    Only regular Bethe finite-size contribution to $b_1$ computed, agreeing with zero-mode part of $d_1$ of string and LL results.\\\hline
    $\mathrm{SL}(2)$ & $J, S\neq 0$ with $J=\alpha L$ and $pS+\mu J=0$. &
    Stable for all allowed windings $\mu\in[0,1)$. Relation \eqref{eq:orbifoldSJvirasoro} implies $\mu(\mu-p)>0$. & $\Tr[\gamma^nX^J]\to$ only part of the physical spectrum for $n=0$ and $J=0\mod\,L$, else virtual reference state. & $e^{iP}=\xi^n$ &
    Complete agreement at one-loop between all three approaches.\\
    \hline
    \end{tabular}%
    }
\caption{Summary of the different sectors of semiclassical string states considered and their corresponding twisted spin-chain descriptions. Here, $\mu=\widetilde m+\frac{n}{L}$ denotes the full winding branch, common to all three sectors. The explicit comparisons presented throughout the paper are performed on the purely fractional branch where $\widetilde m=0$ and $n\in\{0,1,\dots, L-1\}$.}
    \label{tab:sector_summary}
\end{table}

The rest of this paper is organised as follows. In Section \ref{sec: 2}, we construct the three families of semiclassical string solutions dual to operators \eqref{eq:stateopmap}. In Section \ref{sec: 3}, we compute the one-loop correction ($d_1$ coefficient in \eqref{eq:energyexpansion}) to the energies of the solutions of the previous section, highlighting the new stable winding branches due to the $\mathbb{Z}_L$ orbifold. These smoothly connect with different BPS regimes at large $L$ with $n\neq\order{L}$. Details on the corresponding computations are relegated to Appendices \ref{sect:A}, \ref{sect:3spindetails} and \ref{sect:SJdetails}, respectively. Section \ref{sect:4} considers the LL effective theory derived from the string theory sigma-model, reproducing, in all cases, the results of Section \ref{sec: 3}. Finally, Section \ref{sect:5} discusses the finite-size corrections to the twisted Bethe ansätze in the continuum $J\to\infty$ limit for each of the relevant closed bosonic subsectors, showing exact matches with the string and LL results in the SU$(2)$ and SL$(2)$ cases, with partial matching in the SU$(3)$ case. Further details on the twisted SU$(3)$ Bethe equations are given in Appendix \ref{app:2}.
\section{Classical string solutions on the orbifold}\label{sec: 2}
The bosonic sector of the Green-Schwarz type IIB superstring on $\mathrm{AdS}_{5}\times S^{5}/\mathbb{Z}_{L}$ is described by a non-linear sigma model \cite{Metsaev:1998it, Metsaev:2002re} with target space metric written in terms of `angular' coordinates as
\begin{equation}
\begin{split}
\label{eq:unorbifoldedmetrics}
&\dd s^{2}_{\mathrm{AdS}_{5}}=\mathrm{d}\rho^{2}-\cosh^{2}\rho\,\mathrm{d}t^{2}+\sinh^{2}\rho\,(\mathrm{d}\theta^{2}+\sin^{2}\theta\,\mathrm{d}\phi^{2}+\cos^{2}\theta\,\dd\varphi^{2})\,,\\
&\dd s^{2}_{S^{5}}=d\gamma^{2}+\cos^{2}\gamma\,\mathrm{d}\varphi_{3}^{2}+\sin^{2}\gamma\,(\mathrm{d}\psi^{2}+\cos^{2}\psi\,\mathrm{d}\varphi_{1}^{2}+\sin^{2}\psi\,\mathrm{d}\varphi_{2}^{2})\,,
\end{split}
\end{equation}
where $\rho\in[0,\infty)$, $\theta\in[0,\frac{\pi}{2}]$, $\phi,\varphi\in[0,2\pi)$ and $t\in(-\infty,\infty)$ for the AdS$_5$ metric,\footnote{We work with the universal cover of AdS$_5$, where the $t$ direction is decompactified, avoiding closed timelike curves and allowing for a direct identification with the dual gauge theory on $\mathbb{R}\cross S^3$ (see, e.g. \cite{Tseytlin:2010jv}).} and $\gamma,\,\psi\in[0,\frac{\pi}{2}]$, $\varphi_{1,2,3}\in[0,2\pi)$ for $S^5$. Acting with \eqref{orbifoldaction} implies the modified identifications $\varphi_{1,2}\sim\varphi_{1,2}\pm\frac{2\pi}{L}$. We define the coordinates $\Phi=\frac{\varphi_{1}+\varphi_{2}}{2}$ and $\bar{\chi}=\frac{\varphi_{1}-\varphi_{2}}{2}$ such that $\Phi\in[0,2\pi)$ and $\bar{\chi}\sim\bar{\chi}+\frac{2\pi}{L}$ under the $\mathbb{Z}_{L}$ action. Then, a metric on $S^{5}/\mathbb{Z}_{L}$ reads
\begin{equation}
\label{orbifoldmetric}
\dd s^{2}_{S^{5}/\mathbb{Z}_{L}}=\mathrm{d}\gamma^{2}+\cos^{2}\gamma\,\mathrm{d}\varphi_{3}^{2}+\sin^{2}\gamma\,\left(\mathrm{d}\psi^{2}+\mathrm{d}\Phi^{2}+\frac{1}{L^{2}}\mathrm{d}\chi^{2}+\frac{2\cos2\psi}{L}\,\mathrm{d}\Phi\mathrm{d}\chi\right)\,,
\end{equation}
where we further defined $\chi=L\bar{\chi}\in[0,2\pi)$ to rewrite the metric in terms of $2\pi$-periodic angles, making the background dependence on the parameter $L$ explicit. 

The sigma-model action may be represented in terms of embedding coordinates of both AdS$_5$ in $\mathbb{R}^{2,4}$ and $S^5$ in $\mathbb{R}^{6}$ as
\begin{equation}
\label{bosonicsigmamodel}
\begin{split}
&S_B=T\int\mathrm{d}\tau\mathrm{d}\sigma\bigg(\mathcal{L}_{\mathrm{AdS}}+\mathcal{L}_{S}\bigg),\quad \mathcal{L}_{\mathrm{AdS}}=-\frac{1}{2}\eta_{PQ}\partial_{a}Y_{P}\partial^{a}Y_{Q}-\frac{1}{2}\bar{\Lambda}(\eta_{PQ}Y_{P}Y_{Q}+1),\\
&\mathcal{L}_{S}=-\frac{1}{2}(\partial_{a}X_{M})(\partial^{a}X_{M})+\frac{1}{2}\Lambda(X_{M}X_{M}-1),\qquad T\equiv \frac{R^2}{2\pi\alpha'}=\frac{\sqrt{\lambda}}{2\pi}\,,
\end{split}
\end{equation}
where $T$ is the (effective) string tension, given in terms of the curvature radius $R$ of both AdS$_5$ and the orbifolded five-sphere, with $\sqrt{\alpha'}=\ell_s$ the string length. The quantities $\Lambda,\bar{\Lambda}$ are Lagrange multipliers enforcing the sphere and hyperboloid constraints. The indices $P,Q\in[0,5]$ and $M\in[1,6]$ label the embedding coordinates $(Y_P, X_{M})$ of $\mathbb{R}^{2,4}$ and $\mathbb{R}^{6}$, respectively, with $\eta_{PQ}=(-,+,+,+,+,-)$. These are then related to the `angles' of the AdS$_5$ and $S^5/\mathbb{Z}_{L}$ metrics \eqref{eq:unorbifoldedmetrics} and \eqref{orbifoldmetric}, respectively, by
\begin{equation}
\label{complexcoords}
\begin{split}
&Y_{1}+iY_{2}=\sinh\rho\sin\theta\,e^{i\phi}\,,\qquad \mathrm{X}\equiv X_{1}+iX_{2}=\sin\gamma\cos\psi \,e^{i(\Phi+\chi/L)}\,,\\
& Y_{3}+iY_{4}=\sinh\rho\cos\theta\,e^{i\varphi}\,,\quad\;\;\; \mathrm{Y}\equiv X_{3}+iX_{4}=\sin\gamma\sin\psi\,e^{i(\Phi-\chi/L)}\,,\\
&\mathrm{W}\equiv Y_{5}+iY_{0}=\cosh\rho\,e^{it}\,, \quad\;\;\; \mathrm{Z}\equiv X_{5}+iX_{6}=\cos\gamma\,e^{i\varphi_{3}}\,.
\end{split}
\end{equation}
\subsection{$J_1=J_2$ solution in $\mathrm{SU}(2)$ sector}
We review the classical motion of rigid circular closed spinning strings rotating on $\mathbb{R}_{t}\times S^{5}/\mathbb{Z}_{L}\subset \mathrm{AdS}_5\cross S^5/\mathbb{Z}_{L}$ (i.e. at the `centre' of $\mathrm{AdS}_{5}$ where $\rho=0$). The simplest $J_1=J_2$ solution with constant equal radii was constructed in \cite{Ideguchi2004}, and reads
\begin{equation}\begin{split}
\label{classicalsoln}
&\mathrm{X}=\frac{1}{\sqrt{2}}e^{i\left(\omega\tau+\mu\sigma\right)}\,,\quad \mathrm{Y}=\frac{1}{\sqrt{2}}e^{i\left(\omega\tau-\mu\sigma\right)}\,,\quad \mathrm{W}=e^{i\kappa\tau}\,,\quad X_5=X_6=0\,,\quad Y_{1}=\dots=Y_4=0\,,\\
&t=\kappa\tau\,,\quad \rho=0\,,\quad \gamma=\frac{\pi}{2}\,,\quad \psi=\frac{\pi}{4}\,,\quad \varphi_{3}=\mathrm{undefined}\,,\quad \Phi=\omega\tau\,,\quad \chi=L\mu\,\sigma\,,
\end{split}\end{equation}
where
\begin{equation}\label{eq:mudefinition}
    \mu\equiv \widetilde{m}+\frac{n}{L}\,,\qquad \widetilde{m}\in\mathbb{Z}\,,\qquad n\in\{0,1,\dots,L-1\}\,,
\end{equation}
parametrises the allowed fractional windings around the $\chi$ direction as a consequence of the $\mathbb{Z}_{L}$ action, with $n$ denoting a twisted sector label.\footnote{A more general solution of this type need not assume equal windings along $\Phi$, however their values should be consistent with the constraint imposed by orbifolding (cf. \eqref{eq:angmomentaprojection}).} At the classical value $\gamma=\frac{\pi}{2}$, the string size is `extremal' (i.e. it rotates on the maximal size $S^{3}/\mathbb{Z}_{L}\subset S^{5}/\mathbb{Z}_{L}$). These semiclassical states correspond to heavy stretched twisted strings, and should not be identified with the aforementioned light twisted modes on AdS$_5\cross S^1$, which sit at $\gamma=0$. With respect to its unorbifolded counterpart, novel features of this solution are most transparent on the purely fractional winding branch, that is, when $\mu=\frac{n}{L}$ with $\tilde{m}=0$. Since $n$ fixes $\mu$ modulo an integer, this is a choice of winding branch within each twisted sector. Branches with different values of $\widetilde m$ need not have the same stability properties (see Section \ref{sect:2spin1loop}). The non-trivial conformal gauge (Virasoro) constraint reads
\begin{equation}
\label{conformalgaugeconstraints}
\kappa^{2}=\omega^{2}+\mu^2\,,
\end{equation}
being the only expression that relates $\mathrm{AdS}_{5}$ to $S^{5}/\mathbb{Z}_{L}$ parameters. The conserved charges of the classical solution are given by
\begin{equation}\label{eq:conservedquants}
\mathcal{J}_{i}=\frac{J_{i}}{\sqrt{\lambda}}=\int_{0}^{2\pi}\frac{\mathrm{d}\sigma}{2\pi}(X_{2i-1}\dot{X}_{2i}-X_{2i}\dot{X}_{2i-1})\,,\qquad \mathcal{E}=\frac{E}{\sqrt{\lambda}}=\int_{0}^{2\pi}\frac{\mathrm{d}\sigma}{2\pi}\;\dot{t}=\kappa\,,
\end{equation}
where $\mathcal{J}_{1}=\mathcal{J}_{2}=\frac{\omega}{2}\,,\, \mathcal{J}_3=0$, are associated to the three commuting (translational) U$(1)$ isometries of the angles $\varphi_3,\Phi,\chi$ of \eqref{orbifoldmetric}. The energy $E$ corresponds to the AdS$_5$ conserved charge related to a compact generator of SO$(2)\subset\mathrm{SO}(2,4)$. The consistency condition \eqref{eq:angmomentaprojection} is automatically satisfied by our choice of angular momenta. The classical energy of the solution is given by
\begin{equation}\label{eq:SU2classicalenergy}
E=J\sqrt{1+\frac{\lambda\mu^2}{J^2}}=J\left[1+\frac{\lambda}{J^2}\frac{\mu^2}{2}-\frac{\lambda^2}{J^4}\frac{\mu^4}{8}+\dots\right]\,,\qquad J\equiv J_1+J_2\,,
\end{equation}
which is nothing more than a rewriting of \eqref{conformalgaugeconstraints} in terms of the conserved charges. In the second equality, we have expanded in large $J$ with $n,L$ held fixed. One may now read off the values $c_{1}=\frac{\mu^2}{2}$ and $c_{2}=-\frac{\mu^4}{8}$ of the coefficients in \eqref{eq:energyexpansion}, encoding the expected $L$ dependence at the classical level. 
\subsection{$J_1=J_2,\,J_3\neq 0$ solution in $\mathrm{SU}(3)$ sector}
The above $J_1=J_2$ solution may be generalised by turning on a third angular momentum ($J_3\neq 0$) on $S^5/\mathbb{Z}_{L}$ providing the orbifold generalisation of the solution in \cite{Frolov:2003tu}, which reads
\begin{equation}\begin{split}\label{eq:3spinclassicalsoln}
&\mathrm{X}=\frac{\sin\gamma_0}{\sqrt{2}}e^{i
\left(\omega\tau+\mu\sigma\right)}\,,\; \mathrm{Y}=\frac{\sin\gamma_0}{\sqrt{2}}e^{i
\left(\omega\tau-\mu\sigma\right)}\,, \; \mathrm{Z}=\cos\gamma_0\,e^{i\mathrm{k}\tau}\,,\;\mathrm{W}=e^{i\kappa\tau}\,,\; Y_1=\dots=Y_4=0\,,\\
&t=\kappa\tau\,,\quad \rho=0\,,\quad\gamma=\gamma_0\,\quad\psi=\frac{\pi}{4}\,,\quad\varphi_3=\mathrm{k}\tau\,,\quad \Phi=\omega\tau\,,\quad \chi=L\mu\,\sigma\,,\\
\end{split}\end{equation}
where the label $\mathrm{k}$ parametrises the angular velocity in the $Z$-plane.\footnote{The limit $\sin^2\gamma_0\to 0$ moves the string towards the fixed circle of the orbifold action \eqref{orbifoldaction}. However, the $\sin^2\gamma_0=0$ endpoint, which is relevant to strictly localized modes on AdS$_5\cross S^1$, is degenerate in the present parametrization and is not included in the one-loop analysis (see footnote \ref{ftn:35}).} As before, we restrict the analysis to the purely fractional winding branch $\mu=\frac{n}{L}$. For $\gamma_0=\frac{\pi}{2}$ and $\mathrm{k}=0$, the above solution reduces to the one described in \cite{Ideguchi2004}. The equation of motion for $\gamma$ reads
\begin{equation}\label{eq:3spineom}
 (\mu^2-\omega^2+\mathrm{k}^2)\sin 2\gamma_0=0\,,
\end{equation}
which, in the non-trivial $J_3\neq 0$ branch where $\gamma_0\in(0,\tfrac{\pi}{2})$, imposes 
\begin{equation}\label{eq:3spindynamical}
    \omega^2=\mu^2+\mathrm{k}^2\,.
\end{equation}
With this condition, the diagonal Virasoro constraint reads 
\begin{equation}\label{eq:classicalrelations3spin}
\kappa^2=\mathrm{k}^2+2\mu^2\sin^2\gamma_0\,.
\end{equation}
Relative to the $J_1=J_2$ solution, where the equation for $\gamma$ was satisfied automatically for $\gamma_0=\frac{\pi}{2}$, \eqref{eq:3spineom} now becomes dynamical, generating the relation \eqref{eq:3spindynamical}.
The conserved charges of the solution are computed as in \eqref{eq:conservedquants}, giving
\begin{equation}\label{eq:3spinconservedcharges}
\mathcal{J}_{1,2}=\frac{J/2}{\sqrt{\lambda}}=\frac{\omega}{2}\sin^2\gamma_0\,,\quad \mathcal{J}_3=\frac{J_3}{\sqrt{\lambda}}=\mathrm{k}\cos^2\gamma_0\,,\qquad \mathcal{E}=\frac{E}{\sqrt{\lambda}}=\kappa\,.
\end{equation}
One may then deduce the following `auxiliary' relation
\begin{equation}\label{eq:auxeqn}
  \mathrm{k}=\mathcal{J}_3+\frac{\mathcal{J}}{\sqrt{1+\frac{\mu^2}{\mathrm{k}^2}}}\,,\qquad \mathcal{J}=\mathcal{J}_1+\mathcal{J}_2\,,
\end{equation}
relating the different angular momenta on $S^5/\mathbb{Z}_{L}$. The condition \eqref{eq:angmomentaprojection} holds identically here, since $J_3$ is naturally associated to the plane which rotates trivially under \eqref{orbifoldaction}. Rewriting the second relation in \eqref{eq:classicalrelations3spin} in terms of the conserved charges \eqref{eq:3spinconservedcharges} and defining  $\mathcal{K}\equiv\sqrt{\lambda}\mathrm{k}$, gives \eqref{eq:auxeqn} and the classical energy of the $J_1=J_2$, $J_3\neq 0$ solution as
\begin{equation}
    \mathcal{K}=J_3+\frac{J}{\sqrt{1+\frac{\lambda\mu^2}{\mathcal{K}^2}}}\,,\qquad E=\mathcal{K}\sqrt{1+\frac{2\lambda\mu^2}{\mathcal{K}^2}\left(1-\frac{J_3}{\mathcal{K}}\right)}\,.
\end{equation}
In a large $J_{\rm{tot}}\equiv J+J_3$ expansion, one obtains
\begin{equation}\label{eq:3spinclassicalenergy}
\mathcal{K}=J_{\rm{tot}}-\lambda\frac{\mu^2J}{2J_{\rm{tot}}^2}+\dots\,,\quad E=J_{\rm{tot}}\left[1+\frac{\lambda}{J_{\rm{tot}}^2}\frac{\mu^2}{2}\frac{J}{J_{\rm{tot}}}-\frac{\lambda^2}{J_{\rm{tot}}^4}\frac{\mu^4}{8}\frac{J}{J_{\rm{tot}}}+\dots\right]\,,
\end{equation}
as well as the following leading-order relationships
\begin{equation}\label{eq:3spinsinrelation}
 \sin^2\gamma_0=\frac{J}{J_{\rm{tot}}}+\dots\,,\qquad \cos^2\gamma_0=\frac{J_3}{J_{\rm{tot}}}+\dots\,.
\end{equation}
With reference to \eqref{eq:energyexpansion}, one now finds the coefficients $c_1=\frac{\mu^2}{2}\frac{J}{J_{\rm{tot}}}$ and $c_2=-\frac{\mu^4}{8}\frac{J}{J_{\rm{tot}}}$. The solution connects to BPS sectors in two regimes: $J\to0$, where $E\to J_3$, and $\mu\to0$, where $E\to J_{\rm{tot}}$. The latter corresponds to the limit of a point-like BMN geodesic. For purely fractional windings, we later show that for $L\to\infty$ with fixed $n$ not of order $L$, there exists a stable subsector of states, which was absent in the analogous solution on AdS$_5\cross S^5$ and connects smoothly to the BPS state with $E=J_{\rm{tot}}$, proving a novel feature of the classical solution on the orbifold.
\subsection{$SJ$ solution in $\mathrm{SL}(2)$ sector}\label{sect:SJsolution}
The classical solution with angular momentum $J_3$ on $S^5$ and spin $S$ on AdS$_5$ discussed in, e.g. \cite{Arutyunov:2003za, Park:2005ji}, is not classically `sensitive' to the orbifold. Such a solution describes a circular string rotating within AdS$_3\cross S^1_{Z}\subset \mathrm{AdS}_5\cross S^5/\mathbb{Z}_{L}$, where $S^1_{Z}$ denotes the fixed circle of the orbifold action \eqref{orbifoldaction}. The `orbifold-sensitive' analog of the `SJ' solution in \cite{Arutyunov:2003za} is constructed by instead choosing the $S^1$ momentum to be aligned along an orbifolded direction, say\footnote{Note that one could equivalently choose the embedding where $\mathrm{X}=0$ with $\mathrm{Y}=e^{i\left(\omega\tau-\mu\sigma\right)}$ within $S^5/\mathbb{Z}_{L}$. In terms of $S^5/\mathbb{Z}_{L}$ angles, that corresponds to $\gamma=\psi=\frac{\pi}{2}$, $\Phi=\omega\tau$ and $\chi=-L\mu\sigma$. One could also consider turning on a further spin on AdS$_5$ while keeping the $S^5/\mathbb{Z}_{L}$ angular momentum aligned along an orbifolded direction. This solution should then be dual to a gauge theory operator in the SU$(1,2)$ sector.}
\begin{equation}\begin{split}\label{eq:orbifoldSJsolution}
&\mathrm{W}=\cosh\rho_0\, e^{i\kappa\tau}\,,\quad Y_1+iY_2=\sinh\rho_0\,e^{i(w\tau+p\sigma)}\,,\quad \mathrm{X}=e^{i\left(\omega\tau+\mu\sigma\right)}\,,\\
&t=\kappa\tau\,,\quad\rho=\rho_0\,,\quad \theta=\frac{\pi}{2}\,,\quad \varphi=w\tau+p\sigma\,,\quad \gamma=\frac{\pi}{2}\,,\quad \psi=0\,,\quad \Phi=\omega\tau\,,\quad \chi=L\mu\,\sigma\,,
\end{split}\end{equation}
where $p\in\mathbb{Z}$ denotes a winding number in the $(Y_1,Y_2)$ plane of AdS$_3$ and the parameters $w,\omega$ denote rotation frequencies of the string along itself. Solving the equations of motion and Lagrange multiplier constraints implied by \eqref{bosonicsigmamodel} gives
\begin{equation}\label{eq:SL2relations}
w^2=\kappa^2+p^2\,,\qquad \omega^2=\nu^2+\mu^2\,,\qquad \nu^2=-\Lambda\,,\qquad \kappa^2=\bar{\Lambda}\,.
\end{equation}
In addition to $\mathcal{E}$ and $\mathcal{J}$, the solution now features the additional conserved (on-shell) AdS$_5$ spin $\mathcal{S}$ defined by
\begin{equation}
    \mathcal{S}=\frac{S}{\sqrt{\lambda}}=\int_0^{2\pi}\frac{\dd\sigma}{2\pi}\,(Y_{1}\dot{Y}_{2}-Y_{2}\dot{Y}_{1})\,,
\end{equation}
corresponding to a Cartan generator of the $\mathrm{SO}(4)\subset \mathrm{SO}(2,4)$ isometry group of AdS$_5$. The conserved charges associated to this solution read
\begin{equation}
    \mathcal{E}=\frac{E}{\sqrt{\lambda}}=\kappa\cosh^2\rho_0\,,\qquad \mathcal{S}=\frac{S}{\sqrt{\lambda}}=w\sinh^2\rho_0\,,\qquad \mathcal{J}=\frac{J}{\sqrt{\lambda}}=\omega\,,
\end{equation}
with the following `auxiliary' relation between the AdS$_5$ charges
\begin{equation}\label{eq:SJauxrelation}
    \frac{\mathcal{E}}{\kappa}-\frac{\mathcal{S}}{\sqrt{\kappa^2+p^2}}=1\,.
\end{equation}
Using \eqref{eq:SJauxrelation}, the (off-)diagonal Virasoro constraints read, respectively,
\begin{equation}\label{eq:orbifoldSJvirasoro}
   p\mathcal{S}+\mu\mathcal{J}=0\,,\qquad 2\kappa\mathcal{E}-\kappa^2=2\sqrt{\kappa^2+p^2}\mathcal{S}+\mathcal{J}^2+\mu^2\,.
\end{equation}
Consistency with \eqref{orbifoldaction} (cf. \eqref{eq:angmomentaprojection}) imposes that $J=\alpha L$ where $\alpha\in\mathbb{Z}_{>0}$. Let us now define $u\equiv\frac{\mathcal{S}}{\mathcal{J}}=\frac{S}{J}$. Then, one finds that $S=uJ=-\frac{n\alpha}{p}\in\mathbb{Z}$. In a large $J$ expansion, the classical energy reads
\begin{equation}
    E-S-J=\frac{\lambda}{2J}\left(\mu^2+p^2u\right)-\frac{\lambda^2}{8J^3}\left(\mu^4+4\mu^2p^2u+4p^4u^2+p^4u\right)+\dots\,,
\end{equation}
where $u,p$ and $\mu$ are held fixed. Imposing the first relation in \eqref{eq:orbifoldSJvirasoro} implies $u=-\frac{\mu}{p}$, but since $u>0$ and $\mu>0$ by construction, one is forced to take $p$ to be a negative integer. One can then eliminate $u$ from the above, giving
\begin{equation}\label{eq:classicalenergySJ}
    E-S=J\left[1+\frac{\lambda}{J^2}\frac{\mu^2(1+u)}{2u}-\frac{\lambda^2}{J^4}\frac{\mu^4(1+u)(u^2+3u+1)}{8u^3}+\dots\right]\,,
\end{equation}
from which one may deduce that $c_1=\frac{\mu^2(1+u)}{2u}$ and $c_2=-\frac{\mu^4(1+u)(u^2+3u+1)}{8u^3}$ (cf. \eqref{eq:energyexpansion}).

It therefore appears that the large $J$ regime could, a priori, be probed by either taking large integers $\alpha$ with $L$ held fixed, or by keeping $\alpha$ fixed but taking $L\to\infty$. The former corresponds to the usual fixed-background semiclassical large tension expansion with $\mathcal{J},\mathcal{S},L$ fixed, requiring $\alpha\sim\sqrt{\lambda}\gg 1$. The twist $\mu$ enters at the classical level (cf. \eqref{eq:classicalenergySJ}), with the ratio $u=\frac{\mathcal{S}}{\mathcal{J}}=\frac{\mu}{\vert p\vert}$ being discrete. This regime is thus probing a fractional winding subsector of semiclassical states which was absent in the analogous AdS$_5\cross S^5$ solution. In the latter case, keeping $\mathcal{J}$ finite forces to take $L\sim\sqrt{\lambda}$, which reorganises the semiclassical expansion (see comments around \eqref{eq:SU2Lclassicalenergy}). Within the large $L$ regime, keeping $\alpha,p,n$ fixed, with $n$ \textit{not} of order $L$, one has that $S=-\frac{\alpha n}{p}=\order{1}$, so that $\mathcal{S}\to 0$. This regime is thus more akin to a near-BMN/small-spin limit. Furthermore, the coefficients $c_{1,2}$ of \eqref{eq:classicalenergySJ} vanish in this case since $u,\mu=\order{J^{-1}}$. This double-scaling limit differs from the fixed $u,\mu$ large $J$ limit used to organise the original expansion \eqref{eq:classicalenergySJ}, reorganising its $1/J$ hierarchy. Alternatively, one may keep $\mu$ finite as $J,L\to\infty$ by choosing twisted sectors where $n=\order{L}$. In this case, $u=-\frac{\mu}{p}\sim\order{1}$ and $S\sim\order{J}$, so that the standard semiclassical expansion with fixed $u,\mu$ remains applicable. This scaling, however, involves moving through the different twisted sectors $n$ as $L$ is increased. It is therefore different from the fixed $n$ long quiver limit considered above.
\section{One-loop energy correction to classical string solutions}\label{sec: 3}
Having constructed three types of solutions which are classically `sensitive' to the orbifold, we now summarise the computation of the leading quantum corrections to their classical energy. Since these are relatively standard, we relegate the details to Appendices \ref{sect:A}, \ref{sect:3spindetails} and \ref{sect:SJdetails}, giving only the relevant results. With reference to \eqref{eq:energyexpansion}, we determine expressions for $d_1$ in each case, highlighting the novel features introduced by the $\mathbb{Z}_{L}$ quotient.
\subsection{$J_1=J_2$ solution}\label{sect:2spin1loop}
To find the string one-loop correction to \eqref{eq:SU2classicalenergy}, one is to first determine the characteristic (bosonic and fermionic) frequencies by expanding in fluctuations around the classical solution \eqref{classicalsoln} up to and including quadratic order, and then compute an appropriate sum over them. After diagonalising the equations of motion arising from the respective quadratic fluctuation Lagrangians, one obtains a characteristic polynomial whose zeros determine the characteristic frequencies. In the bosonic sector, one finds the following two `coupled' frequencies
\begin{equation}\label{eq:S3CFs}
(w_{r,\pm})^2=r^2+2(\kappa^2-\mu^2)\pm 2\sqrt{(\kappa^2-\mu^2)^2+\kappa^2r^2}\,,\qquad r\in\mathbb{Z}_{\geq 0}\,,
\end{equation}
corresponding to fluctuation modes on $S^3/\mathbb{Z}_{L}\subset S^5/\mathbb{Z}_{L}$, which satisfy the property that
\begin{equation}\label{eq:usefulproperty1}
    (w_{r,+})^2(w_{r,-})^2=r^2(r^2-4\mu^2)\,.
\end{equation}
There is also a set of two `decoupled' frequencies on $S^5/\mathbb{Z}_{L}$ of the form
\begin{equation}\label{eq:DecoupledCF}
\mathrm{w}_r=\sqrt{r^2+\kappa^2-2\mu^2}=\sqrt{r^2+\mathcal{J}^2-\mu^2}\,.
\end{equation}
The four frequencies associated to AdS$_5$ fluctuations are 
\begin{equation}\label{eq:AdSCF}
\omega_{i}^{\mathrm{AdS}}=\sqrt{\kappa^{2}+r^{2}}\,,\qquad i=5,6,7,8\,,
\end{equation}
which are the same as in the $J_1=J_2$ solution on AdS$_5\cross S^5$, since the orbifold acts trivially on the AdS$_5$ factor of the background. The relevant expansions of the above frequencies in different regimes of interest may be found in Appendix \ref{sect:2spinbosonfreqs}.

Expanding the quadratic fermionic part of the Green-Schwarz superstring action \cite{Metsaev:1998it, Metsaev:2002re} (see \eqref{eq:fermionL2}) in fluctuations and diagonalising the corresponding Dirac-type operator in the quadratic Lagrangian by choosing a suitable representation of Gamma matrices, gives two four-fold degenerate (positive) frequencies
\begin{equation}\label{eq:fermionicCFs}
\Omega_{r,+}=\frac{\kappa}{2}+\sqrt{r^2+\kappa^2-\mu^2}\,,\qquad \Omega_{r,-}=\left\vert-\frac{\kappa}{2}+\sqrt{r^2+\kappa^2-\mu^2}\right\vert\,.
\end{equation}

In the case of $r=0$, \eqref{eq:DecoupledCF} are real provided that $\mathcal{J}\geq \vert\mu\vert$. This is readily satisfied in the `fast-string' regime where $\mathcal{J}\gg1$. From \eqref{eq:usefulproperty1}, the sign of the lower branch $(w_{r,-})^2$ is controlled by the sign of $r^2-4\mu^2$. The lowest non-trivial oscillator ($r=1$), results in the branch-independent stability condition $\vert\mu\vert<\tfrac{1}{2}$. At $|\mu|=\frac{1}{2}$, the lower branch becomes a zero-mode, while for $|\mu|>\frac{1}{2}$ it becomes unstable. On the purely fractional branch with $\widetilde{m}=0$, these three regimes correspond to, respectively, $n<\frac{L}{2}$, $n=\frac{L}{2}$ and $n>\frac{L}{2}$. The last two statements apply only when the corresponding values of $n$ exist, with $n=\frac{L}{2}$ requiring even $L$. They should not be interpreted as intrinsic properties of the twisted-sector label. Indeed, for $\frac{L}{2}<n\leq L-1$, the same twisted sector admits the branch with $\tilde{m}=-1$ (cf. \eqref{eq:mudefinition}) and $\mu=\frac{n-L}{L}$, for which $\vert\mu\vert<\frac{1}{2}$ and the solution is stable. Stability is therefore a property of the full winding branch $\mu=\widetilde m+\frac{n}{L}$, rather than of $n$ alone. For even $L$, the two minimal winding branches in the twisted sector with $n=\frac{L}{2}$ have $\mu=\pm\frac{1}{2}$ and are marginal.\footnote{The appearance of such stable subsectors of semiclassical states upon orbifolding is not a novel feature of this type IIB construction. In fact, it was already observed for M2-brane instantons wrapping $S^3/\mathbb{Z}_k\subset S^7/\mathbb{Z}_k$ in \cite{Beccaria:2023ujc}, where the freely acting $\mathbb{Z}_k$ projects out, for $k>2$, the modes responsible for the negative eigenvalues of the unquotiented saddle, modulo a separate treatment of the zero-modes. A closer analogue containing fixed-points is provided by the $(p,q)$ models of \cite{Imamura:2008nn}, whose semiclassical M2-brane instantons were analysed in \cite{Kurlyand:2026yke}. There, the orbifold projection similarly selects a distinguished semiclassical subsector of states, and, for $pk>2$, $qk>2$, removes the relevant cohomological zero modes, yielding isolated BPS saddles. The symmetric $p=q$ case was studied in the type IIA limit in \cite{Gautason:2023igo}.}
In the case of $\widetilde{m}=0$, the $J_1=J_2$ instability is therefore recovered for states in twisted sectors labelled by $n>\frac{L}{2}$.

The one-loop correction to the classical energy \eqref{eq:SU2classicalenergy}, is given by the oscillator sum
\begin{equation}\label{eq:oneloopstructure}
E_1=\frac{1}{2\kappa}\sum_{r\in\mathbb{Z}}\left(\omega_r-\Omega_r\right)\,,\qquad \omega_r\equiv\sum_{s}\omega_{r}^s\,,\qquad \Omega_r\equiv\sum_s\Omega_{r}^s\,,\qquad s\in\{1,2,\dots,8\}\,,
\end{equation}
where $\omega_r$ is the sum over bosonic frequencies \eqref{eq:S3CFs}, \eqref{eq:DecoupledCF} and \eqref{eq:AdSCF} and $\Omega_r$ denotes the sum over the eight fermionic constituents in \eqref{eq:fermionicCFs} \cite{Frolov:2004bh}. In the large $r$ regime, the values of $\omega_r$ and $\Omega_r$ cancel out in the one-loop sum (cf. \eqref{eq:largenbosonicsum} and \eqref{eq:largenfermionicfreq}), rendering $E_1$ UV finite. To compute the sum in \eqref{eq:oneloopstructure}, we follow \cite{Frolov:2002av, Frolov:2003tu, Frolov:2004bh, Schafer-Nameki:2006dtt}. The explicit form of $E_1$ is
\begin{equation}\begin{split}\label{eq:2spin1loopexactsum}
&E_1=2+\sqrt{1-\frac{2\mu^2}{\mathcal{J}^2+\mu^2}}-3\sqrt{1-\frac{\mu^2}{\mathcal{J}^2+\mu^2}}+\sum_{r=1}^{\infty}\mathcal{S}_r(\kappa,\mu)\,,\\
&\mathcal{S}_r=2\sqrt{1+\frac{\left(r+\sqrt{r^2-4\mu^2}\right)^2}{4\left(\mathcal{J}^2+\mu^2\right)}}+2\sqrt{1+\frac{r^2-2\mu^2}{\mathcal{J}^2+\mu^2}}+4\sqrt{1+\frac{r^2}{\mathcal{J}^2+\mu^2}}-8\sqrt{1+\frac{r^2-\mu^2}{\mathcal{J}^2+\mu^2}}\,,
\end{split}\end{equation}
where we made use of \eqref{eq:usefulproperty}. This is the orbifold analog of the one-loop sum derived in \cite{Beisert:2005mq}, with the effective replacement $\tilde{m}\to \mu$ of the windings. The sum over $r$ here is finite as a result of conformal invariance of the string theory sigma-model. In the `fast-string' regime (i.e. at large $\kappa$ with fixed $r$, see \eqref{conformalgaugeconstraints} at fixed $\mu$), one finds, to leading order\footnote{To extract the coefficient $d_1(\mu)=\lim_{\kappa\to\infty}\kappa^2E_1(\kappa,\mu)$, one is to check if the procedures of summing over $r$ and then expanding at large $\kappa$ with $r$ held fixed commute. For $\mu<\frac{1}{2}$, the leading large $\kappa$ coefficient of \eqref{eq:2spin1loopexactsum} is indeed uniformly summable in $r$ and agrees precisely with the results obtained via the analytic contour method of \cite{Schafer-Nameki:2006dtt}. Note that this, in general, need not apply to higher-order coefficients in the large $\kappa$ expansion of $E_1$.}
\begin{equation}\label{eq:oneloopcorrection}
    E_1=\frac{d_1\left(\mu\right)}{\kappa^2}\,,\qquad d_1(\mu)=\frac{1}{2}\left[\mu^2+\sum_{r=1}^{\infty}\bigg(r\sqrt{r^2-4\mu^2}-r^2+2\mu^2\bigg)\right]\,.
\end{equation}
At large $\kappa$ with $x=\frac{r}{\kappa}$ held fixed (see \eqref{eq:bosonicsumx} and \eqref{eq:fermionicsumx}), one observes that $\omega_r^B-\Omega_r^F\sim\order{\kappa^{-3}}$, so that there is no contribution to $d_1(\mu)$ from this regime of frequencies.

\begin{figure}[h]
\centering
\includegraphics[scale=0.4]{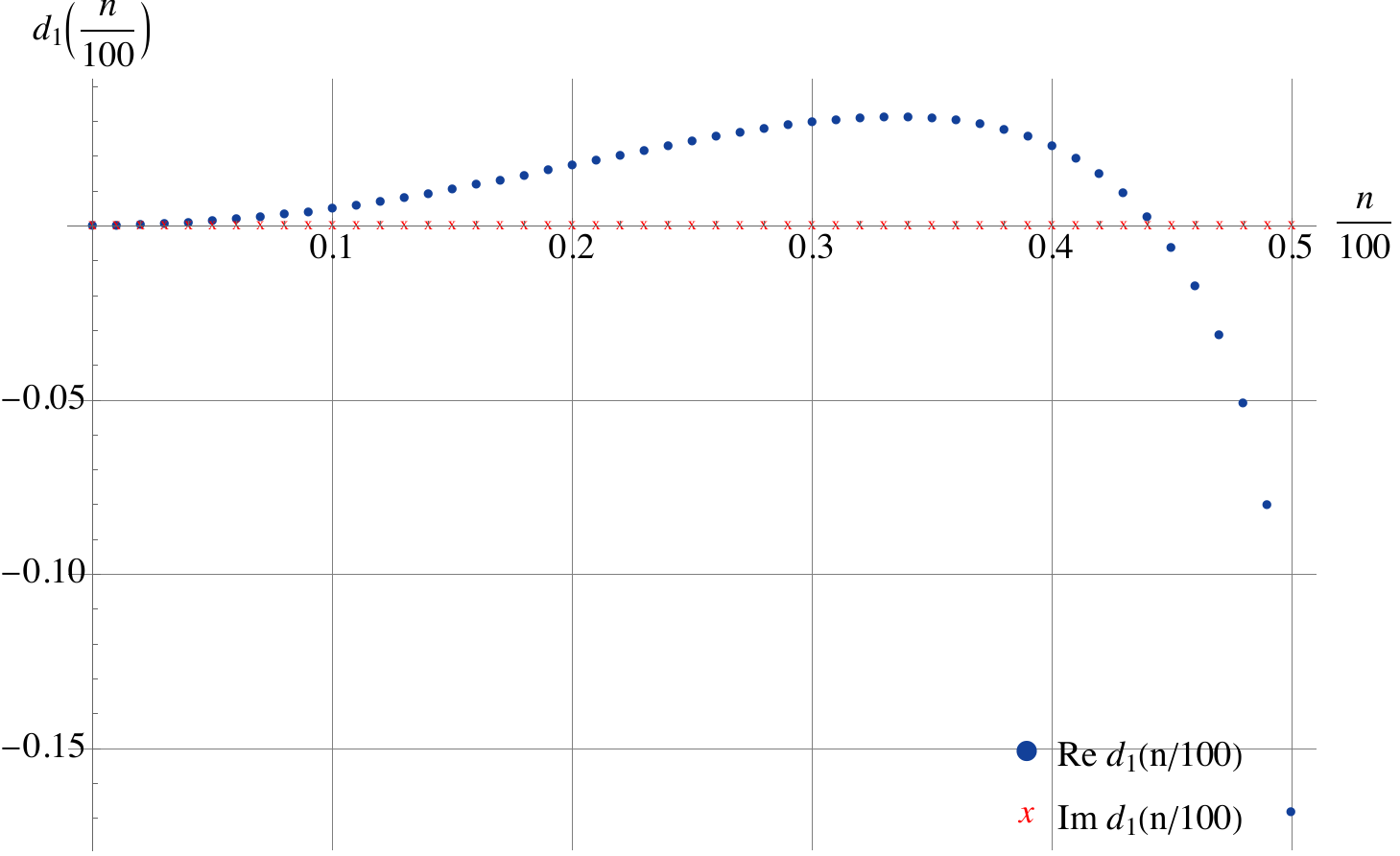}
\caption{Sample plot of numerical integration showing real (blue) and imaginary (red) parts of $d_1(\mu)$ for $L=100$ as a function of the (discrete) winding $0\leq\mu\leq\frac{1}{2}$ on the $\tilde{m}=0$ branch. The function $d_1(\mu)$ vanishes as $\mu\to 0$, where one approaches a near-BPS regime, and its continuous interpolation shows a second zero at the irrational value $\mu\simeq0.44298...$ While the latter root does not correspond to an allowed winding at fixed finite $L$, it has a physical consequence. The adjacent physical states at $n=44$ and $n=45$ lie on opposite sides of the root and thus have one-loop corrections of opposite sign. For $\mu>\frac{1}{2}$, the function $d_1(\mu)$ acquires a non-zero imaginary part, signaling an instability of the semiclassical solution.}
\label{fig:1}
\end{figure}

According to our stability analysis, the above expression is real for $\vert\mu\vert<\frac{1}{2}$. On the winding branch with $\widetilde{m}=0$, this is equivalent to $n<\frac{L}{2}$. In this stable winding regime, all the integer modes $r\geq 1$ remain real and the following integral representation for \eqref{eq:oneloopcorrection} in terms of Bessel functions $I_k(x)$ exists
\begin{equation}\label{eq:2spin1loopintegral}
d_1\left(\mu\right)=\mu^2\left[\frac{1}{2}-\int_{0}^{\infty}\frac{\mathrm{d}t}{t(e^t-1)}\left(I_0\left(2\mu t\right)+I_2\left(2\mu t\right)-\frac{I_1(2\mu t)}{\mu t}\right)\right]\,.
\end{equation}
We have not been able to evaluate the above integral into a simple closed-form expression, however, we can evaluate it numerically. Fixing $L=100$, the plot in Figure \ref{fig:1} shows the real and imaginary parts of $d_1$ as a function of the twist $\mu$.

One may also check that the one-loop correction vanishes in the `point-particle' limit. This limit may be probed either by going to the untwisted sector of the theory (by setting $n=0$ while keeping $L$ fixed), or by considering the regime where $L\to\infty$ with fixed $n$ \textit{not} of order $L$. In the former, the classical solution becomes BMN, since we are effectively undoing the windings \cite{Berenstein:2002jq}, while in the latter, given \eqref{eq:2spin1loopintegral}, one may verify that $\lim _{\mu\to 0}d_1(\mu)=0$ as expected. Thus, on the purely fractional branch with fixed $n\neq\order{L}$, one has $\mu=\frac{n}{L}\to0$, and the corresponding stable states smoothly connect to the BPS sector with $E=J$.
\subsection{$J_1=J_2$, $J_3\neq 0$ solution}
The bosonic (fermionic) characteristic frequencies correspond to the roots of the quartic \eqref{eq:3spinquartic} (the polynomial \eqref{eq:fermionpolynomial}), for which no useful closed-form expressions are available at arbitrary values of the parameters. It is possible, however, to extract their asymptotic expansions across different regimes.\footnote{To extract the large $r$ asymptotics, consider an ansatz of the form $w_r^s=\vert r\vert+A+\frac{B}{\vert r\vert}$ for the bosonic frequencies. Substituting into the quartic \eqref{eq:3spinquartic} and solving for $A$ and $B$ gives the frequencies \eqref{eq:3spinlarger}.} To determine new stable subsectors of the orbifold solution, we examine the expansion of the four bosonic $S^5/\mathbb{Z}_L$ frequencies in the `fast-string' regime, which read
\begin{equation}\begin{split}\label{eq:3spinblargek}
   &w_{r,\pm}^{(1)}=\frac{\vert r\vert}{2\kappa}\sqrt{r^2+2\mu^2(2-3\sin^2\gamma_0)\pm 2\vert\mu\vert\sqrt{4r^2\cos^2\gamma_0+\mu^2\sin^2\gamma_0(9\sin^2\gamma_0-8)}}+\dots\,,\\
   & w_{r,\pm}^{(2)}=2\kappa+\frac{r^2+\mu^2(2-5\sin^2\gamma_0)\pm\sqrt{\mu^2(4r^2\cos^2\gamma_0+\mu^2\sin^4\gamma_0)}}{2\kappa}+\dots\,.
\end{split}\end{equation}
The fermionic counterparts are given in \eqref{eq:3spinflargek}. The lighter branches $w_{r,\pm}^{(1)}$ are potentially dangerous regions for instabilities. Requiring positivity of $w_{r,-}^{(1)}$ for $r=1$ (the mode number imposing the strongest condition) is equivalent to\footnote{The inner square root of $w_{r,-}^{(1)}$ is never negative since for $r^2\geq 1$ and $\mu^2<1$, then $4r^2\cos^2\gamma_0+\mu^2\sin^2\gamma_0(9\sin^2\gamma_0-8)\geq 0$. The result thus follows from demanding positivity of the outer square root. Similarly, positivity of $w_{r,+}^{(1)}$ does not result in any further stability regimes.}
\begin{equation}\label{eq:3spindoublestability}
\left(1-4\mu^2\right)\bigg[1-4\mu^2\left(1-\sin^2\gamma_0\right)\bigg]\geq 0\,.
\end{equation}
For $\vert\mu\vert<\frac{1}{2}$, both the first and second factors are strictly positive for all values of $\sin^2\gamma_0\in(0,1)$, so the solution is stable. This stable subsector is common to the $J_1=J_2$ solution discussed previously, with $\vert\mu\vert=\frac{1}{2}$ being again marginal for even $L$. For $\vert\mu\vert>\frac{1}{2}$, the first factor of \eqref{eq:3spindoublestability} is negative, but the solution `re-stabilises' provided that
\begin{equation}\label{eq:3spinstabilitycond}
q\equiv \sin^2\gamma_0\leq 1-\frac{1}{4\mu^2}\,.
\end{equation}
Recalling \eqref{eq:3spinsinrelation}, the $J_1=J_2$, $J_3\neq 0$ solution on the orbifold is thus stable whenever
\begin{equation}\label{eq:3spinstability}
    \frac{J}{J_{\rm{tot}}}\leq \begin{cases}
    1\,,\qquad\quad\;\;\, \vert\mu\vert<\frac{1}{2}\\
    1-\frac{1}{4\mu^2}\,,\quad \vert\mu\vert>\frac{1}{2}\,,
    \end{cases}
\end{equation}
with the corresponding boundary values describing marginal modes. Interestingly, on the winding branch with $\tilde{m}=0$, choosing $n=L-k$ with $k=\order{1}$, gives, in the large $L$ regime, \(\frac{J}{J_{\rm{tot}}}\leq 1-\frac{L^2}{4(L-k)^2}\to \frac{3}{4}\), which coincides with the stability condition obtained for the analogous solution in smooth AdS$_5\cross S^5$ background \cite{Frolov:2003tu}.

Solving the characteristic equations \eqref{eq:3spinquartic} and \eqref{eq:fermionpolynomial} numerically, allows to compute the one-loop correction \eqref{eq:oneloopstructure} by adopting the method of \cite{Frolov:2004bh}. Given our large $r$ expressions in \eqref{eq:3spinblarger} and \eqref{eq:3spinflarger}, one can verify that here $E_1$ is also UV finite (i.e. the paired sum $\omega_r-\Omega_r$ is of order $\vert r\vert^{-3}$ and thus convergent at large $r$). Note that since the bosons and fermions are moded by the same integer $r$, there is no need to introduce a regulator once we know that the paired sum converges. Therefore, we need to evaluate
\begin{equation}\label{eq:oneloopSU3expression}
E_1=\frac{1}{2\kappa}\sum_{r\in\mathbb{Z}}\left(\omega_r-\Omega_r\right)=\frac{1}{2\kappa}\bigg[(\omega_0-\Omega_0)+2\sum_{r=1}^{\Lambda}(\omega_r-\Omega_r)\bigg]\,,
\end{equation}
where we have introduced a large upper limit $\Lambda$ required for the numerical evaluation, which, as in \cite{Frolov:2004bh}, we take to be $\Lambda=40000$. In a large $\kappa$ expansion, $\kappa^2E_1=d_1(q,\mu)+\kappa^{-2}e_2(q,\mu)$, so that for large enough $\kappa$, the value of $\kappa^2E_1$ will be very close to that of $d_1(q,\mu)$ provided one can show that $\frac{e_2}{\kappa^2}\ll d_1$. For $L=5$, we provide numerical results for the coefficient $d_1(q,\mu)$ for the different allowed values of $q$ and $n$ in Appendix \ref{sect:numerics}.

\begin{figure}[H]
\centering
\includegraphics[scale=0.38]{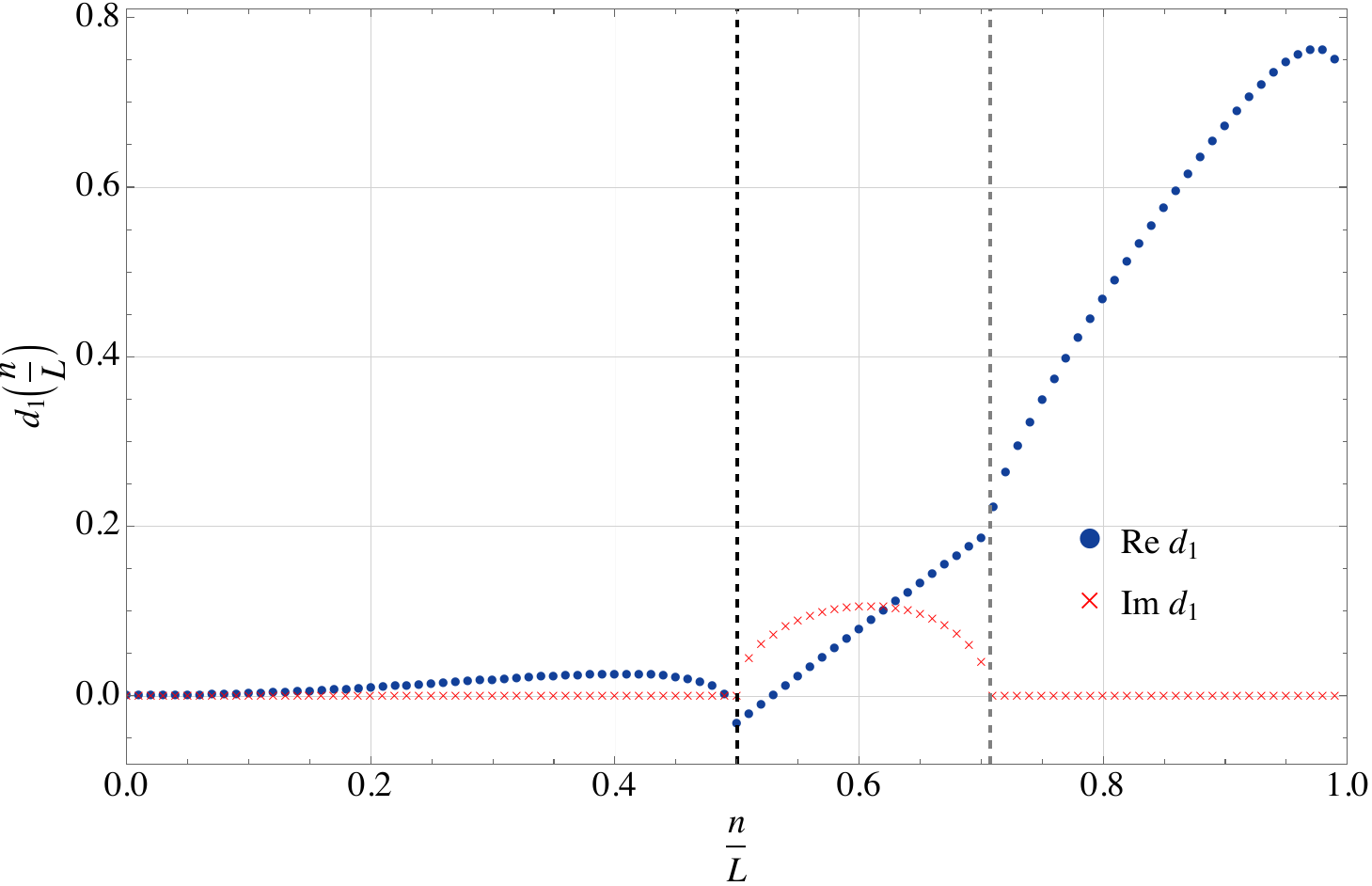}
\caption{Sample plot showing the values of the real and imaginary parts of the one-loop coefficient $d_1(\frac12,\mu)$ of the $J_1=J_2$, $J_3\neq 0$ solution \eqref{eq:3spinclassicalsoln} for $L=100$ and $q=\frac{1}{2}$. The black dashed line at $\mu=\frac{1}{2}$ marks the endpoint of the $\vert\mu\vert<\frac{1}{2}$ stable region on the $\tilde{m}=0$ branch. The grey dashed line is at $\mu=\frac{1}{\sqrt{2}}$, beyond which, for $q=\frac{1}{2}$, the solution re-stabilises. It features a non-zero imaginary part in the unstable `window' where $51\leq n\leq 70$.}
\label{fig:2}
\end{figure}

\begin{figure}[h]
\centering
\includegraphics[scale=0.38]{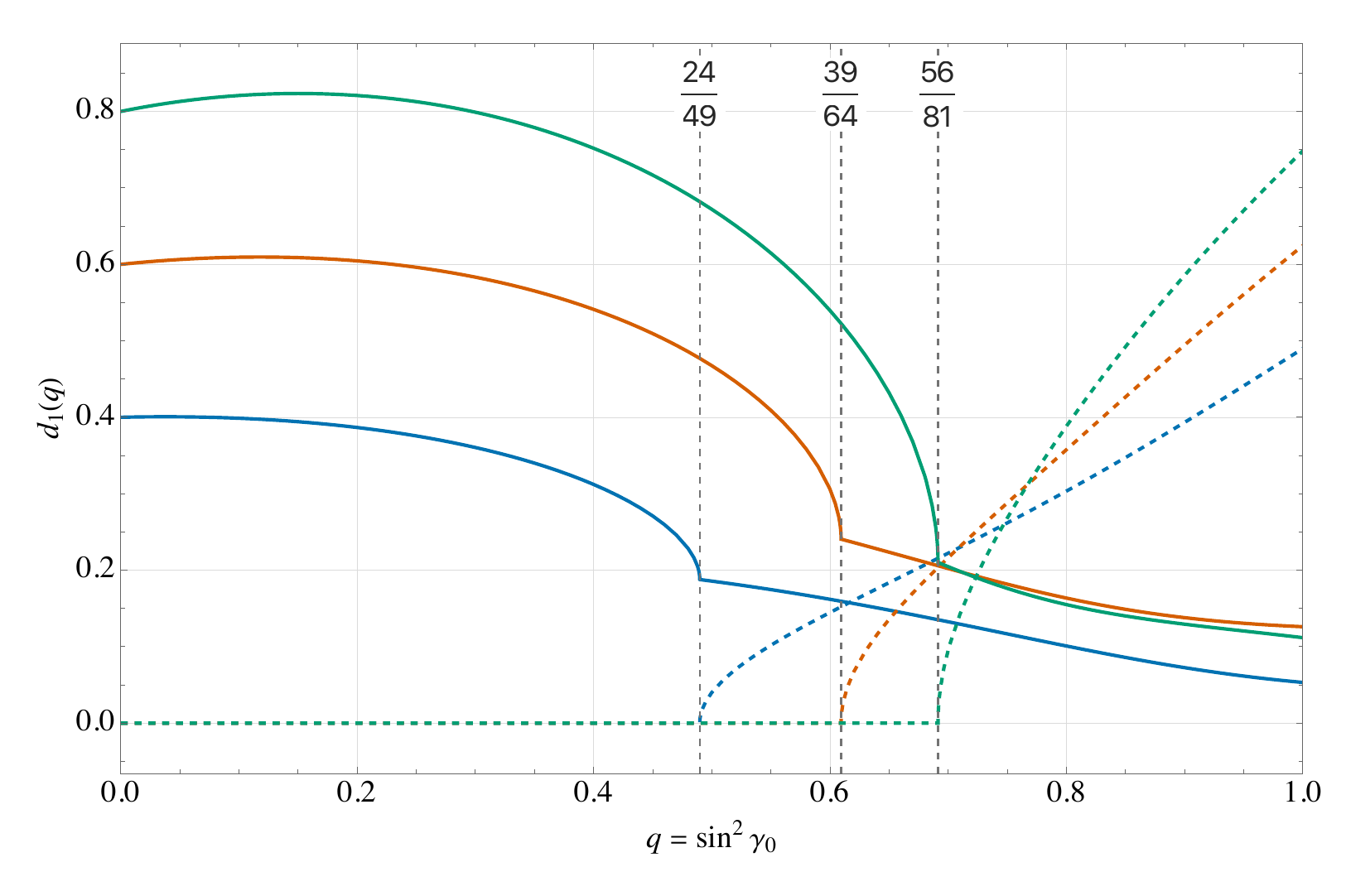}
\caption{Real (solid) and imaginary (dashed) parts of $d_{1}(q,\mu)$ for $L=100$ as a function of $q=\sin^2\gamma_0$. The sample values of $n=70,80,90$, colour-coded in blue, orange and green, respectively, are shown. The vertical dashed lines at $q\in\{\frac{24}{49},\frac{39}{64},\frac{56}{81}\}$ denote the values of $q$ beyond which the displayed winding branch ceases to be stable in the given twisted sector (cf. \eqref{eq:3spinstabilitycond}). All cases show the presence of a non-trivial stable region of states for $\frac{L}{2}<n\leq L-1$ whose range depends on the value of $q$.}
\label{fig:3}
\end{figure}

Let us now specialise to the purely fractional winding branch with $\tilde{m}=0$, where we fix $q=\frac{1}{2}$ with $L=100$, taking $\Lambda=40000$. From \eqref{eq:3spindoublestability}, we expect modes in twisted sectors with $n<50$ to be stable for the given value of $q$, with $n=50$ being marginal. We also expect modes with $\mu\geq\frac{1}{\sqrt{2}}$ (namely those in twisted sectors with $n>70$) to be stable, with an `unstable window' emerging for modes in the twisted sectors labelled by $51\leq n\leq 70$, where \eqref{eq:3spinstability} is not satisfied for the given value of $q$. A plot of $d_1\left(\frac{1}{2},\mu\right)$ in this case is shown in Figure \ref{fig:2}. Fixing $q=\frac{1}{12}$, the solution is unstable provided $\frac{1}{2}<\mu\leq\sqrt{\frac{3}{11}}$ (for $n\in\{51,52\}$), so the range of values of $n$ where the solution is unstable is reduced as $q$ decreases. The stable subsector of states with $n<\frac{L}{2}$ smoothly connects with the BPS sector with $E=J_{\rm{tot}}$ when $L\to\infty$ and fixed $n$ not of order $L$.
Given the stability condition \eqref{eq:3spinstabilitycond}, it is also useful to plot the dependence of the real and imaginary parts of $d_1(q,\mu)$ on $q=\sin^2\gamma_0$ for fixed $\mu\in\big(\frac{1}{2},\frac{L-1}{L}\big]$. For $L=100$, choosing sample values $n=70,80$ and $90$, this dependence is shown in Figure \ref{fig:3}.
\subsection{$SJ$ solution} 
\begin{figure}[h]
\centering
\includegraphics[scale=0.45]{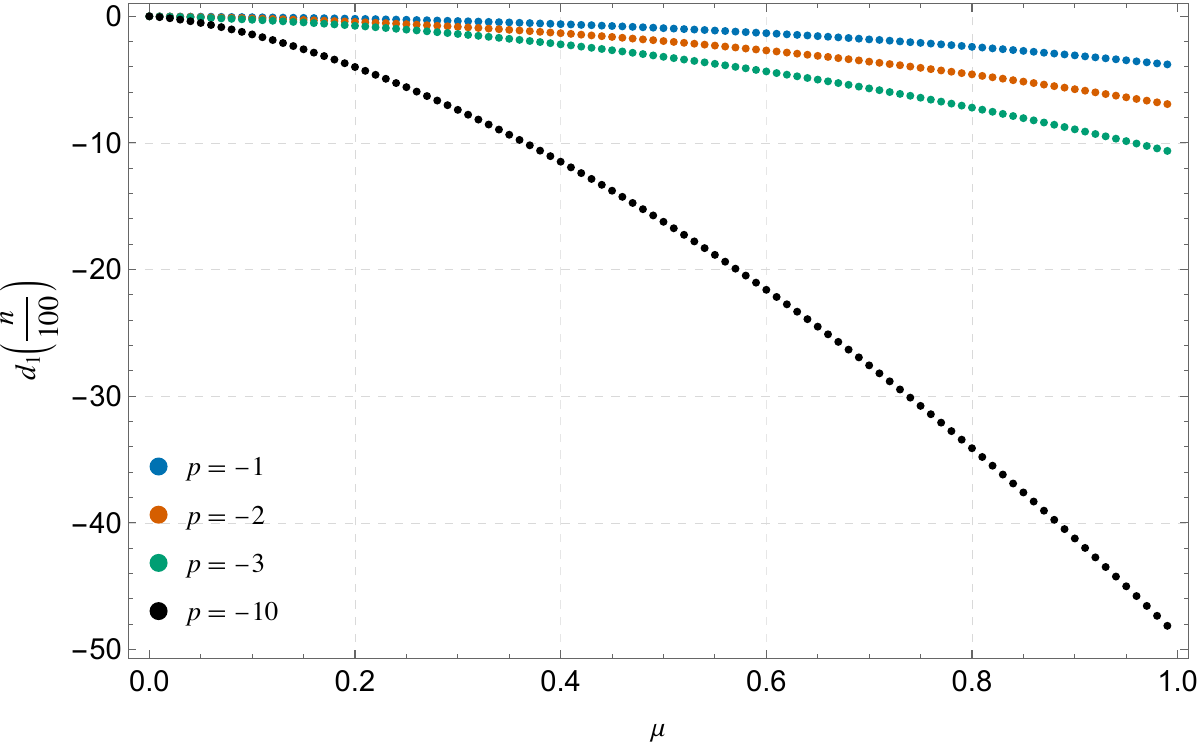}
\caption{Dependence on $\mu$ of \eqref{eq:SJd1coeff} for fixed $L=100$ and $p\in\{-1,-2,-3,-10\}$.}
\label{fig:4}
\end{figure}

In this case, the different bosonic and fermionic characteristic frequencies are obtained in \ref{sect:SJdetails}, with their different asymptotic behaviors detailed in \ref{sect:SJfrequencies}.
As before, we need to evaluate the sum \eqref{eq:oneloopstructure}. The zero-mode contribution is given by the sum of bosonic and fermionic frequencies at $r=0$, which are
\begin{equation}\begin{split}
&\omega_0=4\sqrt{\omega^2-\mu^2}+2\kappa+2\sqrt{\kappa^2+p^2\cosh^2\rho_0}\,,\\
&\Omega_0=8\left[\frac{p^2\kappa^2\sinh^22\rho_0}{4(\mu^2+p^2\sinh^2\rho_0)^{1/2}}+\frac{\mu^2\kappa^2}{\kappa^2+p^2}\left(\frac{\kappa^2+p^2-\omega^2}{\kappa^2+\mu^2-\omega^2}\right)^2\right]^{1/2}\,.
\end{split}\end{equation}
The rest of the one-loop correction involves the infinite sum
\begin{equation}\label{eq:SJE1oneloopsum}
\resizebox{\textwidth}{!}{$\displaystyle
\sum_{r=1}^{\infty}\left[
2\sqrt{r^2+\kappa^2}
+4\sqrt{r^2+\mathcal{J}^2-\mu^2}
+\frac{1}{2}\sum_{I=1}^4\operatorname{sign}(I)\omega_{I,r}^{\mathrm{AdS}_3}
-4\left(
\sqrt{(r+\mathrm{f}_2)^2+\mathrm{f}_1^2}
+\sqrt{(r-\mathrm{f}_2)^2+\mathrm{f}_1^2}
\right)
\right]\,,
$}
\end{equation}
where the quantities $\omega_{I,r}^{\rm{AdS}_3}$ refer to the roots of the quartic \eqref{eq:ads3quartic} and $\mathrm{f}_i$ are defined in \eqref{eq:fdefinition}.
One is then interested in the large $\mathcal{J}$ expansion of $E_1(\mathcal{J},\mathcal{S},p,\mu)$ with fixed $u=\frac{\mathcal{S}}{\mathcal{J}}$, $p$ and $\mu$, which may be probed, as argued in \cite{Park:2005ji}, by expanding the frequencies at large $\mathcal{J}$ and then performing the sum over $r$ at each order in $1/\mathcal{J}$. This may not be consistent in general, however, for the $1/\mathcal{J}^2$ term one finds the convergent sum
\begin{equation}
    E_1\to-\frac{1}{2\mathcal{J}^2}\left[\mu(\mu-p)+\sum_{r=1}^{\infty}\left(r^2+2\mu(\mu-p)-r\sqrt{r^2+4\mu(\mu-p)}\right)\right]+\order{\mathcal{J}^{-4}}\,,
\end{equation}
so that $d_1(p,\mu)$ in this case is given by
\begin{equation}\begin{split}\label{eq:SJd1coeff}
 d_1(p,\mu)&=\frac{\mu(p-\mu)}{2}-\frac{1}{2}\sum_{r=1}^{\infty}\left(r^2+2\mu(\mu-p)-r\sqrt{r^2+4\mu(\mu-p)}\right)\\
 &=-\frac{1}{2}\int_{0}^{2\sqrt{\mu(\mu-p)}}\dd t\, t\coth(\pi t)\sqrt{4\mu(\mu-p)-t^2}\,,
\end{split}\end{equation}
where the integral representation holds for $\mu\in[0,1)$. Introducing the cut-off $\Lambda=40000$, the dependence of $d_1(p,\mu)$ on $\mu$ is shown for fixed $L=100$ and several values of $p$, in Figure \ref{fig:4}. As in AdS$_5\cross S^5$, the solution is stable for all values of the winding \cite{Park:2005ji}. New stable twisted semiclassical states thus occur for purely fractional windings $\mu$, whose energies smoothly connect to those of BPS states with $E=J+S$ in the near-BMN ($\mu\to 0$) regime.

\section{Energy correction from Landau-Lifshitz model}\label{sect:4}
The LL sigma-model action describes the low-energy spectrum of the ferromagnetic Heisenberg spin-chain \cite{Kruczenski:2003gt, Fradkin:2013anc}. From the gauge theory side, this action may be derived from the one-loop dilatation operators in each of the subsectors of interest of the $\mathcal{N}=2$ quiver theory, which may be regarded as Hamiltonians for three nearest-neighbour spin-chains, and may be represented by path integrals over coherent states. One is then to assign a local coherent state to every site of the chain, parametrising the corresponding target space of the model. For a specific class of low-energy configurations for which the coherent-state variables vary slowly across adjacent sites, one may define a `semiclassical' limit of the path integral as $J\to\infty$ with fixed $\frac{\lambda}{J^2}$. Taking the continuum limit, each site label is rescaled to a continuous coordinate on $S^1$, the lattice sum becomes an integral and nearest-neighbour differences reduce to spatial derivatives.

Another way to obtain the LL effective action is from the bosonic part of the Green-Schwarz string action by first isolating a `fast' coordinate, fixing static gauge, and expanding in derivatives of the `slow' coordinates. While this effective theory includes only a subset of the modes present in the string theory result, quantising the fluctuations around the corresponding classical solution of this model allows to exactly capture the leading quantum correction to the classical energy of the semiclassical strings described in Section \ref{sec: 2}  \cite{Minahan:2005mx, Beisert:2005he}.
\subsection{$\mathrm{SU}(2)$ sector}
The LL effective action arising from the continuum coherent state limit of the orbifolded SU$(2)$ spin-chain was discussed in \cite{Astolfi:2008yw}. Here, we obtain it from \eqref{bosonicsigmamodel} by splitting the two complex coordinates of $S^3/\mathbb{Z}_{L}$ into a `fast' coordinate $\Phi$ associated to $J$, and two `slow' coordinates $U_{i}$, determining the `transverse' profile of the string. This amounts to the reparametrisation
\begin{equation}
    \mathrm{X}=U_1e^{i\Phi}\,,\qquad \mathrm{Y}=U_2e^{i\Phi}\,,\qquad U_iU^{*}_i=1\,,
\end{equation}
where $U_1=\cos\psi \,e^{i\frac{\chi}{L}}$ and $U_2=\sin\psi\,e^{-i\frac{\chi}{L}}$. One then finds that
\begin{equation}
    \dd \mathrm{X}\dd \mathrm{X}^{*}+\dd \mathrm{Y}\dd \mathrm{Y}^*=(\dd\Phi+C)^2+DU_iDU_i^*\,,\;\; C=-iU_i^*\dd U_i=\frac{\cos2\psi}{L}\dd\chi\,,\quad DU_i=\dd U_i-iCU_i\,.
\end{equation}
Defining $U=(U_1,U_2)^T$, such that $U^{\dagger}U=1$, and introducing the $S^2\simeq \mathbb{CP}^1$ unit vector $\vec{n}=U^{\dagger}\vec{\sigma} U=\left(\sin2\psi\cos\frac{2\chi}{L},-\sin2\psi\sin\frac{2\chi}{L},\cos2\psi\right)$ where $\psi\in[0,\frac{\pi}{2}]$, $\chi\in[0,2\pi)$, and $\vec{\sigma}=\sigma_i$ are the Pauli matrices, the metric on $S^3/\mathbb{Z}_{L}$ (cf. \eqref{orbifoldmetric}) can be written as
\begin{equation}\label{eq:S3ZLmetricSU2LL}
    \dd s^2_{S^3/\mathbb{Z}_L}=(\dd\Phi+C)^2+\frac{1}{4}\dd\vec{n}^2\,,
\end{equation}
i.e. as a `fast' $S^1$ fibration parametrised by $\Phi$, over a `slow' $S^2$ base. Locally, this is the same Hopf decomposition as for the round $S^3$, with the orbifold action appearing at the level global identifications of the $S^2$ angle $\frac{2\chi}{L}\in[0,\frac{4\pi}{L})$ \cite{Astolfi:2008yw}. As a result, the local LL Lagrangian is unchanged, but the allowed closed string configurations can include purely fractional windings $\mu$. One then obtains\footnote{The Lagrangian density is equivalent to $\mathcal{L}=-iU_i^*\partial_0 U_i-\frac{\lambda}{2J^2}\vert D_{1}U_i\vert^2+\order{\lambda^2/J^4}$, to leading order. To see this, one assumes that evolution of $U_i$ in $t$ is slow, that is, that $\partial_0U_i\sim\order{\lambda/J^2}$, with $\partial_1U_i\sim\order{1}$.}
\begin{equation}\begin{split}\label{eq:SU2LLaction}
S_{\rm{LL}}&=\frac{J}{2\pi}\int\dd t\int_{0}^{2\pi}\dd\sigma\left[\vec{C}(\vec{n})\cdot\partial_0\vec{n}-\frac{\lambda}{8J^2}(\partial_1\vec{n})^2\right]+\order{\lambda^2/J^4}\\
&=\frac{J}{2\pi}\int\dd t\int_{0}^{2\pi}\dd\sigma\,\left[\frac{2\rho(\partial_0\chi)}{L}-\frac{\lambda}{2J^2}\left(\frac{(\partial_1\rho)^2}{1-4\rho^2}+\frac{1-4\rho^2}{L^2}(\partial_1\chi)^2\right)\right] +\order{\lambda^2/J^4}\,,
\end{split}\end{equation}
where in the second line we defined the coordinate $\rho\equiv\frac{\cos2\psi}{2}\in[-\frac{1}{2},\frac{1}{2}]$, which is useful to expand around classical solutions with $\psi\neq 0$ \cite{Minahan:2005mx}. The $J_1=J_2$ solution \eqref{classicalsoln} corresponds to the classical LL configuration where $ \chi=L\mu\sigma\,,\,\rho=0$. Its classical energy may be computed from the leading LL Hamiltonian
\begin{equation}
    \mathcal{H}_{\rm{LL}}=\frac{\lambda}{2J}\int_{0}^{2\pi}\frac{\dd\sigma}{2\pi}\left[\frac{(\partial_1\rho)^2}{1-4\rho^2}+\frac{1-4\rho^2}{L^2}(\partial_1\chi)^2\right]=\frac{\lambda\mu^2}{2J}=\frac{\lambda}{J}c_1\,,
\end{equation}
which agrees with the $\order{\lambda/J}$ term in \eqref{eq:SU2classicalenergy}. Considering periodic fluctuations
\begin{equation}
    \chi\to L\mu\sigma+\frac{L\tilde{\chi}(t,\sigma)}{\sqrt{J}}\,,\qquad\qquad \rho\to0+\frac{\tilde{\rho}(t,\sigma)}{\sqrt{J}}\,,
\end{equation}
around the classical LL solution, results in the quadratic action
\begin{equation}
\resizebox{\textwidth}{!}{$\displaystyle
S^{(2)}_{\mathrm{LL}}
=
\int\dd t\int_{0}^{2\pi}\frac{\dd\sigma}{2\pi}
\left(
2\tilde{\rho}\partial_0\tilde{\chi}
-\mathcal{H}_{\mathrm{LL}}^{(2)}\right)
=
\int\dd t\int_{0}^{2\pi}\frac{\dd\sigma}{2\pi}
\left[
2\tilde{\rho}\partial_0\tilde{\chi}
-\frac{\lambda}{2J^2}
\left(
(\partial_1\tilde{\chi})^2
+(\partial_1\tilde{\rho})^2
-4\mu^2\tilde{\rho}^2
\right)
\right]
\,,
$}
\end{equation}
from which the following linearised equations of motion
\begin{equation}
    \partial_0\tilde{\rho}-\frac{\lambda}{2J^2}\partial_1^2\tilde{\chi}=0\,,\qquad \partial_0\tilde{\chi}+\frac{\lambda}{2J^2}\left(\partial_1^2\tilde{\rho}+4\mu^2\tilde{\rho}\right)=0\,,
\end{equation}
follow. Diagonalising them, we find the following characteristic frequencies 
\begin{equation}
    \omega_r=\frac{\lambda}{2J^2}r\sqrt{r^2-4\mu^2}\,,
\end{equation}
which, as before, are real provided that $\vert\mu\vert<\frac{1}{2}$. Imposing the commutation relations $[\tilde{\chi}(t,\sigma),\tilde{\chi}(t,\sigma')]=0=[\tilde{\rho}(t,\sigma),\tilde{\rho}(t,\sigma')]$ and $[\tilde{\chi}(t,\sigma),\tilde{\rho}(t,\sigma')]=i\pi\delta(\sigma-\sigma')-\frac{i}{2}$, where the zero-mode $r=0$ has been removed, the correction to the classical LL energy is
\begin{equation}
E_1=\frac{1}{2}\sum_{r=-\infty}^{\infty}\vert\omega_r\vert=\frac{\lambda}{2J^2}\sum_{r=1}^{\infty}r\sqrt{r^2-4\mu^2}\,,\qquad r\neq0\,,
\end{equation}
which diverges. As pointed out in \cite{Beisert:2005mq} however, one may use $\zeta$-function regularisation to generate the following finite result\footnote{At large $r$, the energy correction behaves like $E_1=\frac{\lambda}{2J^2}(r^2-2\mu^2+\order{r^{-2}})$. The regularisation procedure involves subtracting and adding the divergent asymptotic terms of the sum and then using the fact that $\sum_r r^2=\zeta(-2)=0$ and $\sum_r 1=\zeta(0)=-\frac{1}{2}$. This then gives that $\sum_r(r^2-2\mu^2)=\mu^2$, leading to the finite result above.}
\begin{equation}\begin{split}\label{eq:SU2LLcorrection}
    E_{1,\rm{LL}}^{\rm{SU}(2)}&=\frac{\lambda}{2J^2}\left[\mu^2+\sum_{r=1}^{\infty}\left(r\sqrt{r^2-4\mu^2}-r^2+2\mu^2\right)\right]\\
    &=\frac{\lambda}{J^2}\mu^2\left[\frac{1}{2}-\int_{0}^{\infty}\frac{\mathrm{d}t}{t(e^t-1)}\left(I_0\left(2\mu t\right)+I_2\left(2\mu t\right)-\frac{I_1(2\mu t)}{\mu t}\right)\right]\,,
\end{split}\end{equation}
exactly matching \eqref{eq:2spin1loopintegral}.
\subsection{$\mathrm{SU}(3)$ sector}
The solution \eqref{eq:3spinclassicalsoln} propagates non-trivially on the subspace $\mathbb{R}_{t}\cross S^5/\mathbb{Z}_{L}$, so we reparametrise
\begin{equation}
\mathrm{X}=U_1e^{i\Phi}\,,\qquad  \mathrm{Y}=U_2e^{i\Phi}\,,\qquad \mathrm{Z}=U_3e^{i\Phi}\,,\qquad   \sum_{i=1}^3\vert U_i\vert^2=1\,,
\end{equation}
where \(U_1=\sin\gamma\cos\psi\,e^{i\frac{\chi}{L}}\,,\, U_2=\sin\gamma\sin\psi\,e^{-i\frac{\chi}{L}}\,,\, U_3=\cos\gamma e^{i(\varphi_3-\Phi)}\), denote the `slow' coordinates \cite{Hernandez:2004uw, Stefanski:2004cw}. Defining the vector $U=(U_1,U_2,U_3)^T$, the $S^5/\mathbb{Z}_{L}$ metric reads
\begin{equation}\begin{split}
&\dd \mathrm{X}\dd \mathrm{X}^*+\dd \mathrm{Y}\dd \mathrm{Y}^*+\dd \mathrm{Z}\dd \mathrm{Z}^*=(\dd\Phi+C)^2+DU^{\dagger}DU\,,\; C=\sin^2\gamma\cos2\psi\dd\bar{\chi}+\cos^2\gamma\,(\dd\varphi_3-\dd\Phi)\,,\\
&DU^{\dagger}DU=\dd\gamma^2+\sin^2\gamma\left(\dd\psi^2+
\sin^22\psi\dd\bar{\chi}^2\right)+\sin^2\gamma\cos^2\gamma\big(\dd(\varphi_3-\Phi)-\cos2\psi\dd\bar{\chi}\big)^2.
\end{split}\end{equation}
The last expression is, locally, the usual Fubini-Study metric on $\mathbb{CP}^2$ with the modification that the orbifold action implies the global identification $\bar{\chi}\in[0,\frac{2\pi}{L})$. Therefore, we again expect that the local LL Lagrangian will be the same as in the unorbifolded case, modulo the modified periodicity of $\bar{\chi}$.

At leading order, we thus find
\begin{equation}\begin{split}\label{eq:SU3LLaction}
S_{\rm{LL}}&=J_{\rm{tot}}\int\dd t\int_0^{2\pi}\frac{\dd\sigma}{2\pi}\left[C_0-\frac{\lambda}{2J_{\rm{tot}}^2}D_1U^{\dagger}D_1U\right]\\
&=\frac{J_{\rm{tot}}}{2\pi}\int\dd t\int_0^{2\pi}\dd\sigma\bigg\{\frac{\sin^2\gamma\cos2\psi}{L}\partial_0\chi+\cos^2\gamma\,\partial_0(\varphi_3-\Phi)-\frac{\lambda}{2J_{\rm{tot}}^2}\bigg[(\partial_1\gamma)^2\\
&+\sin^2\gamma\left((\partial_1\psi)^2+\frac{\sin^22\psi}{L^2}(\partial_1\chi)^2\right)+\sin^2\gamma\cos^2\gamma\left(\partial_1(\varphi_3-\Phi)-\frac{\cos2\psi}{L}\partial_1\chi\right)^2\bigg]\bigg\}\,.
\end{split}\end{equation}
The classical LL configuration corresponding to \eqref{eq:3spinclassicalsoln} is $ \gamma=\gamma_0\,,\; \psi=\frac{\pi}{4}\,,\; \chi=L\mu\sigma$ and $\varphi_3-\Phi=-\frac{\lambda}{2J_{\rm{tot}}^2}\mu^2t$,\footnote{Note that $\varphi_3-\Phi=\frac{\mathrm{k}-\omega}{\kappa}t=-\frac{\mu^2}{2\mathrm{k}^2}t+\order{\mathrm{k}^{-4}}=-\frac{\lambda\mu^2}{2J_{\rm{tot}}^2}t+\dots$ at large $J_{\rm{tot}}$.}
and its energy is directly computed from the leading Hamiltonian
\begin{equation}
    \mathcal{H}_{\rm{LL}}=\frac{\lambda}{4\pi J_{\rm{tot}}}\int_0^{2\pi}\dd\sigma\,D_1U^{\dagger}D_1U=\frac{\lambda}{2J_{\rm{tot}}}\mu^2\sin^2\gamma_0=\frac{\lambda\mu^2J}{2J_{\rm{tot}}^2}\,,
\end{equation}
reproducing the leading large $J_{\rm{tot}}$ term in the classical energy expansion \eqref{eq:3spinclassicalenergy}.

Considering periodic fluctuations around the classical LL solution of the form
\begin{equation}
    \gamma\to\gamma_0+\frac{\tilde{\gamma}(t,\sigma)}{\sqrt{J_{\rm{tot}}}}\,,\quad \psi\to\frac{\pi}{4}+\frac{\tilde{\psi}(t,\sigma)}{\sqrt{J_{\rm{tot}}}}\,,\quad \chi\to L\mu\sigma+\frac{L\tilde{\chi}(t,\sigma)}{\sqrt{J_{\rm{tot}}}}\,,\quad \varphi_3-\Phi\to-\frac{\lambda\mu^2t}{2J_{\rm{tot}}^2}+\frac{h(t,\sigma)}{\sqrt{J_{\rm{tot}}}}\,,
\end{equation}
one may then compute the quadratic fluctuation action, which reads
\begin{equation}\begin{split}
S_{\rm{LL}}^{(2)}&=\int\dd t\int_0^{2\pi}\frac{\dd\sigma}{2\pi}\bigg[-2s_0c_0\tilde{\gamma}\partial_0h-2s_0^2\tilde{\psi}\partial_0\tilde\chi-\frac{\lambda}{2J_{\rm{tot}}^2}\bigg((\partial_1\tilde\gamma)^2+s_0^2(\partial_1\tilde\psi)^2+s_0^2(\partial_1\tilde\chi)^2\\
&\qquad\qquad\qquad\qquad\qquad +s_0^2c_0^2(\partial_1h)^2+4\mu s_0c_0\tilde\gamma\,(\partial_1\tilde\chi)+4\mu s_0^2c_0^2\tilde\psi\,(\partial_1h)-4\mu^2s_0^4\tilde\psi^2\bigg)\bigg]\,,
\end{split}\end{equation}
where $s_0\equiv\sin\gamma_0$ and $c_0\equiv\cos\gamma_0$. 

Upon diagonalisation of the linearised equations of motion, one obtains the two (positive) characteristic frequencies
\begin{equation}
\omega_{r,\pm}=\frac{\lambda}{2J_{\rm{tot}}^2}\mathcal{W}_{r,\pm}\equiv\frac{\lambda}{2J_{\rm{tot}}^2}\vert r\vert\sqrt{r^2+2\mu^2(2-3s_0^2)\pm 2\vert\mu\vert\sqrt{4r^2c_0^2+\mu^2s_0^2(9s_0^2-8)}}\,,
\end{equation}
which match the light branches $w_{r,\pm}^{(1)}$ in \eqref{eq:3spinblargek}, computed in the `fast-string' regime.

After regularising the divergent asymptotic large $r$ terms by $\zeta$-function regularisation, the leading quantum correction to the classical LL energy is given by
\begin{equation}\label{eq:SU3LLcorrection}
    E_{1,\rm{LL}}^{\rm{SU}(3)}=\sum_{r=1}^{\infty}(\omega_{r,+}+\omega_{r,-})\to \frac{\lambda}{J_{\rm{tot}}^2}\left[\frac{\mu^2s_0^2}{2}+\frac{1}{2}\sum_{r=1}^{\infty}\bigg(\mathcal{W}_{r,+}+\mathcal{W}_{r,-}-2(r^2-\mu^2s_0^2)\bigg)\right]\,.
\end{equation}
This is the orbifold analog of the SU$(3)$ expression found in \cite{Beisert:2005mq}, which matches \eqref{eq:oneloopSU3expression} in the `fast-string' regime at leading order in large $\kappa$.
\subsection{$\mathrm{SL}(2)$ sector}
We now adapt the constructions of \cite{Stefanski:2004cw, Park:2005ji} to the solution \eqref{eq:orbifoldSJsolution}. This case is structurally different to the previous two, since the dependence on $L$ now enters via the `fast' $S^1_{\rm{X}}$ coordinate, as opposed to along an LL target coordinate. The local LL sigma-model action will be identical to that obtained from the analogous $SJ$ solution on AdS$_5\cross S^5$, having a non-compact $\frac{\mathrm{SU}(1,1)}{\mathrm{U}(1)}$ target space, with the orbifold dependence entering through \eqref{eq:orbifoldSJvirasoro}.

We reparametrise the solution as
\begin{equation}
 \mathrm{X}=e^{i\delta}\,,\quad \mathrm{W}\equiv \mathrm{Y}_0=V_0 e^{it}\,,\quad \mathrm{Y
}\equiv\mathrm{Y}_1=V_1e^{it}\,,\quad V_i^{*}V_j\beta^{ij}=-1\,,\quad \beta^{ij}=\mathrm{diag}(-1,1)\,,
\end{equation}
where $\delta=\Phi+\frac{\chi}{L}$, $V_0=\cosh\rho$, $V_1=\sinh\rho\,e^{i\eta}$ and $\eta=\varphi-t=p\sigma+(\frac{w}{\kappa}-1)t$ (cf. \eqref{eq:orbifoldSJsolution}). The AdS$_3\cross S^1_{\rm{X}}$ metric then reads
\begin{equation}
    \dd s^2_{\rm{AdS}_{3}\cross S^1}=\dd\delta^2-(\dd t+B)^2+DV_i^*DV^{i}\,,\quad B=iV_i^*\dd V^{i}=-\sinh^2\rho\,\dd\eta\,,\quad DV_i=\dd V_i-iBV_i\,.
\end{equation}
The effective action for the `slow' transverse coordinates $V_i$ reads\footnote{The Lagrangian density may be written as $\mathcal{L}_{\rm{LL}}=-iV_i^{*}\partial_0V^i-\frac{\lambda}{2J^2}D_1V_i^{*}D_1V^i$.}
\begin{equation}\begin{split}\label{eq:LLSL2action}
    S_{\rm{LL}}&=\frac{J}{2\pi}\int\dd t\int_{0}^{2\pi}\dd\sigma\left[\partial_0\eta\sinh^2\rho-\frac{\lambda}{2J^2}\left((\partial_1\rho)^2+(\partial_1\eta)^2\sinh^2\rho\cosh^2\rho\right)\right]+\order{\lambda^2/J^4}\\
    &=\frac{J}{2\pi}\int\dd t\int_0^{2\pi}\dd\sigma\left[v\,\partial_0\eta-\frac{\lambda}{2J^2}\left(\frac{(\partial_1v)^2}{4v(1+v)}+v(1+v)(\partial_1\eta)^2\right)\right]+\order{\lambda^2/J^4}\,,
\end{split}\end{equation}
where, in the last equality, we have rewritten the action in terms of the `canonical' variable $v\equiv\sinh^2\rho\in[0,\infty)$. Let us define $V\equiv (V_0,V_1)^T\in\mathbb{C}^{1,1}$, so that $V^{\dagger}\beta V=-1$. The non-compact target space of this sigma-model is simply $\{V\in\mathbb{C}^{1,1}\,:\,V^{\dagger}\beta V=-1\}/\mathrm{U}(1)=\frac{\mathrm{SU}(1,1)}{U(1)}$.\footnote{In terms of $(v,\eta)$, the metric on $\mathrm{SU}(1,1)/\mathrm{U}(1)$ can be written as $\dd s^2=\frac{\dd v^2}{4v(1+v)}+v(1+v)\dd\eta^2$, where $v\in[0,\infty)$ and $\eta\in[0,2\pi)$. A re-definition $z= e^{i\eta}\sqrt{\frac{v}{1+v}}$ allows to rewrite it in usual `disk' coordinates, where $\dd s^2=\frac{\dd z\dd\bar{z}}{(1-\vert z\vert^2)^2}$.} The SL$(2)$ solution corresponds to a classical LL configuration given by $\rho=\rho_0$ and $\eta=\eta_0=p\sigma+\frac{\lambda}{2J^2}p^2(1+2u)t$, and its classical energy may be computed from the leading LL integrated Hamiltonian
\begin{equation}
    \mathcal{H}_{\rm{LL}}=\frac{\lambda}{2J}\int_0^{2\pi}\frac{\dd\sigma}{2\pi}\bigg[(\partial_1\rho)^2+(\partial_1\eta)^2\sinh^2\rho\cosh^2\rho\bigg]=\frac{\lambda}{2J}p^2u(1+u)=\frac{\lambda}{2J}\frac{\mu^2(1+u)}{u}\,,
\end{equation}
where the final equality follows after imposing \eqref{eq:orbifoldSJvirasoro}. As expected, the classical LL energy correctly reproduces the corresponding term in the expansion \eqref{eq:classicalenergySJ}. 

We now want to study the periodic fluctuations
\begin{equation}
    \eta\to\eta_{\rm{0}}+\frac{\tilde{\eta}(t,\sigma)}{\sqrt{J}}\,,\qquad v\to u+\frac{\tilde{v}(t,\sigma)}{\sqrt{J}}\,
\end{equation}
around the LL classical solution described above. Expanding \eqref{eq:LLSL2action} to quadratic order gives
\begin{equation}\label{eq:SL2LLquadraticaction}
\resizebox{\textwidth}{!}{$\displaystyle
    S_{\rm{LL}}^{(2)}=\int\dd t\int_{0}^{2\pi}\frac{\dd\sigma}{2\pi}\left[\tilde{v}\dot{\tilde{\eta}}-\frac{\lambda}{2J^2}\left(u(1+u)(\tilde{\eta}')^2+\frac{(\tilde{v}')^2}{4u(1+u)}+2p(1+2u)\tilde{v}\tilde{\eta}'+p^2\tilde{v}^2\right)\right]\,.
    $}
\end{equation}
Once again, diagonalising the corresponding linearised equations of motion that follow from \eqref{eq:SL2LLquadraticaction}, one finds the following characteristic frequencies
\begin{equation}
    \omega_r^{\pm}=\frac{\lambda}{J^2}\left[(p-2\mu)r\pm\frac{\vert r\vert}{2}\sqrt{r^2+4\mu(\mu-p)}\right]\,,
\end{equation}
which are always real since $\mu(\mu-p)>0$. The correction to the classical energy is
\begin{equation}\label{eq:SL2LLcorrection}
\resizebox{\textwidth}{!}{$\displaystyle
\begin{aligned}
E_1
=
\frac{\lambda}{2J^2}
\sum_{r=1}^{\infty}r\sqrt{r^2+4\mu(\mu-p)}
&\to
\frac{\lambda}{2J^2}
\bigg[
\mu(p-\mu)
+
\sum_{r=1}^{\infty}
\left(
r\sqrt{r^2+4\mu(\mu-p)}
-r^2
-2\mu(\mu-p)
\right)
\bigg]
\\
&=
-\frac{\lambda}{2J^2}
\int_0^{2\sqrt{\mu(\mu-p)}}
\dd t\,t\coth(\pi t)
\sqrt{4\mu(\mu-p)-t^2}\,.
\end{aligned}
$}
\end{equation}
where the divergent sum has again been regularised by $\zeta$-function regularisation. The above matches the correction coefficient \eqref{eq:SJd1coeff}, corresponding to the leading part of the large $J$, fixed $u$ string one-loop energy.
\section{Finite-size corrections from orbifolded Bethe ansatz}\label{sect:5}
In the closed bosonic subsectors of interest, the local one-loop dilatation operators are the same as their $\mathcal N=4$ counterparts. The corresponding spin-chains differ globally through twisted boundary conditions to account for the $\mathbb{Z}_L$ projection \cite{Beisert:2005he, Ideguchi2004, Skrzypek:2022cgg, Pomoni:2019oib}, the orbifold condition \eqref{eq:angmomentaprojection} and twisted momentum constraints. Furthermore, the Bethe `vacua' may be auxiliary reference states rather than nonzero gauge theory operators. At the level of the Bethe equations in the continuum $J\to\infty$ limit, these effects amount to an effective replacement of certain integers by fractional quantities depending on $\mu$. 

At one-loop, the spectrum of anomalous dimensions for the SU$(2)$ sector operators \eqref{eq:stateopmap}, dual to semiclassical strings \eqref{classicalsoln}, is governed by the solutions to the twisted Bethe equations
\begin{equation}\label{eq:twistedSU2betheeqn}
 \xi^{-2n}\left(\frac{u_k+i/2}{u_k-i/2}\right)^J=\prod_{\substack{j=1 \\ j\neq k}}^{J/2}\frac{u_k-u_j+i}{u_k-u_j-i}\,,\qquad \xi\equiv e^{i\frac{2\pi}{L}}\,,
\end{equation}
where $j,k\in[1,\dots \frac J2]$ label the magnon number and $J$ refers to the total length of the spin-chain. The one-loop anomalous dimension is
\begin{equation}\label{eq:SU21loopad}
    \Delta-J=\frac{\lambda}{8\pi^2}\sum_{k=1}^{J/2}\frac{1}{u_k^2+1/4}\,.
\end{equation}
The insertion of the twist matrix inside the trace together with \eqref{eq:scalarZL} result in the `twisted' momentum constraint\footnote{Solutions to \eqref{eq:twistedSU2betheeqn} which do not satisfy \eqref{eq:SU2twistedpconstraint} still correspond to well-defined eigenstates of the spin-chain. Only those satisfying the twisted cyclicity condition correspond to physical single-trace operators of the quiver gauge theory.}
\begin{equation}\label{eq:SU2twistedpconstraint}
    e^{iP}=\prod_{k=1}^{J/2}\frac{u_k+i/2}{u_k-i/2}=\xi^n\,,\qquad P\equiv \sum_{k=1}^{J/2}p_k\,.
\end{equation}
In the continuum limit $(J\to\infty)$, one defines the (finite) scaled magnon rapidities $x_k=\frac{u_k}{J}$. Taking logarithms and expanding the LHS of \eqref{eq:twistedSU2betheeqn} at large $J$, as well as choosing the same logarithm branch $r_k\equiv - m$ for each $k$, we find
\begin{equation}\label{eq:SU2thermodynamic}
    2\pi q_{\rm{SU(2)}}+\frac{1}{x_k}=\frac{1}{i}\sum_{j\neq k}\log\frac{x_k-x_j+\frac iJ}{x_k-x_j-\frac{i}{J}}\,,\qquad q_{\rm{SU(2)}}\equiv m-2\mu\,,
\end{equation}
which differs from the expression in \cite{Beisert:2005mq} by the effective replacement $m\to q_{\rm{SU(2)}}$. For the case of purely fractional windings, we choose $m=0$ as in \cite{Astolfi:2008yw}. Note that one does not expand the RHS in $1/J$ so as to account for contributions from `anomalous' nearby Bethe roots where $x_j-x_k\sim\order{J^{-1}}$.  Introducing the resolvent
\begin{equation}
    G(x)=\frac{1}{J}\sum_{k=1}^{J/2}\frac{1}{x-x_k}\,,\qquad G(x\to\infty)=\frac{1}{2x}+\order{x^{-2}}\,,
\end{equation}
at leading $J\to\infty$ order, one has $P=-G(0)=2\pi\mu$, and $\Delta-J=-\frac{\lambda}{8\pi^2J}G'(0)$. The next steps follow exactly as in \cite{Beisert:2005mq}, where one has the same integral representation for the RHS of \eqref{eq:SU2thermodynamic}, rewrites the Bethe equations in terms of $G(x)$ and solves for the resolvent perturbatively. At leading order, $G(x)$ satisfies the same quadratic equation as in \cite{Beisert:2005mq} with $m\to -2\mu$, hence one finds that
\begin{equation}
    G(x)=-2\pi\mu+\frac{1}{2x}\left(1-\sqrt{1+16\pi^2\mu^2x^2}\right)\,,
\end{equation}
where the square root has branch points at $ x_\pm=\pm \frac{i}{4\pi\mu}$. The square-root sheet and the associated cut configuration are chosen so that $G(x)$ is regular at the origin, has the required large $x$ asymptotics, and defines a physical Bethe-root density.\footnote{At finite $J$, the poles of the resolvent give the Bethe roots, which condense into cuts as $J\to\infty$. The discontinuity of $G(x)$ across a given cut then gives the density of Bethe roots.} The result of \cite{Beisert:2005mq} thus follows, but adapted to the orbifold solution with purely fractional windings, giving
\begin{equation}\label{eq:SU2finitesizecorrection}
b_1(\mu)=2\mu^3\int_{-1}^1\dd x\,x\sqrt{1-x^2}\cot(2\pi\mu x)\,,
\end{equation}
which can be put into the form \eqref{eq:2spin1loopintegral} and \eqref{eq:SU2LLcorrection}.\footnote{For that, one uses the fact that the integrand \eqref{eq:SU2finitesizecorrection} is even in $x$, as well as the identity $\cot(2\pi\mu x)=\frac{1}{2\pi\mu x}-\frac{2}{\pi}\int_0^{\infty}\dd t\,\frac{\sinh(2\mu x t)}{e^t-1}$, which holds for $\mu\in[0,\frac{1}{2}]$.} Note that although the Bethe roots are complex, stability of the finite-size correction depends on whether the poles of the integrand lie outside of the integration region. The cotangent has poles whenever $2\mu x=s$, for $s\in\mathbb{Z}_{\neq 0}$ (the pole at $s=0$ is harmless since near $x=0$, $x\cot(2\pi\mu x)\sim \frac{1}{2\pi\mu}$). The first non-zero pair of poles lies at $x=\pm\frac{1}{2\mu}$ (corresponding to $s=\pm1$), reaching the endpoint of the integration domain at $\mu=\frac{1}{2}$ and moving into its interior when $\mu> \frac{1}{2}$. Remarkably, this coincides exactly with the worldsheet stability threshold. The lowest ($r=1$) bosonic mode becomes unstable for $\mu>\frac{1}{2}$ (cf. \eqref{eq:usefulproperty1}). The appearance of poles inside the Bethe integral therefore provides a direct Bethe-ansatz counterpart of the semiclassical string instability. Accordingly, the finite-size correction \eqref{eq:SU2finitesizecorrection} remains real for $|\mu|<\frac{1}{2}$ (on the winding branch with $\tilde{m}=0$, this reduces to $2n<L$, in agreement with the stability bound of Section \ref{sect:2spin1loop}).

The SL$(2)$ sector is spanned by single trace operators of the form specified in \eqref{eq:stateopmap}. The one-loop Hamiltonian acts by redistributing derivatives among neighboring spin-chain sites, namely $D^{s_i'}XD^{s_{i+1}'}X\to D^{s_i}XD^{s_{i+1}}X$, such that $s_i'+s_{i+1}'=s_i+s_{i+1}$. Since this does not change the $\mathbb{Z}_{L}$ charge of a given site, the local Bethe equations remain the same as in the unorbifolded case. Due to \eqref{eq:scalarZL}, the twist enters at the level of the momentum constraint, which reads
\begin{equation}
    e^{iP}=\prod_{k=1}^S\frac{u_k+i/2}{u_k-i/2}=\xi^n\,,\qquad P=\sum_{k=1}^Sp_k\,,
\end{equation}
and imposes the `level-matching' condition $pu+\mu=0$ obtained in \eqref{eq:orbifoldSJvirasoro}. For the rational SL$(2)$ solution considered, all Bethe roots lie along the real axis, and the finite-size correction of \cite{Beisert:2005mq} follows after specialising to the orbifold solution with purely fractional windings (i.e. by setting the standard SL$(2)$ parameter $M^2\equiv p^2u(1+u)\to\mu(\mu-p)$) giving
\begin{equation}\label{eq:SL2finitesizecorrection}
    b_1(p,\mu)=-2[\mu(\mu-p)]^{3/2}\int_{-1}^1\dd x\,x\sqrt{1-x^2}\coth\left(2\pi x\sqrt{\mu(\mu-p)}\right)\,.
\end{equation}
The above result is indeed equivalent to \eqref{eq:SJd1coeff} and \eqref{eq:SL2LLcorrection} after the change of variables $t=2x\sqrt{\mu(\mu-p)}$. The correction \eqref{eq:SU2finitesizecorrection} can be obtained by analytic continuation of \eqref{eq:SL2finitesizecorrection}, which involves the replacement of the `filling fraction' $\frac{S}{J}=-\frac{\mu}{p}\to-\frac{J_2}{J_1+J_2}$, and by noting that the $J_1=J_2$ fractional winding state imposes $p=2\mu$.

We now consider the subset of SU$(3)$ sector states spanned by the operators \eqref{eq:stateopmap}. The choice of different representation labels for the same physical SU$(3)$ `highest weight' state \cite{Minahan:2002ve, Engquist:2003rn}, corresponds to a choice of `vacuum' on top of which excitations are built. The physical anomalous dimension should be independent of the choice of representation, although the explicit form of the Bethe equations, the assignment of twist phases, and the momentum constraint will depend on this choice. In either case, the Bethe ansatz is nested, containing two sets of rapidities $\{u_{1,k},u_{2,k}\}$. The first set counts excitations away from the chosen `vacuum', while the second set resolves their internal `type'. The explicit form of the Bethe equations, as well as the corresponding momentum constraints in each case, can be found in Appendix \ref{app:2}.
    
In \cite{Freyhult:2004iq,Freyhult:2005fn}, the regular finite-size correction for rational $J_3\neq 0$ states was computed. There, the dimensionless one-loop coefficient $\gamma_1$ is defined through (cf. \eqref{eq:conformaldimexpansion})
\begin{equation}
\Delta =J_{\rm tot} \left[ 1+\frac{\lambda}{J_{\rm tot}^{\,2}}\gamma_1 +O(\lambda^2) \right],\qquad \gamma_1\equiv\gamma_1^{\rm{reg}}+\gamma_1^{\rm{anom.}}= a_1+\frac{b_1}{J_{\rm{tot}}}+\order{J_{\rm{tot}}^{-2}}\,,
\end{equation}
where
\begin{equation}
\gamma_1^{\rm reg}=\frac{1}{2}
    \left[
        \ell^2\alpha(1-\alpha)
        +2\beta m\ell(1-\alpha)
        +m^2\beta(1-\beta)
    \right]\left(1+\frac{1}{J_{\rm tot}}\right),
\label{eq:SU3regularcoefficient}
\end{equation}
is, in their notation, the regular contribution to $\gamma_1$ from Bethe roots whose separations are of order 1, with $\gamma_1^{\rm{anom.}}$ denoting the anomalous contribution due to Bethe roots with order $1/J_{\rm tot}$ separations. The quantities $\alpha,\beta$ refer to the two filling fractions of the nested SU$(3)$ Bethe ansatz, while the integers $\ell$ and $m$ denote the common logarithmic branch numbers carried by each type of Bethe root. The regular contribution to $\gamma_1$ was shown to reproduce the zero-mode part of the full one-loop correction to the classical energy computed on the string side \cite{Frolov:2003tu}. 

One may then adapt the result of \cite{Freyhult:2005fn} to the $J_3\neq 0$ solution described in Section \ref{sec: 2} by appropriately tuning the parameters of \eqref{eq:SU3regularcoefficient}, giving
\begin{equation}
    \Delta-J_{\rm{tot}}=\frac{\lambda}{J_{\rm{tot}}}\frac{\mu^2}{2}\sin^2\gamma_0\left(1+\frac{1}{J_{\rm{tot}}}\right)\,,\; \begin{cases}
    \alpha=\frac{J/2+J_3}{J_{\rm{tot}}}\,,\,\beta=\frac{J_3}{J_{\rm{tot}}}\,,\,l=-2\mu\,,\;m=\mu\,,\, J/2\geq J_3\,,\\
    \alpha=\frac{J}{J_{\rm{tot}}}\,,\,\beta=\frac{J/2}{J_{\rm{tot}}}\,,\,l=\mu\,,\,m=-2\mu\,,\, J_3\geq J/2\,,
    \end{cases}
\end{equation}
for each choice of ordered basis, after making use of \eqref{eq:3spinsinrelation}. The first term is the leading $J_{\rm{tot}}\to\infty$ anomalous dimension and matches
the $\order{\lambda/J_{\rm tot}}$ term in the classical string energy \eqref{eq:3spinclassicalenergy}. The $\order{\lambda/J_{\rm{tot}}^2}$ term matches the zero-mode part of \eqref{eq:oneloopSU3expression} in the `fast-string' regime (cf. \eqref{eq:3spinblargek}, \eqref{eq:3spinflargek} and the AdS$_5$ contributions \eqref{eq:bosonicCFfixedn} after setting $r=0$) and the zero-mode part of the SU$(3)$ LL result \eqref{eq:SU3LLcorrection}.
\section{Concluding remarks}
In this paper, we investigated three families of semiclassical type IIB spinning string configurations on AdS$_5\cross S^5/\mathbb{Z}_L$ orbifold backgrounds whose energies are holographically related to the scaling dimensions of `twisted' single-trace operators in the SU$(2)$, SU$(3)$ and SL$(2)$ sectors, respectively, of the dual $\mathcal{N}=2$ quiver gauge theory (see \eqref{eq:stateopmap}). Despite the lack of maximal supersymmetry, we observed an exact agreement between the string theory coefficients $d_1$ of \eqref{eq:energyexpansion} and the gauge theory coefficients $b_1$ of \eqref{eq:conformaldimexpansion} in the SU$(2)$ and SL$(2)$ sectors, as well as in the SU$(3)$ sector to the extent we checked.

On the string side, the presence of the $\mathbb{Z}_{L}$ quotient allows for states with fractional windings $\mu=\tilde{m}+\frac{n}{L}$ along orbifolded directions of the background. The twisted sector label $n$ fixes only the fractional part of $\mu$, meaning that stability is therefore intrinsic to a given winding branch rather than to $n$ alone. The explicit computations in this paper were performed on the purely fractional branch $\widetilde m=0$, for which $\mu=\frac{n}{L}$. 

Quantising the quadratic fluctuations around each classical solution, we determined the corresponding one-loop coefficients $d_1$ in each case and identified regions of stability in the full winding parameter $\mu$. For the $J_1=J_2$ solution, strict stability requires $|\mu|<\frac{1}{2}$, with $|\mu|=\frac{1}{2}$ being marginal. Thus, on the branch $\widetilde m=0$, the solution is unstable for $n>\frac{L}{2}$. The $\mathbb{Z}_L$ quotient consequently enlarges the set of available stable semiclassical winding branches relative to the smooth AdS$_5\cross S^5$ background. In all three cases, in the purely fractional branches with fixed $n\neq\order{L}$, their energies connect smoothly to those of the appropriate BPS states as $\mu\to0$.

Motivated by recent localisation results on the dual gauge theory \cite{Beccaria:2023qnu, Korchemsky:2025eyc}, it is of interest to consider the scaling $L\gg\sqrt{\lambda}\gg 1$ from a semiclassical perspective. To this end, we consider the $J_1=J_2$ solution in the SU$(2)$ sector and emphasise that $L$ is a discrete, fixed, dimensionless parameter of the background introduced upon orbifolding, and is thus not associated to any conserved charges of the classical solution \eqref{classicalsoln}. The semiclassical limit corresponds to the regime where one expands in large effective string tension $T\gg 1$, with the classical scaled charges $\mathcal{E},\,\mathcal{J}_i$ in \eqref{eq:conservedquants} held fixed, and, crucially, with fixed background parameter $L$ \cite{Tseytlin:2002gz, Tseytlin:2010jv}, which is completely independent from the effective string tension.\footnote{This is different in the case of ABJM theory, which at large $N$ and fixed $k$ is dual to M-theory on AdS$_4\cross S^7/\mathbb{Z}_k$, where the $\mathbb{Z}_{k}$ action on $S^7$ is smooth. The (dimensionless) effective M2-brane tension is $\mathrm{T}_2=\frac{\sqrt{2Nk}}{\pi}$ (see \cite{Giombi:2024itd} and references therein). At large $N,k$ with $\frac{N}{k}=$ fixed, the dual description is in terms of type IIA theory on AdS$_4\cross \mathbb{CP}^3$, and the effective tension of the fundamental type IIA string may be obtained from wrapping the M2-brane along the 11d circle, giving $\mathrm{T}_{\rm{IIA}}=\frac{\pi}{2k}\mathrm{T}_2=\sqrt{\frac{\lambda_{\rm{ABJM}}}{2}}$ with $\lambda_{\rm{ABJM}}=\frac{N}{k}$. In our type IIB orbifold, the parameter $L$ never divides the fundamental type IIB string tension. Taking $L\sim\sqrt{\lambda}$, means that the $L^{-2}$ and $L^{-1}$ terms of the classical background \eqref{orbifoldmetric} scale with the same powers of $\lambda$ (i.e. $\lambda^{-1}$ and $\lambda^{-1/2}$) as worldsheet quantum corrections, invalidating the usual semiclassical fixed-background expansion at large $T$ and fixed $L$.} Within this regime, the twist $\mu$ enters only at the level of \eqref{eq:SU2classicalenergy}. If one, however, were to define a fixed scaled orbifold parameter $\mathcal{L}\gg 1$ analogous to $\mathcal{J}$ and $\mathcal{E}$, that would imply the scaling $L\gg \frac{R^2}{\alpha'}\sim\sqrt{\lambda}$. In this large $L$ regime, the classical energy becomes
\begin{equation}\label{eq:SU2Lclassicalenergy}
    E=J\sqrt{1+\frac{n^2}{J^2\mathcal{L}^2}}\,,\qquad \mathcal{L}\equiv \frac{L}{\sqrt{\lambda}}\,,
\end{equation}
and $\mu=\frac{n}{\sqrt{\lambda}\mathcal{L}}\to 0$,\footnote{As $\mu\to 0$, the induced metric, which is flat and proportional to $\mu^2$, degenerates.} so that the twist enters at subleading orders in large $\lambda$ with respect to \eqref{eq:SU2classicalenergy}. In this `semiclassical' expansion, one would thus find
\begin{equation}
E=J\left[1+\frac{n^2}{2J^2\mathcal{L}^2}-\frac{n^4}{8J^4\mathcal{L}^4}+\dots\right]\,,
\end{equation}
where the second term is parametrically of the same order as a two-loop worldsheet quantum correction to the classical energy. However, from \eqref{eq:SU2classicalenergy} one sees that this contribution originates at tree-level from the classical background. Its `apparent' order in the above large $T$ expansion has been shifted by the choice of scaling $L\gg \sqrt{\lambda}\gg 1$, which reorganises the semiclassical expansion.\footnote{The `winding-state' regime of \cite{Astolfi:2006is, Astolfi:2008yw} is a double-scaling limit where $J,L\to\infty$ with $\frac{J}{L^2}$ fixed. In the orbifolded LL description, the target space becomes, to leading order in this limit, the cylinder $\mathbb{R}\cross S^1$, with the $S^1$ radius being proportional to $\frac{\sqrt{J}}{L}$, around which the winding state wraps. This regime defines a novel LL scaling sector, and thus does not demand a `sensible' fixed-background semiclassical expansion. Imposing the usual semiclassical scaling would lead to $L\sim\lambda^{1/4}$, which still reorganises the expansion.} Making sense of the localisation regime $L\gg\sqrt{\lambda}\gg 1$ from a string theory perspective remains, however, an interesting open problem that is left for future work.

We also derived the corresponding Landau-Lifshitz effective theories from the relevant bosonic parts of the string sigma-model by isolating a `fast' coordinate and expanding in derivatives of the remaining `slow' fields. The leading terms in the $J\gg 1$ expansion of the classical energies of each string solution are captured, in each case, upon quantisation of these low-energy models and after assuming a particular regularisation scheme dictated by the `microscopic' theory one is trying to approximate. In the SU$(2)$ and SU$(3)$ cases the local LL Lagrangians are the same as those obtained from strings in AdS$_5\cross S^5$, with $\mathbb{Z}_L$ effects appearing as appropriate global identifications of a target space coordinate. In contrast, in the SL$(2)$ case, the orbifold dependence enters through the `fast' $S^1$ angle, leaving the LL target space unchanged.

On the gauge theory side, the same LL actions follow from a suitable continuum coherent state limit of the corresponding spin-chain Hamiltonians, providing a common effective description of both the gauge and string theory dynamics. The computation of the coefficients $b_1$ was performed by analysing the leading finite-size corrections to the one-loop orbifolded Bethe equations in the continuum $J\to\infty$ limit. The Bethe equations are locally equivalent to their $\mathcal{N}=4$ counterparts, with orbifold effects entering in the form of phases due to twisted boundary conditions and twisted momentum constraints. Taking into account these modifications, in the SU$(2)$ and SL$(2)$ cases, we adapted the arguments of \cite{Beisert:2005mq} to reproduce the appropriate $d_1$ coefficients computed on the string side. A particularly striking aspect of this agreement is that, in the SU$(2)$ sector, the worldsheet stability threshold $\vert\mu\vert=\frac{1}{2}$ is reproduced directly in the Bethe description, where the first cotangent poles reach the integration contour at the same value and move into its interior precisely when the semiclassical string develops an unstable mode. For the SU$(3)$ sector, the finite-size correction computation including contributions from `anomalous' Bethe roots was not performed for the orbifolded system. Nevertheless, adapting the known `non-anomalous' contributions of \cite{Freyhult:2005fn} yields agreement with the zero-mode part of the full $J_1=J_2$, $J_3\neq 0$ string theory result for $d_1$. Thus, at first order in $\lambda/J^{2}$ and including the leading finite-size correction, we find complete agreement amongst all three approaches in the $\mathrm{SU}(2)$ and $\mathrm{SL}(2)$ sectors, together with the corresponding agreement in the $\mathrm{SU}(3)$ sector to the extent that has been checked.

The present results extend the semiclassical string program on AdS$_5\cross S^5/\mathbb{Z}_L$ backgrounds initiated in \cite{Ideguchi2004}, and suggest several natural directions for future work. First, it would be important to determine whether agreement between the three descriptions still holds through order $\lambda^2/J^4$, in analogy with the corresponding analysis on AdS$_5\cross S^5$. This would give a broader test for planar integrability in the twisted sectors of the 4d $\mathcal{N}=2$ circular quiver gauge theory of interest. Another direction involves the construction and semiclassical quantisation of more general non-circular string solutions, including `spiky' string configurations, whose discrete rotational symmetry can be made compatible with the $\mathbb{Z}_L$ orbifold action. Finally, recent progress on spin-chains for ADE quiver theories \cite{Bath:2026gcu} motivates extending the present analysis to semiclassical string solutions on AdS$_5\cross S^5/\Gamma$ backgrounds, where $\Gamma$ is a finite subgroup of $\mathrm{SU}(2)$ in the ADE classification. This would require constructing and quantising semiclassical strings in the corresponding non-Abelian twisted sectors and comparing their quantum corrections to the finite-size corrections predicted by the ADE spin-chains.\newline

\textbf{Acknowledgements:} It is a pleasure to thank Juan Miguel Nieto García, Stefan A. Kurlyand, Dennis le Plat and Torben Skrzypek for useful comments on the draft and discussions on related topics, as well as the participants of the workshop `Let's twist again: Integrability in Orbifold theories' for many inspiring discussions. I would also like to thank my supervisor, Arkady A. Tseytlin, for collaboration at earlier stages of this project and for valuable comments on the manuscript. The work of CBM is supported by the STFC DTP research studentship grant ST/Y509231/1.

\appendix
\section{One-loop computation of $J_1=J_2$ solution}\label{sect:A}
In this Appendix we provide details on the computation of the quantum correction to the classical energy of the $J_1=J_2$ solution of Section \ref{sec: 2}, as well as details on the asymptotic expansions of the different characteristic frequencies for large mode number $r$, large $\kappa$ with fixed $r$ and large $\kappa$ with $x=\frac{r}{\kappa}$ held fixed.
\subsection{Bosonic frequencies}\label{sect:A1}
To compute the one-loop correction to the classical energy of the solution in \eqref{classicalsoln}, we follow \cite{Frolov:2003tu}. We first fix the static gauge condition
\begin{equation}\label{eq:staticgauge}
    t=\kappa\tau\,,\quad\chi=L\mu\sigma\,,\quad\Rightarrow\quad \tilde{t}=\tilde{\chi}=0\,,
\end{equation}
at the level of the fluctuations, where a tilde over a coordinate denotes its fluctuation around the classical solution. We then expand the classically equivalent Nambu-Goto action
\begin{equation}\label{eq:NGaction}
    S_B=-T\int\mathrm{d}\tau\mathrm{d}\sigma\sqrt{-\det h_{ab}}\,,\qquad h_{ab}=G_{mn}(x)\partial_ax^m\partial_bx^n\,,
\end{equation}
where $h_{ab}$ are the components of the induced metric on the worldsheet. At the classical value $\gamma_0=\frac{\pi}{2}$ (corresponding to the centre of the `two-sphere' part, $\mathrm{d}\gamma^2+\cos^2\gamma\,\mathrm{d}\varphi_3^2$, of the $S^5/\mathbb{Z}_L$ metric), the coordinates $(\gamma,\varphi_3)$ of \eqref{orbifoldmetric} are not suitable to study fluctuations around the solitonic solution \eqref{classicalsoln}. We remedy this bad choice of local coordinates by re-introducing the two embedding Cartesian coordinates $X_{5,6}$ of \eqref{complexcoords}, whose value on the classical solution is $X_{5}=X_{6}=0$. In terms of these, the metric \eqref{orbifoldmetric} now reads
\begin{equation}
\begin{split}
\label{neworbifoldmetric}
&\mathrm{d}s^{2}_{S^{5}/\mathbb{Z}_{L}}=\mathrm{d}X_{5}^{2}+\mathrm{d}X_{6}^{2}+\frac{(X_{5}\mathrm{d}X_{5}+X_{6}\mathrm{d}X_{6})^{2}}{1-(X_{5}^{2}+X_{6}^{2})}+\big[1-(X_{5}^{2}+X_{6}^{2})\big]\mathrm{d}s_{S^{3}/\mathbb{Z}_{L}}^{2}\,,\\
& \mathrm{d}s_{S^{3}/\mathbb{Z}_{L}}^{2}=\mathrm{d}\psi^{2}+\mathrm{d}\Phi^{2}+\frac{1}{L^{2}}\mathrm{d}\chi^{2}+\frac{2\cos2\psi}{L}\,\mathrm{d}\Phi\mathrm{d}\chi\,.
\end{split}
\end{equation}
After fixing \eqref{eq:staticgauge}, the relevant physical fluctuations of the fields on $S^5/\mathbb{Z}_L$ are thus
\begin{equation}
    \Phi\to\omega\tau+\frac{1}{\lambda^{1/4}}\tilde{\phi}(\tau,\sigma)\,,\quad \psi\to\frac{\pi}{4}+\frac{1}{\lambda^{1/4}}\tilde{\psi}(\tau,\sigma)\,,\quad X_{5,6}\to 0+\frac{1}{\lambda^{1/4}}\tilde{X}_{5,6}(\tau,\sigma)\,.
\end{equation}
We first deal with the coupled $\tilde{\psi},\tilde{\phi}$ sector in $S^3/\mathbb{Z}_{L}$ and thus set $\tilde{X}_{5,6}=0$. The metric determinant up to and including quadratic order in fluctuations is
\begin{equation}
\sqrt{-\det h}
={}
\mu^2
-\omega\,\partial_{\tau}\tilde{\phi}
-\frac{\kappa^2}{2\mu^2}\bigg[(\partial_{\tau}\tilde{\phi})^2-(\partial_{\sigma}\tilde{\phi})^2\bigg]
-\frac{1}{2}(\partial_{\tau}\tilde{\psi})^2
+\frac{1}{2}(\partial_{\sigma}\tilde{\psi})^2-\frac{2\kappa^2}{\mu}\tilde{\psi}\,\partial_{\sigma}\tilde{\phi}
+2\omega^2\tilde{\psi}^{\,2}\, .
\end{equation}
Dropping the constant and total derivative terms, one obtains the quadratic Lagrangian
\begin{equation}\label{eq:quadLS3}
    \mathcal{L}^{(2)}_{\tilde{\psi},\tilde{\phi}}=-\frac{1}{2}(\partial_a\tilde{\psi})^2-\frac{\kappa^2}{2\mu^2}(\partial_a\tilde{\phi})^2+\frac{2\kappa^2}{\mu}\tilde{\psi}\partial_{\sigma}\tilde{\phi}-2(\kappa^2-\mu^2)\tilde{\psi}^2\,.
\end{equation}
To deal with the decoupled $\tilde{X}_{5,6}$ sector, we set $\tilde{\psi}=\tilde{\phi}=0$. The classical background reads
\begin{equation}
\mathrm{d}s^2 = -\mathrm{d}t^2+ \mathrm{d}X_{5}^2+\mathrm{d}X_{6}^2+\left(1-(X_{5}^2+X_{6}^2)\right)\left(\mathrm{d}\Phi^2+\frac{1}{L^2}\mathrm{d}\chi^2\right)\,.
\end{equation}
To quadratic order, the metric determinant is 
\begin{equation}
    \sqrt{-\det h}=\mu^2+\frac{1}{2}\bigg[\vert\partial_{\sigma}\tilde{Z}\vert^2-\vert\partial_{\tau}\tilde{Z}\vert^2-\left(\mu^2-\omega^2\right)\vert\tilde{Z}\vert^2\bigg]\,,
\end{equation}
where we have introduced the notation $\tilde{Z}=\tilde{X}_{5}+i\tilde{X}_6$ with $\vert \tilde{Z}\vert^2=\tilde{X}_{5}^2+\tilde{X}_6^2$ for convenience. The corresponding fluctuation Lagrangian becomes
\begin{equation}\label{eq:quadLZ}
    \mathcal{L}^{(2)}_{\tilde{Z}}=-\frac{1}{2}\vert\partial_{a}\tilde{Z}\vert^2-\frac{1}{2}(\kappa^2-2\mu^2)\vert\tilde{Z}\vert^2\,.
\end{equation}
The full (bosonic) quadratic fluctuation Lagrangian on $S^5/\mathbb{Z}_{L}$ is simply the combination of \eqref{eq:quadLS3} and \eqref{eq:quadLZ}. After the change of variables $\alpha=-\frac{\kappa}{\mu}\tilde{\phi}\,,\, \beta=\tilde{\psi}\,$, it reads
\begin{equation}\label{eq:2spinL2}
\mathcal{L}^{(2)}_{S^5/\mathbb{Z}_L}=-\frac{1}{2}\vert\partial_{a}\tilde{Z}\vert^2-\frac{1}{2}(\kappa^2-2\mu^2)\vert\tilde{Z}\vert^2-\frac{1}{2}(\partial_a\alpha)^2-\frac{1}{2}(\partial_{a}\beta)^2-2\kappa\beta\,\partial_{\sigma}\alpha-2(\kappa^2-\mu^2)\beta^2\,.
\end{equation}
The equations of motion that arise are
\begin{equation}\begin{split}\label{eq:2spineoms}
\square\beta+2\kappa\partial_{\sigma}\alpha+4(\kappa^2-\mu^2)\beta=0\,,\quad \square\alpha-2\kappa\,\partial_{\sigma}\beta=0\,,\quad \square \tilde{X}_{5,6}+(\kappa^2-2\mu^2)\tilde{X}_{5,6}=0\,.
\end{split}\end{equation}
Looking for plane-wave solutions of the form
\begin{equation}\label{eq:bosonicplanewaves}
    \xi(\tau,\sigma)=\sum_{r=-\infty}^{\infty}\sum_{s=1}^{8}\xi_{r}^s\,e^{i(w_{r}^s\tau+r\sigma)}\,,
\end{equation}
where $s$ labels different characteristic frequencies $w_r^s$ for a given mode number $r\in\mathbb{Z}$, for the coupled $(\alpha,\beta)$ system, one obtains the following quartic equation for the characteristic frequencies 
\begin{equation}
\det\begin{pmatrix}
       r^2-(w_r^s)^2&-i2\kappa r\\
       i2\kappa r & r^2-(w_r^s)^2+4(\kappa^2-\mu^2)
   \end{pmatrix}=0\,,
\end{equation}
which may be rewritten as
\begin{equation}
\Delta^2+4\omega^2\Delta-4\kappa^2r^2=0\,,\qquad \Delta\equiv r^2-(w_r^s)^2\,.
\end{equation}
The solutions to the above equation give the frequencies in \eqref{eq:S3CFs}. Note that
\begin{equation}\label{eq:usefulproperty}
    (w_{r,+}+w_{r,-})^2=4\kappa^2+\left(\vert r\vert+\sqrt{r^2-4\mu^2}\right)^2\,.
\end{equation}
The decoupled $\tilde{X}_{5,6}$ fluctuations result in the (positive) doubly-degenerate characteristic frequencies \eqref{eq:DecoupledCF}.

There are an additional four massive bosonic fluctuations coming from $\mathrm{AdS}_{5}$. Starting with the $\mathrm{AdS}$ part of \eqref{bosonicsigmamodel} and choosing static gauge as in \eqref{eq:staticgauge}, we are able to fix one of the fluctuations of $\mathrm{AdS}_{5}$. We wish to obtain the gauge-fixed Lagrangian to quadratic order in fluctuations of the remaining fields. Since our solution is located at $\rho=0$, where the angular coordinates of the $\mathrm{AdS}$ metric become degenerate, it is convenient to introduce four Cartesian-type coordinates $\eta_{k}$ ($k=1,2,3,4)$ in terms of which the $\mathrm{AdS}_{5}$ metric becomes
\begin{equation}
\mathrm{d}s^{2}_{\mathrm{AdS_{5}}}=-\frac{(1+\frac{1}{4}\eta^{2})^{2}}{(1-\frac{1}{4}\eta^{2})^{2}}\mathrm{d}t^{2}+\frac{1}{(1-\frac{1}{4}\eta^{2})^{2}}\mathrm{d}\eta_{k}\mathrm{d}\eta_{k}\,,\quad \eta^{2}=\sum_{k=1}^{4}\eta_{k}\eta_{k}\,.
\end{equation}
Noting that the region where $\rho=0$ corresponds to $\eta_{k}=0$, we may expand the above metric as $\eta_{k}\to0+\frac{1}{\lambda^{1/4}}\tilde{\eta}_{k}$, resulting in
\begin{equation}
ds^{2}_{\mathrm{AdS_{5}}}\bigg|_{\eta_{k}\to 0}=-(1+\tilde{\eta}^{2})\mathrm{d}t^{2}+(1+\frac{1}{2}\tilde{\eta}^{2})\mathrm{d}\tilde{\eta}_{k}\mathrm{d}\tilde{\eta}_{k}\,.
\end{equation}
Expanding also the $\mathrm{AdS}$ part of \eqref{bosonicsigmamodel} to quadratic order in $\tilde{\eta}_{k}$, we obtain
\begin{equation}
\mathcal{L}_{\mathrm{AdS}}^{(2)}=-\frac{1}{2}(\partial_{a}\tilde{\eta}_{k})^{2}-\frac{1}{2}\kappa^{2}\tilde{\eta}^{2}_{k}\,,\qquad \square\tilde{\eta}_k+\kappa^2\tilde{\eta}_k=0\,.
\end{equation}
The equations of motion describe four massive bosonic fluctuations with $m_{\mathrm{AdS}}^{2}=\kappa^{2}$. This then results in (positive) characteristic frequencies \eqref{eq:AdSCF}.
\subsection{Fermionic frequencies}
The quadratic part in the fermions of the AdS$_5\times S^5/\mathbb{Z}_{L}$ Green-Schwarz superstring action is (following \cite{Metsaev:1998it,Metsaev:2002re})
\begin{equation}\label{eq:fermionL2}
\mathcal{L}_{F}=i(\eta^{ab}\delta^{IJ}-\varepsilon^{ab}s^{IJ})\overline{\theta}^{I}\varrho_{a}D_{b}\theta^{J}\,,
\end{equation}
where
\begin{equation}
\begin{split}
&\varrho_{a}=\Gamma_{A}e_{a}^{A}\,,\quad e_{a}^{A}=E_{M}^{A}(X)\partial_{a}X^{M}\,,\quad D_{a}\theta^{I}=(\delta^{IJ}\tilde{D}_{a}-\frac{i}{2}\varepsilon^{IJ}\Gamma_{*}\varrho_{a})\theta^{J}\,,\\
&s^{IJ}=\mathrm{diag}(1,-1)\,,\quad \varepsilon^{IJ}=\begin{pmatrix}0 & 1 \\-1 & 0\end{pmatrix}\,,\quad \Gamma_{*}=i\Gamma_{01234}\,,\quad\Gamma_{*}^{2}=1\,,
\end{split}
\end{equation}
and $I,J=1,2$ label the 10d Majorana-Weyl spinors $\theta^{I}$. Here, $\varrho_{a}$ are projections of the 10d Dirac matrices $\Gamma_A$ onto the worldsheet, with $X_{M}$ the target space string coordinates corresponding to $\mathrm{AdS}_{5}$ $(M=0,1,2,3,4)$ and $S^{5}/\mathbb{Z}_{L}$ $(M=5,6,7,8,9)$ parts of the metric respectively. The covariant derivative $\tilde{D}_{a}$ is
\begin{equation}\label{eq:pullbacks}
\tilde{D}_{a}=\partial_{a}+\frac{1}{4}\omega_{a}^{AB}\Gamma_{AB}\,,\qquad \omega_{a}^{AB}=\partial_{a}X^{M}\omega_{M}^{AB}\,,
\end{equation}
where $\omega_{a}^{AB}$ denote projections of the 10d Lorentz connection $\omega_{M}^{AB}$ onto the worldsheet. Choosing $\kappa$-symmetry gauge where $\theta^{1}=\theta^{2}\equiv\theta$, we may rewrite quadratic Lagrangian \eqref{eq:fermionL2} as
\begin{equation}
\label{kfixedfermionicaction}
\mathcal{L}_{F}=i2\overline{\theta}D_{F}\,\theta\,,\qquad D_{F}=-\varrho^{a}\tilde{D}_{a}-\frac{i}{2}\varepsilon^{ab}\varrho_{a}\Gamma_{*}\varrho_{b}\,.
\end{equation}
We choose $A=0,5,6,7,8,9$ to label the (tangent space) coordinates $\{t,\gamma,\varphi_{3},\psi,\varphi_{1},\varphi_{2}\}$ used to describe the motion of the classical rotating string. An appropriate set of vielbeins is
\begin{equation}
\begin{split}\label{eq:fermionONframe}
&E_{t}^{0}=\mathrm{d}t,\quad E_{\gamma}^{5}=\mathrm{d}\gamma,\quad E_{\varphi_{3}}^{6}=\cos\gamma\mathrm{d}\varphi_{3},\quad E_{\psi}^{7}=\sin\gamma\mathrm{d}\psi,\\
&E^{8}_{\Phi+\chi/L}=\sin\gamma\cos\psi\left(\mathrm{d}\Phi+\frac{1}{L}\mathrm{d}\chi\right),\quad E^{9}_{\Phi-\chi/L}=\sin\gamma\sin\psi\left(\mathrm{d}\Phi-\frac{1}{L}\mathrm{d}\chi\right),
\end{split}
\end{equation}
and given the classical solution in \eqref{classicalsoln}, one can show that\footnote{We have defined $\tilde{\Gamma}_{8}\equiv\cos\psi_{0}\Gamma_{8}+\sin\psi_{0}\Gamma_{9}=\frac{1}{\sqrt{2}}(\Gamma_{8}+\Gamma_{9})$ and $\tilde{\Gamma}_{9}\equiv\cos\psi_{0}\Gamma_{8}-\sin\psi_{0}\Gamma_{9}=\frac{1}{\sqrt{2}}(\Gamma_{8}-\Gamma_{9})$, which obey $\tilde{\Gamma}_{8,9}^{2}=\mathbb{1}$ and $\{\tilde{\Gamma}_{8,9},\Gamma_{A}\}=0$ since they are just linear combinations of 10d Dirac matrices.}
\begin{equation}\label{eq:2spin2dmatrices}
\begin{split}
&\varrho_{0}=\kappa\Gamma_{0}+\omega\tilde{\Gamma}_{8}\,,\qquad \varrho_{1}=\mu\,\tilde{\Gamma}_{9}\,,\qquad \varrho_{(a}\varrho_{b)}=\mu^2\eta_{ab}\,,\qquad \eta_{ab}=\mathrm{diag}(-1,1)\,.
\end{split}
\end{equation}
The non-zero components of the projected spin connection $\omega_{a}^{AB}$ are computed by using Cartan's structure equation $\mathrm{d}E^A+\omega^{A}_B\wedge E^B=0$, where $\omega^{AB}=-\omega^{BA}$. Projecting back onto the worldsheet requires the use of the pull-back \eqref{eq:pullbacks}. The non-zero components of $\omega_a^{AB}$ are thus
\begin{equation}
\omega_{1}^{78}=\omega_{1}^{79}=\frac{\mu}{\sqrt{2}}\,,\qquad \omega_{0}^{78}=-\omega_{0}^{79}=\frac{\omega}{\sqrt{2}}\,.
\end{equation}
The pulled-back covariant derivatives \eqref{eq:pullbacks} are\footnote{We have used the fact that $\Gamma_{78}-\Gamma_{79}=\sqrt{2}\,\Gamma_{7\tilde{9}}$ and $\Gamma_{78}+\Gamma_{79}=\sqrt{2}\,\Gamma_{7\tilde{8}}$. An additional factor of two comes from the sum over antisymmetric pairs $\omega^{AB}_{a}$.}
\begin{equation}\label{eq:2spinpullbackD}
   \tilde{D}_0=\partial_{0}+\frac{\omega}{2} \Gamma_{7\tilde{9}}\,,\qquad \tilde{D}_1=\partial_{1}+\frac{\mu}{2}\Gamma_{7\tilde{8}}\,.
\end{equation}
Note that these are already $\sigma$-independent, so that in this case there is no need for a $\sigma$-dependent rotation changing the periodicity of the 10d Majorana-Weyl fermions, unlike in the treatment of \cite{Frolov:2003tu}. Using the fact that $[\Gamma_*,\Gamma_0]=0$, as well as that $\{\Gamma_{0},\tilde{\Gamma}_9\}=0$, the `Wess-Zumino' term in \eqref{kfixedfermionicaction} may be written as
\begin{equation}
\frac{i}{2}\varepsilon^{ab}\varrho_a\Gamma_{*}\varrho_b=\omega\mu\,\Gamma_{89}\Gamma_{01234}\,.
\end{equation}
The full fermionic kinetic operator $D_F$ of \eqref{kfixedfermionicaction} may thus be written in the form
\begin{equation}\label{simplifiedDF}
  D_F=(\kappa\Gamma_0+\omega\tilde{\Gamma}_8)\partial_{0}-\mu\,\tilde{\Gamma}_9\partial_{1}+\frac{\kappa\omega}{2}\Gamma_0\Gamma_{7\tilde{9}}-\frac{\kappa^2}{2}\Gamma_7\Gamma_{89}-\omega\mu\,\Gamma_{89}\Gamma_{01234}\,.
\end{equation}
The above kinetic operator can be put in a simpler form by performing a constant spinor rotation in the $(0\tilde{8})$ plane of the form
\begin{equation}
    \theta=S_{0\tilde{8}}\Psi\,,\qquad S_{0\tilde{8}}=e^{-\frac{1}{2}p\Gamma_{0\tilde{8}}}\,, \qquad \cosh p= \frac{\kappa}{\mu}\,,\qquad \sinh p=\frac{\omega}{\mu}\,.
\end{equation}
The rotation acts on the 10d Dirac matrices in the following way
\begin{equation}
    S_{0\tilde{8}}^{-1}\Gamma_0S_{0\tilde{8}}=\cosh p\,\Gamma_0-\sinh p\,\tilde{\Gamma}_8\,,\quad S_{0\tilde{8}}^{-1}\tilde{\Gamma}_8S_{0\tilde{8}}=\cosh p\,\tilde{\Gamma}_8-\sinh p\,\Gamma_0\,,\quad  S_{0\tilde{8}}^{-1}\tilde{\Gamma}_9S_{0\tilde{8}}=\tilde{\Gamma}_9\,,
\end{equation}
so that the operator \eqref{simplifiedDF} becomes
\begin{equation}\label{eq:simplifiedDF}
    D_F=\mu\bigg[\Gamma_0\,\partial_{0}-\tilde{\Gamma}_9\left(\partial_{1}+\frac{\kappa}{2}\Gamma_{7\tilde{8}}\right)+\omega\Gamma_{0\tilde{9}}\tilde{\Gamma}_8\Gamma_{1234}\bigg]\,.
\end{equation}
If one now defines $\tau_0\equiv\Gamma_0$, $\tau_1\equiv \tilde{\Gamma}_9$, $\tau_3\equiv\tau_0\tau_1=\Gamma_{0\tilde{9}}$ as well as $A_0=0$ and $A_1=\frac{\kappa}{2}\Gamma_{7\tilde{8}}$, the above operator can be written in the form
\begin{equation}
    D_F=\mu\bigg[-\tau^0(\partial_0+A_0)-\tau^1(\partial_1+A_1)+\omega\tau_3\tilde{\Gamma}_8\Gamma_{1234}\bigg]\,,
\end{equation}
which when inserted into \eqref{kfixedfermionicaction}, may be interpreted as describing a theory with a collection of 8 two-dimensional massive Majorana fermions on a flat 2d background coupled to constant-valued non-Abelian gauge fields $A_{0,1}$. The mass term $\tau_3$ should be attributed to the coupling of the Green-Schwarz fermions to the five-form flux of the orbifold background $F_5=4(\mathrm{vol}_{\mathrm{AdS_5}}+\mathrm{vol}_{S^5/\mathbb{Z}_{L}})$ \cite{Frolov:2003tu}. In solving the Dirac equation associated to
\begin{equation}
    \mathcal{L}_F^{(2)}=i2\bar{\Psi}\bigg[\Gamma_0\partial_0-\tilde{\Gamma}_9\bigg(\partial_1+\frac{\kappa}{2}\Gamma_{7\tilde{8}}\bigg)+\omega\Gamma_{0\tilde{9}}\tilde{\Gamma}_8\Gamma_{1234}\bigg]\Psi\,,
\end{equation}
where we have re-scaled $\Psi$ by $\mu^{-1/2}$ to absorb the overall factor of $\mu$ in \eqref{eq:simplifiedDF},\footnote{This re-scaling is valid so long as $\mu\neq 0$.} one must look for plane-wave solutions of the form
\begin{equation}\label{eq:fermionplanewaves}
    \Psi(\tau,\sigma)=\sum_{r=-\infty}^{\infty}\sum_{s=1}^8\psi_r^{s}\,e^{i\left(\Omega_r^s\tau+r\sigma\right)}\,,\qquad r\in\mathbb{Z}\,.
\end{equation}
To obtain the characteristic equation for the fermionic frequencies, one must fix a basis of 10d Dirac matrices. Since only the matrices $\Gamma_0,\tilde{\Gamma}_9,\Gamma_{7\tilde{8}}$ and $\mathcal{A}\equiv\Gamma_{0\tilde{9}}\tilde{\Gamma}_8\Gamma_{1234}$ appear, and we only need the algebraic relations between them, a $4\times4$ realisation of these is sufficient.\footnote{The algebra satisfied by these four matrices is \begin{equation}\begin{split}
 &(\Gamma_0)^2=(\Gamma_{7\tilde{8}})^2=-\mathbb{1}_4\,,\;(\tilde{\Gamma}_9)^2=\mathcal{A}^2=\mathbb{1}_4\,,\\
 &\{\Gamma_0,\tilde{\Gamma}_9\}=0=\{\Gamma_{7\tilde{8}},\mathcal{A}\}\,,\;[\Gamma_0,\Gamma_{7\tilde{8}}]=0=[\tilde{\Gamma}_9,\Gamma_{7\tilde{8}}]\,,\;[\Gamma_0,\mathcal{A}]=0=[\tilde{\Gamma}_9,\mathcal{A}]\,.
\end{split}\end{equation}} A convenient choice that realises their algebra is thus
\begin{equation}\label{eq:gammabasis}
\Gamma_0=i\sigma_2\otimes \mathbb{1}_2\,,\quad 
\tilde{\Gamma}_{9}=\sigma_1\otimes\mathbb{1}_2\,,\quad \Gamma_{7\tilde{8}}=i\mathbb{1}_2\otimes\sigma_3\,,\quad \mathcal{A}=\mathbb{1}_2\otimes\sigma_1\,,
\end{equation}
where $\sigma_i$ denote the $2\times 2$ Pauli matrices. Using the ansatz in \eqref{eq:fermionplanewaves}, the Dirac equation becomes
\begin{equation}
 (\mathcal{F}_r^s)\,\psi_r^s=0\,,\qquad   (\mathcal{F}_r^s)=i\Omega_r^s\Gamma_0-ir\tilde{\Gamma}_9-\frac{\kappa}{2}\tilde{\Gamma}_9\Gamma_{7\tilde{8}}+\omega\mathcal{A}\,,
\end{equation}
where, in the basis \eqref{eq:gammabasis}, $\mathcal{F}_r^s$ becomes
\begin{equation}
  ( \mathcal{F}_r^s)= \begin{pmatrix}
        0&\omega&i(\Omega_r^s-r-\frac{\kappa}{2})&0\\
        \omega&0&0& i(\Omega_r^s-r+\frac{\kappa}{2})\\
        -i(\Omega_r^s+r+\frac{\kappa}{2})&0&0&\omega\\
        0&-i(\Omega_r^s+r-\frac{\kappa}{2})&\omega&0
    \end{pmatrix}\,.
\end{equation}
The characteristic equation is simply
\begin{equation}
    \det(\mathcal{F}_r^s)=\frac{\kappa^4}{16}+\bigg[r^2+\omega^2-(\Omega_r^s)^2\bigg]^2-\frac{\kappa^2}{2}\bigg[r^2+\omega^2+(\Omega_r^s)^2\bigg]=0\,,
\end{equation}
leading to the four-fold degenerate set of two (positive) roots in \eqref{eq:fermionicCFs}.
\subsection{Different regimes of frequencies}\label{sect:2spinbosonfreqs}
We have found a total of eight bosonic characteristic frequencies (two from the modes on $S^3/\mathbb{Z}_L$ \eqref{eq:S3CFs}, another two from the decoupled fluctuations $\tilde{X}_{5,6}$ \eqref{eq:DecoupledCF} and four from AdS$_5$ \eqref{eq:AdSCF}) which differ only from their $S^5$ counterparts by the effective replacement $\tilde{m}\to \mu$ of the windings. We now consider different regimes for the bosonic characteristic frequencies on both $\mathrm{AdS}_{5}$ and $S^{5}
/\mathbb{Z}_{L}$. In the large $r$ regime, all three sets of frequencies admit expansions of the form
\begin{equation}
w_{r,\pm}=\vert r\vert\pm\kappa+\frac{\kappa^2-2\mu^2}{2\vert r\vert},\quad \mathrm{w}_r=\vert r\vert+\frac{\kappa^2-2\mu^2}{2\vert r\vert}\,,\quad \omega_{i}^{\mathrm{AdS}}=\vert r\vert+\frac{\kappa^{2}}{2\vert r\vert}\,.
\end{equation}
up to and including $\mathcal{O}(\vert r\vert^{-1})$. The overall large $r$ contribution from the bosonic sector gives
\begin{equation}\label{eq:largenbosonicsum}
w_{r,+}+w_{r,-}+2\mathrm{w}_r+4\omega^{\mathrm{AdS}}=8\vert r\vert+\frac{4\omega^2}{\vert r\vert}+\mathcal{O}(\vert r\vert^{-2}).
\end{equation}
The characteristic frequencies also admit a large $\kappa$ expansion with $r$ held fixed, taking the form
\begin{equation}\begin{split}\label{eq:bosonicCFfixedn}
&w_{r,+}= 2\kappa+\frac{r^{2}-2\mu^2}{2\kappa}\,,\quad w_{r,-}=\frac{\vert r\vert}{2\kappa}\sqrt{r^2-4\mu^2}\,,\quad \mathrm{w}_r=\kappa+\frac{r^{2}-2\mu^2}{2\kappa}\,,\quad \omega_{i}^{\mathrm{AdS}}=\kappa+\frac{r^{2}}{2\kappa}\,,
\end{split}\end{equation}
up to and including $\mathcal{O}(\kappa^{-1})$. Lastly, we will also be interested in the large $\kappa$ expansion with $x=\frac{r}{\kappa}$ held fixed. In this case, we have
\begin{equation}\label{eq:2spinlargekfixedx}
w_{r,\pm}=\kappa\left(\sqrt{1+x^2}\pm 1\right)-\frac{\mu^2}{\sqrt{1+x^2}}\frac{1}{\kappa}\,,\quad \mathrm{w}_{r}=\kappa\sqrt{1+x^2}-\frac{\mu^2}{\sqrt{1+x^2}}\frac{1}{\kappa}\,,\quad \omega_{i}^{\mathrm{AdS}}=\kappa\sqrt{1+x^{2}}\,,
\end{equation}
plus subleading $\mathcal{O}(\kappa^{-3})$ corrections. Adding up all 8 bosonic frequencies in this limit (which we collectively denote by $\omega_{r}^s$) one finds
\begin{equation}\label{eq:bosonicsumx}
\sum_{s=1}^{8}\omega_{r}^s= 8\kappa\sqrt{1+x^{2}}-\frac{4\mu^2}{\sqrt{1+x^{2}}}\frac{1}{\kappa}+\mathcal{O}(\kappa^{-3})\,.
\end{equation}
For the fermionic frequencies, the large $r$ expansion of the four sets of roots takes the form
\begin{equation}\label{eq:largenfermionicfreq}
\Omega_{r,\pm}=\pm\frac{\kappa}{2}+\vert r\vert+\frac{\omega^2}{2\vert r\vert}+\mathcal{O}(\vert r\vert^{-3})\,,\qquad 4(\Omega_{r,+}+\Omega_{r,-})=8\vert r\vert+\frac{4\omega^2}{\vert r\vert}+\mathcal{O}(\vert r\vert^{-3})\,,
\end{equation}
The final regime corresponds to expanding at large $\kappa$ with $r$ held fixed. As in the bosonic case, one observes branching. The frequencies admit the following expansions
\begin{equation}\label{eq:fermionicCFlfixedn}
\Omega_{r,+}=\frac{3\kappa}{2}+\frac{r^2-\mu^2}{2\kappa}+\mathcal{O}(\kappa^{-3})\,,\quad \Omega_{r,-}=\frac{\kappa}{2}+\frac{r^2-\mu^2}{2\kappa}+\mathcal{O}(\kappa^{-3})\,,
\end{equation}
as long as $\kappa$ is large enough so that $\sqrt{r^2+\kappa^2-\mu^2}>\frac{\kappa}{2}$.
Finally, expanding at large $\kappa$ with $x=\frac{r}{\kappa}$ held fixed we obtain
\begin{equation}
    \Omega_{r,+}=\kappa\left(\sqrt{1+x^2}+\frac{1}{2}\right)-\frac{\mu^2}{2\kappa\sqrt{1+x^2}}\,,\quad \Omega_{r,-}=\kappa\left(\sqrt{1+x^2}-\frac{1}{2}\right)-\frac{\mu^2}{2\kappa\sqrt{1+x^2}}\,,
\end{equation}
plus subleading pieces of order $\mathcal{O}(\kappa^{-3})$. Adding up the fermionic characteristic frequencies in this regime gives
\begin{equation}\label{eq:fermionicsumx}
   4(\Omega_{r,+}+\Omega_{r,-})=8\kappa\sqrt{1+x^2}-\frac{4\mu^2}{\sqrt{1+x^2}}\frac{1}{\kappa}+\dots\,,
\end{equation}
which coincides exactly with the bosonic expression \eqref{eq:bosonicsumx} to order $1/\kappa$.
\section{One-loop computation of $J_1=J_2,\,J_3\neq0$ solution}\label{sect:3spindetails}
We now detail the computation of the one-loop energy correction to the classical energy \eqref{eq:3spinclassicalenergy}.
\subsection{Bosonic frequencies}
Fixing static gauge \eqref{eq:staticgauge}, the physical fluctuations on $S^5/\mathbb{Z}_{L}$ are
\begin{equation}
    \Phi\to\omega\tau+\frac{1}{\lambda^{1/4}}\tilde{\phi}(\tau,\sigma)\,,\;\psi\to\frac{\pi}{4}+\frac{1}{\lambda^{1/4}}\tilde{\psi}(\tau,\sigma)\,,\;\varphi_3\to\mathrm{k}\tau+\frac{1}{\lambda^{1/4}}\tilde{\varphi}(\tau,\sigma)\,,\; \gamma\to\gamma_0+\frac{1}{\lambda^{1/4}}\tilde{\gamma}(\tau,\sigma)\,.
\end{equation}
The determinant of the induced metric up to and including quadratic order is
\begin{equation}\begin{split}
& \sqrt{-\det h}=\mu^2\sin^2\gamma_0-\bigg[\mathrm{k}\cos^2\gamma_0\partial_{\tau}\tilde{\varphi}+\omega\sin^2\gamma_0\partial_{\tau}\tilde{\phi}\bigg]+\frac{1}{2}(h_{\sigma\sigma}^{(2)}-h_{\tau\tau}^{(2)})\\
&\qquad\qquad\quad+\frac{1}{2\mu^2\sin^2\gamma_0}\bigg[(h_{\tau\sigma}^{(1)})^2-h_{\tau\tau}^{(1)}h_{\sigma\sigma}^{(1)}-\frac{1}{4}\left(h_{\sigma\sigma}^{(1)}-h_{\tau\tau}^{(1)}\right)^2\bigg]\,,
\end{split}\end{equation}
where
\begin{equation}\begin{split}
&h^{(1)}_{\tau\tau}=\mu^2\sin2\gamma_0\,\tilde{\gamma}+2\mathrm{k}\cos^2\gamma_0\,\partial_{\tau}\tilde{\varphi}+2\omega\sin^2\gamma_0\,\partial_{\tau}\tilde{\phi}\,,\qquad h_{\sigma\sigma}^{(1)}=\mu^2\sin 2\gamma_0\,\tilde{\gamma}\,,\\
&h_{\tau\sigma}^{(1)}=\mathrm{k}\cos^2\gamma_0\,\partial_{\sigma}\tilde{\varphi}+\omega\sin^2\gamma_0\partial_{\sigma}\tilde{\phi}-2\omega\sin^2\gamma_0\mu\tilde{\psi}\,,\\
&h_{\tau\tau}^{(2)}=(\partial_{\tau}\tilde{\gamma})^2+\cos^2\gamma_0(\partial_{\tau}\tilde{\varphi})^2+\sin^2\gamma_0\big((\partial_{\tau}\tilde{\phi})^2+(\partial_{\tau}\tilde{\psi})^2\big)\\
&\qquad-2\sin2\gamma_0\big(\mathrm{k}\partial_{\tau}\tilde{\varphi}-\omega\partial_{\tau}\tilde{\phi}\big)\tilde{\gamma}+\mu^2\cos2\gamma_0\,\tilde{\gamma}^2\,,\\
&h_{\sigma\sigma}^{(2)}=(\partial_{\sigma}\tilde{\gamma})^2+\cos^2\gamma_0(\partial_{\sigma}\tilde{\varphi})^2+\sin^2\gamma_0\big((\partial_{\sigma}\tilde{\phi})^2+(\partial_{\sigma}\tilde{\psi})^2\big)-4\mu\sin^2\gamma_0\,\tilde{\psi}\partial_{\sigma}\tilde{\phi}+\mu^2\cos2\gamma_0\,\tilde{\gamma}^2\,.
\end{split}
\end{equation}
After dropping the constant term, the total derivatives, and using \eqref{eq:classicalrelations3spin}, the quadratic Lagrangian in fluctuations of $S^5/\mathbb{Z}_{L}$ reads
\begin{equation}\begin{split}
    \mathcal{L}_{S^5/\mathbb{Z}_{L}}^{(2)}=&-\frac{1}{2}(\partial_{a}\tilde{\gamma})^2-\frac{\sin^2\gamma_0}{2}(\partial_{a}\tilde{\psi})^2-\frac{\mathrm{a}}{2}(\partial_{a}\tilde{\varphi})^2-\frac{\mathrm{b}}{2}(\partial_{a}\tilde{\phi})^2+\mathrm{c}\left(\partial_{\tau}\tilde{\phi}\partial_{\tau}\tilde{\varphi}-\partial_{\sigma}\tilde{\phi}\partial_{\sigma}\tilde{\varphi}\right)\\
    &+\mathrm{p}\,\tilde{\gamma}\partial_{\tau}\tilde{\varphi}+\mathrm{q}\,\tilde{\gamma}\partial_{\tau}\tilde{\phi}+\mathrm{r}\,\tilde{\gamma}^2+\mathrm{s}\,\tilde{\psi}\partial_{\sigma}\tilde{\varphi}+\mathrm{t}\,\tilde{\psi}\partial_{\sigma}\tilde{\phi}-2\omega^2\sin^2\gamma_0\,\tilde{\psi}^2\,,
\end{split}\end{equation}
where we have defined the coefficients
\begin{equation}\begin{split}
&\mathrm{a}\equiv\sin^2\gamma_0\left(1+\frac{\omega^2}{\mu^2}\right)\,,\quad \mathrm{b}\equiv \cos^2\gamma_0\left(1+\frac{\mathrm{k}^2\cot^2\gamma_0}{\mu^2}\right)\,,\quad \mathrm{c}\equiv \frac{\mathrm{k}\omega\cos^2\gamma_0}{\mu^2}\,,\quad \mathrm{p}\equiv 2\omega\sin2\gamma_0\,,\\
&\mathrm{q}\equiv2\mathrm{k}\cot\gamma_0\cos2\gamma_0\,,\quad \mathrm{r}\equiv 2\mu^2\cos^2\gamma_0\,,\quad \mathrm{s}\equiv 2\left(\mu+\frac{\omega^2}{\mu}\right)\sin^2\gamma_0\,,\quad  \mathrm{t}\equiv\frac{2\mathrm{k}\omega\cos^2\gamma_0}{\mu}\,.
\end{split}\end{equation}
The main difference with respect to \eqref{eq:2spinL2}, is that for $\cos\gamma_0\in(0,1)$ the fluctuations $(\gamma,\varphi_3)$ now couple to those in the $(\psi,\Phi)$ sector. Looking for plane-wave solutions of the form \eqref{eq:bosonicplanewaves}, where the vector $\vec{\xi}_r^s=(\tilde{\gamma}_r,\tilde{\psi}_r,\tilde{\phi}_r,\tilde{\varphi}_r)$ is defined for each mode number $r\in\mathbb{Z}$, the linearised equations of motion from the above Lagrangian take the form
\begin{equation}
 (\mathrm{B}_r^s)\,\vec{\xi}^s_r=0\,,\qquad  \mathrm{B}_r^s=\begin{pmatrix}
      -\Delta+2\mathrm{r}  & 0& i\mathrm{p}(w_r^s) & i\mathrm{q}(w_r^s)&\\
      0& -\sin^2\gamma_0(\Delta+4\omega^2)& i\mathrm{s}r&i\mathrm{t}r &\\
      -i\mathrm{p}(w_r^s)&-i\mathrm{s}r &-\mathrm{a}\Delta & -\mathrm{c}\Delta&\\
      -i\mathrm{q}(w_r^s)& -i\mathrm{t}r& -\mathrm{c}\Delta&-\mathrm{b}\Delta & 
    \end{pmatrix}\,.
\end{equation}
The characteristic frequencies are obtained by solving the equation $\det (\mathrm{B}_r^s)=0$, which reads 
\begin{equation}\begin{split}\label{eq:3spinquartic}
&\Delta^4+4\left(2\omega^2-\mu^2\sin^2\gamma_0\right)\Delta^3+\bigg[16\omega^2\left(\omega^2-\mu^2\sin^2\gamma_0\right)+8r^2\left(\mu^2(\sin^2\gamma_0-2)-\mathrm{k}^2\right)\bigg]\Delta^2\\
-&16\mathrm{k}^2r^2\left(2\omega^2+\mu^2\sin^2\gamma_0\right)\Delta+16\kappa^2\mathrm{k}^2r^4=0\,,
\end{split}\end{equation}
for the non-degenerate branch of the classical solution, where $\sin\gamma_0\in(0,1)$ and $\mu\neq 0$.\footnote{Note that we have dropped an overall factor of $\frac{\kappa^2}{\mu^2}\sin^2\gamma_0\cos^2\gamma_0$ arising from the determinant.}

We can only give explicit expressions for the four bosonic frequencies on $S^5/\mathbb{Z}_L$ in specific regimes. One of them is the BMN limit, where $\mu\to 0$, which corresponds to $L\to\infty$ with fixed $n$ not of order $L$. In this regime, \eqref{eq:3spinquartic} factorises and one finds 
\begin{equation}\label{eq:bosonicBMNlimitCF}
    \bigg[\Delta^2+4\mathrm{k}^2(\Delta-r^2)\bigg]^2=0\quad\to\quad  w_{r,
    \pm}^s=\pm\mathrm{k}+\sqrt{r^2+\mathrm{k}^2}\,,
\end{equation}
giving a (positive) doubly-degenerate pair of frequencies which match those of \cite{Frolov:2003tu}.  Another exactly solvable regime is the `$J_1=J_2$ endpoint' where $\gamma_0\to\frac{\pi}{2}$, so that
\begin{equation}
    \det \mathrm{B}_r^s\to\bigg[\Delta^2+4\mathrm{k}^2(\Delta-r^2)\bigg]\bigg[\Delta^2+4\left(\mathrm{k}^2+\mu^2\right)\Delta-4\left(\mathrm{k}^2+2\mu^2\right)r^2\bigg]=0\,.
\end{equation}
The characteristic frequencies read
\begin{equation}
    w^{(1)}_{r,\pm}=\pm\mathrm{k}+\sqrt{r^2+\kappa^2-2\mu^2}\,,\quad (w_{r,\pm}^{(2)})^2=r^2+2\left(\kappa^2-\mu^2\right)\pm2\sqrt{\left(\kappa^2-\mu^2\right)^2+\kappa^2 r^2}\,.
\end{equation}
The latter set corresponds precisely to those in \eqref{eq:S3CFs} for the $J_1=J_2$ solution, while the former set coincides with \eqref{eq:DecoupledCF} up to a factor of $\pm\mathrm{k}$. This `additional' factor may be understood as a coordinate artifact. Indeed, at $\gamma_0=\frac{\pi}{2}$, the neutral $Z$-plane has zero-radius and the polar angle $\varphi_3$ becomes ill-defined. One thus needs to introduce the Cartesian-type fluctuation coordinates $\tilde{Z}=\tilde{X}_5+i\tilde{X}_6=e^{i\mathrm{k}\tau}(-\tilde{\gamma}\sin\gamma_0+i\tilde{\varphi}\cos\gamma_0)\equiv e^{i\mathrm{k}\tau}U$. The quantity $U$ is thus the Cartesian fluctuation viewed in a basis rotated with angular velocity $\mathrm{k}$. Substituting into the Cartesian equation of motion \eqref{eq:2spineoms}, one finds $(\partial_{\tau}^2-\partial_{\sigma}^2+i2\mathrm{k}\partial_{\tau})U=0$. Looking for plane-wave solutions for $U$, one indeed finds the frequencies \eqref{eq:DecoupledCF} with an extra factor of $\pm\mathrm{k}$ due to the frame rotation. The stability of the solution in the large $\mathrm{k}$ regime is discussed in Appendix \ref{sect:3spinbosonfreqs}, where we find agreement with the $S^5$ stability analysis of \cite{Frolov:2003tu} in the large $L$ regime with $n=\order{L}$.
\subsection{Fermionic frequencies}
Starting from the quadratic Green-Schwarz action in fermions after fixing $\kappa$-symmetry gauge \eqref{kfixedfermionicaction}, choosing the orthonormal frame of \eqref{eq:fermionONframe}, and recalling the classical solution \eqref{eq:3spinclassicalsoln}, one may compute the 2d matrices
\begin{equation}
    \varrho_0=\kappa\Gamma_0+\omega\sin\gamma_0\tilde{\Gamma}_8+\mathrm{k}\cos\gamma_0\Gamma_6\,,\quad \varrho_1=\mu\sin\gamma_0\tilde{\Gamma}_9\,,\quad \varrho_{(a}\varrho_{b)}=\mu^2\sin^2\gamma_0\,\eta_{ab}\,,
\end{equation}
where we adopted the definitions of $\tilde{\Gamma}_{8,9}$ quoted below \eqref{eq:2spin2dmatrices}.\footnote{The discrepancy with respect to the $\sigma$-dependent definitions of $\tilde{\Gamma}_{8,9}$ in \cite{Frolov:2003tu} may be understood by noting that our solution is mapped to theirs via the definitions X$_1\equiv\frac{X_1+X_3}{\sqrt{2}}$, X$_2\equiv\frac{X_2+X_4}{\sqrt{2}}$, X$_3\equiv\frac{X_2-X_4}{\sqrt{2}}$ and X$_4\equiv\frac{X_3-X_1}{\sqrt{2}}$ (i.e. by an O$(4)$ rotation). In terms of these new coordinates, one then finds $\mathrm{X}_1+i\mathrm{X}_2=\sin\gamma_0\cos\left(\mu\sigma\right)e^{i\omega\tau}$ and $\mathrm{X}_3+i\mathrm{X}_4=\sin\gamma_0\sin\left(\mu\sigma\right)e^{i\omega\tau}$, matching their solution where $\psi=\mu\sigma$, which then requires a $\sigma$-dependent rotation.} The non-zero components of the projected spin-connection $\omega_{a}^{AB}$ are
\begin{equation}
  \omega_{0}^{56}=\mathrm{k}\sin\gamma_0\,,\quad \omega_{0}^{58}=-\omega\cos\gamma_0\,,\quad \omega_{0}^{79}=\omega\,,\quad \omega_{1}^{59}=-\mu\cos\gamma_0\,,\quad \omega_{1}^{78}=\mu\,.
\end{equation}
The fermionic operator $D_F$ can thus be written as
\begin{equation}\begin{split}
   D_F&=(\kappa\Gamma_0+\mathrm{k}\cos\gamma_0\Gamma_6+\omega\sin\gamma_0\tilde{\Gamma}_8)\partial_0-\mu\sin\gamma_0\tilde{\Gamma}_9\partial_1+\frac{\kappa\mathrm{k}\sin\gamma_0}{2}\Gamma_0
   \Gamma_{56}\\
   &-\frac{\kappa\omega\cos\gamma_0}{2}\Gamma_0
   \Gamma_{5\tilde{8}}+\frac{\kappa\omega}{2}\Gamma_0\Gamma_{7\tilde{9}}+\frac{\mathrm{k}\omega}{2}\Gamma_{56}\tilde{\Gamma}_{8}+\frac{\mathrm{k}\omega\cos\gamma_0}{2}\Gamma_6\Gamma_{7\tilde{9}}-\frac{\sin\gamma_0}{2}(\omega^2+\mu^2)\Gamma_7\Gamma_{\tilde{8}\tilde{9}}\\
   &+\mu\sin\gamma_0\tilde{\Gamma}_{9}\left(\mathrm{k}\cos\gamma_0\Gamma_6+\omega\sin\gamma_0\tilde{\Gamma}_{8}\right)\Gamma_{01234}\,.
\end{split}\end{equation}
We may simplify the above operator further by performing a constant spinor rotation in the $(6\tilde{8})$ plane and a boost in the $(0\tilde{8})$ plane $(\theta\to S_{0\tilde{8}} S_{6\tilde{8}}\Psi)$, respectively, of the form
\begin{equation}
S_{6\tilde{8}}=e^{-\frac{p}{2}\Gamma_{6\tilde{8}}}\,,\quad \cos p=\frac{\mathrm{k}\cos\gamma_0}{a}\,,\quad \sin p=\frac{\omega\sin\gamma_0}{a}\,,\quad a\equiv \sqrt{\mathrm{k}^2+\mu^2\sin^2\gamma_0}\,,
\end{equation}
\begin{equation}
S_{0\tilde{8}}=e^{-\frac{q}{2}\Gamma_{0\tilde{8}}},\quad \cosh q=\frac{\kappa}{\mu\sin\gamma_0}\,,\quad \sinh q=\frac{a}{\mu\sin\gamma_0}\,,
\end{equation}
where $a^2$ denotes the norm of the spatial part of $\varrho_0$ (i.e. without the first $\kappa\Gamma_0$ factor). After a re-scaling $\Psi\to (\mu\sin\gamma_0)^{-1/2}\Psi$ of the fermions, the quadratic Lagrangian
\begin{equation}\begin{split}\label{eq:3spinL2fermion}
&\mathcal{L}_F^{(2)}=i2\bar{\Psi}\bigg[-\Gamma_0(\partial_0+A_0)+\tilde{\Gamma}_{9}(\partial_1+A_1)-a\Gamma_{0\tilde{9}}\Gamma_6\Gamma_{1234}\bigg]\Psi\,,\\
&A_0=\frac{1}{2a}\bigg(\kappa\mu\cos\gamma_0\Gamma_6+\omega\mathrm{k}\tilde{\Gamma}_{8}\bigg)\Gamma_5\,,\quad A_1=-\frac{1}{2a}\bigg(\kappa\omega\Gamma_6+\mu\mathrm{k}\cos\gamma_0\tilde{\Gamma}_8\bigg)\Gamma_7\,,
\end{split}\end{equation}
is obtained. We now look for plane-wave solutions of the form \eqref{eq:fermionplanewaves} to the Dirac equation associated to the above Lagrangian. To obtain the characteristic equation, first note that the matrix $\Gamma_{1234}$ commutes with the Clifford algebra generated by the rest of the Gamma matrices appearing in \eqref{eq:3spinL2fermion}, so we can project the fermions onto eigenspaces defined by $\Gamma_{1234}\Psi=\pm \Psi$, giving the same determinant. Choosing the $+$ projection, one then fixes an explicit six-dimensional representation for the Dirac matrices appearing in the projected Lagrangian.\footnote{A convenient representation is \begin{equation}\begin{split}
&\Gamma_0=i\sigma_1\otimes\mathbb{1}_2\otimes\mathbb{1}_2\,,\quad \Gamma_5=\sigma_2\otimes\mathbb{1}_2\otimes\mathbb{1}_2\,,\quad \Gamma_6=\sigma_3\otimes\sigma_1\otimes\mathbb{1}_2\,\quad \Gamma_7=\sigma_3\otimes\sigma_2\otimes\mathbb{1}_2\,,\\
&\tilde{\Gamma}_8=\sigma_3\otimes\sigma_3\otimes\sigma_1\,,\quad \tilde{\Gamma}_9=\sigma_3\otimes\sigma_3\otimes\sigma_2\,,\quad \Gamma_0^2=-\mathbb{1}_8\,,\quad \Gamma_{\rm{spatial}}^2=\mathbb{1}_8\,,\quad \{\Gamma_A,\Gamma_B\}=0\,,\;A\neq B\,.
\end{split}\end{equation}
In terms of this representation, the matrices appearing in the Lagrangian \eqref{eq:3spinL2fermion} are
\begin{equation}\begin{split}
&\Gamma_{065}=\mathbb{1}_2\otimes\sigma_1\otimes\mathbb{1}_2\,,\quad \Gamma_{0\tilde{8}5}=\mathbb{1}_2\otimes\sigma_3\otimes\sigma_1\,,\quad \Gamma_{\tilde{9}67}=i\sigma_{3}\otimes\mathbb{1}_2\otimes\sigma_2\,,\quad \Gamma_{\tilde{9}\tilde{8}7}=-i\sigma_3\otimes\sigma_2\otimes\sigma_3\,,\\
&\Gamma_{0\tilde{9}6}=-\sigma_1\otimes\sigma_2\otimes\sigma_2\,,
\end{split}\end{equation} where $\sigma_i$ denote the 2d Pauli matrices.} The Dirac equation then has an $8\cross 8$ mode matrix given by
\begin{equation}\begin{split}
\mathrm{F}_r^s&=\Omega_r^s\sigma_1\otimes\mathbb{1}_2\otimes\mathbb{1}_2+ir\sigma_3\otimes\sigma_3\otimes\sigma_2-\frac{\mu\kappa\cos\gamma_0}{2a}\mathbb{1}_2\otimes\sigma_1\otimes\mathbb{1}_2-\frac{\omega\mathrm{k}}{2a}\mathbb{1}_2\otimes\sigma_3\otimes\sigma_1\\
&\quad-i\frac{\kappa\omega}{2a}\sigma_3\otimes\mathbb{1}_2\otimes\sigma_2+i\frac{\mu\mathrm{k}\cos\gamma_0}{2a}\sigma_3\otimes\sigma_2\otimes\sigma_3\pm a\sigma_1\otimes\sigma_2\otimes\sigma_2\,.
\end{split}\end{equation}
The characteristic frequencies are given by the roots of $F_8(\Omega_r^s)$, where 
\begin{equation}\label{eq:fermionpolynomial}
\resizebox{\textwidth}{!}{$
\begin{aligned}
&F_8(\Omega^s_r)\equiv
\left(\Omega_r^s\right)^8
-\bigg[
4r^2
+6\mathrm{k}^2
+\left(5-\cos2\gamma_0\right)\mu^2
\bigg]\left(\Omega_r^s\right)^6
\\
&+
\bigg[
6\left(r^4+\mu^4\right)
+14r^2\mathrm{k}^2
+9\mathrm{k}^4
+\left(4r^2+14\mathrm{k}^2\right)\mu^2
+2\left(
5r^2+2\mathrm{k}^2+\mu^2
\right)
\mu^2\sin^2\gamma_0
+\frac{3}{2}\mu^4\sin^4\gamma_0
\bigg]\left(\Omega_r^s\right)^4\\
&-2
\bigg[
2\left(r^6+\mu^6\right)
+5r^4\mathrm{k}^2
+5r^2\mathrm{k}^4
+2\mathrm{k}^6
+\left(
-2r^4
-2r^2\mathrm{k}^2
+5\mathrm{k}^4
\right)\mu^2
+\left(
-2r^2
+5\mathrm{k}^2
\right)\mu^4
\\
&
+\left(
7r^4
+\mathrm{k}^4
+2r^2\mu^2
+12r^2\mathrm{k}^2
-\mu^4
\right)
\mu^2\sin^2\gamma_0-\frac{1}{2}
\left(
\mu^2
-7r^2
+\frac{\mathrm{k}^2}{2}
\right)
\mu^4\sin^4\gamma_0
+\frac{\mu^6}{4}\sin^6\gamma_0
\bigg]\left(\Omega_r^s\right)^2
\\
&+
\left(r^2-\mu^2\right)^2
\left[
r^4
+2r^2\left(
\mathrm{k}^2-\mu^2
\right)
+\left(
\mathrm{k}^2+\mu^2
\right)^2
\right]+2\mu^2
\left(r^2-\mu^2\right)
\left[
3r^4
+2r^2\left(
\mathrm{k}^2
-2\mu^2
\right)
+\left(
\mathrm{k}^2+\mu^2
\right)^2
\right]\sin^2\gamma_0
\\
&+
\frac{\mu^4}{2}
\left[
3\mu^4
+19r^4
+5\mathrm{k}^2\mu^2
+2\mathrm{k}^4
+r^2\left(
5\mathrm{k}^2
-14\mu^2
\right)
\right]\sin^4\gamma_0+
\frac{\mu^6}{2}
\left(
3r^2
-\mu^2
-\mathrm{k}^2
\right)\sin^6\gamma_0
+\frac{\mu^8}{16}\sin^8\gamma_0\,.
\end{aligned}
$}
\end{equation}
Since $F_8(\Omega_r^s)$ is even in $\Omega_r^s$, its roots occur in pairs $\pm\Omega_r^s$. The four doubly-degenerate characteristic roots entering the one-loop sum are selected by following the roots continuously from the large-$\kappa$ branches displayed in \eqref{eq:3spinflargek}. They are therefore signed characteristic roots rather than the positive magnitudes of the zeros of $F_8$. Stability requires the characteristic roots to remain real, it does not require every signed root to be positive. 

In general, however, one cannot obtain analytic solutions to the above equation for generic values of the parameters. For the case of $\mathrm{k}=0$, one finds
\begin{equation}
\resizebox{\textwidth}{!}{$
\begin{split}\label{eq:k0fermionfreqs}
& F_8=\left[(\Omega_r^s)^4-\left(2r^2+(2+\sin^2\gamma_0)\mu^2\right)(\Omega_r^s)^2+r^4+(3\sin^2\gamma_0-2)\mu^2r^2+\left(1-\frac{\sin^2\gamma_0}{2}\right)^2\mu^4\right]^2\,,\\
&(\Omega_{r,\pm})^2=r^2+\mu^2\left(1+\frac{\sin^2\gamma_0}{2}\right)\pm\mu\sqrt{2r^2(2-\sin^2\gamma_0)+2\mu^2\sin^2\gamma_0}\,,
\end{split}
$}
\end{equation}
resulting in a $4+4$ degeneracy in the two sets of roots. Another interesting regime is that of the unstable solution with $\gamma_0=\frac{\pi}{2}$, where one finds
\begin{equation}\begin{split}
&F_8=\bigg[\left(\Omega_r^s+\frac{\mathrm{k}}{2}\right)^2-\chi_{+}\bigg]\bigg[\left(\Omega_r^s+\frac{\mathrm{k}}{2}\right)^2-\chi_{-}\bigg]\bigg[\left(\Omega_r^s-\frac{\mathrm{k}}{2}\right)^2-\chi_{+}\bigg]\bigg[\left(\Omega_r^s-\frac{\mathrm{k}}{2}\right)^2-\chi_{-}\bigg]\,,\\
&\Omega_{r,\pm}=\pm\frac{1}{2}\left(\mathrm{k}+\sqrt{4r^2+5\mathrm{k}^2+6\mu^2\pm4\sqrt{\left(\mathrm{k}^2+2\mu^2\right)\left(r^2+\mathrm{k}^2+\mu^2\right)}}\right)\,,
\end{split}\end{equation}
resulting in eight different fermionic characteristic frequencies, which for $\mathrm{k}=0$ result in the $\gamma_{0}=\frac{\pi}{2}$ case of \eqref{eq:k0fermionfreqs}.\footnote{In the above expression for $F_8$ we have defined $\chi_{\pm}\equiv r^2+\frac{5}{4}\mathrm{k}^2+\frac{3}{2}\mu^2\pm\sqrt{\left(\mathrm{k}^2+2\mu^2\right)\left(r^2+\mathrm{k}^2+\mu^2\right)}$ for convenience.} Let us now consider the large $L$ regime with fixed $n$ not of order $L$ (i.e. $\tfrac{n}{L}\to 0$), which is a limit that connects with the BPS regime, giving
\begin{equation}\begin{split}
&F_8=\left(\mathrm{k}^2+r^2-(\Omega_r^s)^2\right)^2\left(r^2-2\mathrm{k}\Omega_r^s-(\Omega_r^s)^2\right)\left(r^2+2\mathrm{k}\Omega_r^s-(\Omega_r^s)^2\right)\\
&\Omega_r^{1,2,3,4}=\sqrt{r^2+\mathrm{k}^2}\,,\qquad \Omega_r^{5,6}=\sqrt{r^2+\mathrm{k}^2}+\mathrm{k}\,,\qquad \Omega_r^{7,8}=\sqrt{r^2+\mathrm{k}^2}-\mathrm{k}\,,
\end{split}\end{equation}
coinciding exactly with the bosonic doubly-degenerate pair \eqref{eq:bosonicBMNlimitCF} and the AdS$_5$ frequencies \eqref{eq:AdSCF}, as expected.
\subsection{Numerics of the one-loop correction}\label{sect:numerics}
Given the one-loop result \eqref{eq:oneloopSU3expression}, we wish to extract some numerical values for the coefficient $d_1(q,\mu)$. Let us fix $L=5$, so that $\mu\in\{\frac{1}{5},\frac{2}{5},\frac{3}{5},\frac{4}{5}\}$, then, following \eqref{eq:3spinstability}, one expects that for $n=1,2$ the solution is stable for all values of $q\in(0,1)$, but that for $n=3,4$ the solution is stable for $q\in(0,\frac{11}{36}]$ and $q\in(0,\frac{39}{64}]$ respectively. This is unlike the $J_1=J_2$ case, where twisted sectors with $n>\frac{L}{2}$ resulted in unstable states. Performing the numerical computation for $\kappa=50,100,200$, one uses the linear fit $\kappa^2E_1=d_1(q,n)+\kappa^{-2}e_2(q,n)$ in order to find the following values of the one-loop coefficient $d_1(q,n)=\lim_{\kappa\to\infty}\kappa^2E_1$ as the intercept of the line (i.e. the term that survives the strict semiclassical limit $\kappa\to\infty$). The values of $d_1(q,n)$ are quoted in Table \ref{tab:1}.\footnote{The endpoint $q=0$ is excluded because the angular fluctuation variables
and the associated static-gauge parametrisation degenerate there (see comment below \eqref{eq:3spinquartic}). \label{ftn:35}}

\begin{table}[ht]
\centering
\footnotesize
\begin{tabular}{|c|c|c|c|c|}
\hline
\(q\)&\(d_1\left(q,\frac{1}{5}\right)\)&\(d_1\left(q,\frac{2}{5}\right)\)&\( d_1\left(q,\frac{3}{5}\right)\)&\( d_1\left(q,\frac{4}{5}\right)\)\\
\hline
\(\frac{1}{12}\)&\(0.0017\)&\(0.0063\)&\(0.1922\)&\(0.6088\)\\
\(\frac{2}{12}\)&\(0.0032\)&\(0.0116\)&\(0.1775\)&\(0.6078\)\\
\(\frac{3}{12}\)&\(0.0048\)&\(0.0162\)&\(0.1508\)&\(0.5961\)\\
\(\frac{4}{12}\)&\(0.0064\)&\(0.0200\)&\(0.0976+0.0357i\)&\(0.5719\)\\
\(\frac{5}{12}\)&\(0.0078\)&\(0.0230\)&\(0.0895+0.0753i\)&\(0.5315\)\\
\(\frac{6}{12}\)&\(0.0093\)&\(0.0253\)&\(0.0780+0.1055i\)&\(0.4675\)\\
\(\frac{7}{12}\)&\(0.0108\)&\(0.0268\)&\(0.0630+0.1345i\)&\(0.3511\)\\
\(\frac{8}{12}\)&\(0.0120\)&\(0.0276\)&\(0.0445+0.1647i\)&\(0.2164+0.1647i\)\\
\(\frac{9}{12}\)&\(0.0134\)&\(0.0276\)&\(0.0228+0.1980i\)&\(0.1817+0.2881i\)\\
\(\frac{10}{12}\)&\(0.0148\)&\(0.0268\)&\(-0.0018+0.2359i\)&\(0.1533+0.4037i\)\\
\(\frac{11}{12}\)&\(0.0160\)&\(0.0252\)&\(-0.0279+0.2801i\)&\(0.1351+0.5168i\)\\
\(\frac{12}{12}\)&\(0.0172\)&\(0.0227\)&\(-0.0532+0.3317i\)&\(0.1261+0.6245i\)\\
\hline
\end{tabular}
\caption{Numerical values of $d_1(q,\mu)$ coefficient for $L=5$ and $q\in(0,1]$.}
\label{tab:1}
\end{table}
From Table \ref{tab:1}, we see that for $n<\frac{5}{2}$ (i.e. $n=1,2$) then the solution is stable for all values $q\in(0,1]$. As advertised from the stability condition \eqref{eq:3spinstabilitycond}, for $n>\frac{5}{2}$, the value of $\kappa^2E_1$ develops non-zero imaginary parts beyond $q>\frac{11}{36}\simeq 0.3056$ and $q>\frac{39}{64}\simeq 0.6094$ for $n=3,4$ respectively. Note that for $q\in\{\frac{4}{12},\dots,\frac{7}{12}\}$, the solution re-stabilises as one moves from the $n=3$ to $n=4$ twisted sectors, in agreement with \eqref{eq:3spinstabilitycond}. In the case of $q=\frac14$, with $n=3$, one finds the following linear function
\begin{equation}
   \kappa^2E_1=0.1508-0.0853 \kappa^{-2}\,,
\end{equation}
showing that the coefficient of $\kappa^{-2}$ is of the same order as $d_1$ so that one may reliably approximate $d_1$ by the value of $\kappa^2 E_1$. At $q=1$, the SU$(3)$ solution reduces to the SU$(2)$ solution, so that the final row of \ref{tab:1} provides a direct consistency check. Indeed, a direct evaluation of \eqref{eq:2spin1loopintegral} for $\mu=\frac{1}{5},\frac{2}{5}$ gives, respectively, $d_1(\frac{1}{5})=0.0172$ and $d_1(\frac{2}{5})=0.0228$, agreeing with the respective table entries.
\subsection{Different regimes of frequencies}\label{sect:3spinbosonfreqs}
Although we do not know exact expressions for the characteristic frequencies, it is possible to extract their asymptotic expansions in different regimes. In the large $r$ regime, the asymptotics of the four $S^5/\mathbb{Z}_L$ frequencies read
\begin{equation}\label{eq:3spinlarger}
    w_{r,\pm}^{\pm}=\vert r\vert\pm\mathrm{k}^2+\mu^2(2-\sin^2\gamma_0)\pm\sqrt{\mu^2\left(4\mathrm{k}^2\cos^2\gamma_0+\mu^2(2-\sin^2\gamma_0)\right)}+\frac{\mathrm{k}^2}{2\vert r\vert}+\dots\,.
\end{equation}
Accounting for the AdS$_5$ frequencies, the overall sum of the eight bosonic frequencies in this regime gives
\begin{equation}\label{eq:3spinblarger}
    4\omega^{\rm{AdS}}+w_{r,+}^++w_{r,+}^-+w_{r,-}^++w_{r,-}^-=8\vert r\vert+\frac{4(\mathrm{k}^2+\mu^2\sin^2\gamma_0)}{\vert r\vert}+\dots\,.
\end{equation}
The regime of large $\kappa$ with $x=\frac{r}{\kappa}$ held fixed probes modes with $r\sim \kappa$, which become `dense' in the semiclassical large-spin limit. The asymptotics are
\begin{equation}\label{eq:3spinblargek}
\resizebox{\textwidth}{!}{$
    w_{r,\pm}^{\pm}=\kappa(\pm1+\sqrt{1+x^2})\pm\frac{\mu\vert x\vert\cos\gamma_0}{\sqrt{1+x^2}}+\frac{\mu^2}{\kappa}\frac{\left[\cos^2\gamma_0-2\sin^2\gamma_0(1+x^2)\pm (1-2\sin^2\gamma_0)(1+x^2)^{3/2}\right]}{2(1+x^2)^{3/2}}
    $}\,,
\end{equation}
to first few orders. The overall contribution from bosonic frequencies in this regime is
\begin{equation}\label{eq:3spinbosonlargekfixedx}
 4\omega^{\rm{AdS}}+w_{r,+}^++w_{r,+}^-+w_{r,-}^++w_{r,-}^-=8\kappa\sqrt{1+x^2}-\frac{\mu^2}{\kappa}\frac{2\left(2\sin^2\gamma_0(1+x^2)-\cos^2\gamma_0\right)}{(1+x^2)^{3/2}}+\dots\,.
\end{equation}
We now consider the same expansion regimes for the roots of \eqref{eq:fermionpolynomial}. Again, here we do not know closed-form expressions for the roots but may obtain their asymptotics in the same way as before. In the large $r$ regime, one finds four sets of doubly-degenerate frequencies
\begin{equation}
    \Omega_r=\vert r\vert\pm\sqrt{\left(1-\frac{\sin^2\gamma_0}{2}\right)\mu^2+\frac{\mathrm{k}^2}{2}\pm\frac{\mathrm{k}}{2}\sqrt{\mathrm{k}^2+2\mu^2\sin^2\gamma_0}}+\frac{\mathrm{k}^2+\mu^2\sin^2\gamma_0}{2\vert r\vert}+\dots\,,
\end{equation}
corresponding to different choices of $\pm$ signs. Adding these up one obtains
\begin{equation}\label{eq:3spinflarger}
    \sum_{s=1}^{8}\Omega_r^s=8\vert r\vert+\frac{4(\mathrm{k}^2+\mu^2\sin^2\gamma_0)}{\vert r\vert}+\dots
\end{equation}
At large $\kappa$ with $r$ held fixed, we obtain a set of doubly-degenerate frequencies which split into three branches. From `lightest' to `heaviest', these are given by
\begin{equation}\begin{split}\label{eq:3spinflargek}
&\Omega_{r,1}=\frac{ r^2-\mu^2\cos^2\gamma_0}{2\kappa}\,,\quad\Omega_{r,2/3}=\kappa+\frac{r^2+\mu^2(1-2\sin^2\gamma_0)\pm\mu\sqrt{4r^2\cos^2\gamma_0+\mu^2\sin^4\gamma_0}}{2\kappa}\,,\\
&\Omega_{r,4}=2\kappa+\frac{r^2+\mu^2(1-3\sin^2\gamma_0)}{2\kappa}\,.
\end{split}\end{equation}
At large $\kappa$ with $x=\frac{r}{\kappa}$ held fixed, one finds
\begin{equation}\begin{split}
    &\Omega_{r,+}=\kappa(1+\sqrt{1+x^2})-\frac{\mu^2}{2\kappa\sqrt{1+x^2}}\left[(1+2\sqrt{1+x^2})\sin^2\gamma_0-\sqrt{1+x^2}\right]+\dots\,,\\
    &\Omega_{r,-}=\kappa(-1+\sqrt{1+x^2})+\frac{\mu^2}{2\kappa\sqrt{1+x^2}}\left[(-1+2\sqrt{1+x^2})\sin^2\gamma_0-\sqrt{1+x^2}\right]+\dots\,,\\
    &\Omega_{r,0}^{\pm}=\kappa\sqrt{1+x^2}\pm\mu\frac{\vert x\vert\cos\gamma_0}{\sqrt{1+x^2}}+\frac{\mu^2}{2\kappa(1+x^2)^{3/2}}\left[1-(2+x^2)\sin^2\gamma_0\right]+\dots\,,
\end{split}\end{equation}
where each of the four frequencies is doubly-degenerate. Summing the above gives
\begin{equation}\label{eq:3spinlargekfixedxsumf}
    2(\Omega_{r,+}+\Omega_{r,-}+\Omega_{r,0}^++\Omega_{r,0}^-)=8\kappa\sqrt{1+x^2}-\frac{1}{\kappa}\frac{2\mu^2\left(2\sin^2\gamma_0(1+x^2)-\cos^2\gamma_0\right)}{(1+x^2)^{3/2}}+\dots\,,
\end{equation}
which, just like in the $S^5$ case, coincides exactly with the bosonic expression \eqref{eq:3spinbosonlargekfixedx} up to and including $\order{\kappa^{-1}}$.
\section{One-loop computation of $SJ$ solution}\label{sect:SJdetails}
We now detail the computation of the one-loop energy correction of the $SJ$ solution discussed in Section \ref{sec: 2}.
\subsection{Bosonic frequencies}
The polar angles associated the transverse $Y$ and $Z$ planes on $S^5/\mathbb{Z}_{L}$ are ill-defined on the classical solution \eqref{eq:orbifoldSJsolution}, so one needs to introduce the embedding coordinates $\mathrm{Y}=X_3+iX_4$ and $\mathrm{Z}=X_5+iX_6$, which allow to rewrite the metric \eqref{orbifoldmetric} as
\begin{equation}\label{eq:2spinS5SJmetric}
    \dd s^2_{S^5/\mathbb{Z}_{L}}=\frac{1-R^2}{L^2}\dd\hat{\varphi}_1^2+\sum_{I=3}^6\dd X_{I}^2+\frac{1}{1-R^2}\left(\sum_{I=3}^6X_I\dd X_I\right)^2\,,
\end{equation}
where $R^2\equiv X_{3}^2+X_{4}^2+X_5^2+X_6^2$ and $\hat{\varphi}_1=L\varphi_1\in[0,2\pi)$ is a re-scaled version of the polar angle defined in \eqref{eq:unorbifoldedmetrics}. Imposing the static gauge condition $t=\kappa\tau\,,\, \hat{\varphi}_1=L\omega\tau+n\sigma$, the four physical transverse fluctuations are $X_I(\tau,\sigma)\to 0+\lambda^{-1/4}\tilde{X}_{I}(\tau,\sigma)$. To quadratic order, the metric \eqref{eq:2spinS5SJmetric} becomes
\begin{equation}
\dd s^2_{S^5/\mathbb{Z}_{L}}=\frac{1}{L^2}\left(1-\sum_{I}\tilde{X}_I^2\right) \dd\hat{\varphi}_1^2+\sum_{I}\dd\tilde{X}_{I}^2\,.
\end{equation}
The determinant of the induced metric up to and including quadratic order in fluctuations is
\begin{equation}
   \sqrt{-\det h}=\frac{1}{2}\sum_{I}\left[(\partial_{\sigma}\tilde{X}_I)^2-(\partial_{\tau}\tilde{X}_I)^2+\left(\omega^2-\mu^2\right)\tilde{X}_{I}^2\right]\,,
\end{equation}
so that the quadratic fluctuation Lagrangian reads
\begin{equation}
    \mathcal{L}^{(2)}_{S^5/\mathbb{Z}_{L}}=-\frac{1}{2}\sum_{I}\left[(\partial_{a}\tilde{X}_I)^2+\nu^2\tilde{X}_{I}^2\right]\,,
\end{equation}
which represents a set of four real 2d fields with mass $\sqrt{\omega^2-\mu^2}$. The equations of motion that follow are $\left(\square+\omega^2-\mu^2\right)\tilde{X}_I=0$. Looking for plane-wave solutions of the form $\tilde{X}_I\sim\sum_r\tilde{X}_{I,r}e^{i(w_r\tau+r\sigma)}$, one finds the following set of four-fold degenerate characteristic frequencies
\begin{equation}\label{eq:SJS5freqs}
    \left(\Delta+\omega^2-\mu^2\right)^4=0\,,\qquad w_{r}=\sqrt{r^2+\omega^2-\mu^2}\,,\qquad \Delta\equiv r^2-w_r^2\,.
\end{equation}
Since the orbifold acts trivially on AdS$_5$, we refer the reader to Section 4.1 of \cite{Park:2005ji} for AdS fluctuations, with the only difference being the implicit dependence on $\mu$ which enters through the classical relations \eqref{eq:orbifoldSJvirasoro}. We expect two massive modes corresponding to the two transverse directions to the AdS$_3\subset\mathrm{AdS}_5$ subspace $(\theta=\frac{\pi}{2})$ where the classical solution sits. These fluctuations contribute with frequencies
\begin{equation}\label{eq:SJAdSfreqs}
w_{1,2}^{\rm{AdS}_{\perp}}=\sqrt{r^2+\kappa^2}\,.
\end{equation}
The remaining physical fluctuations living inside the AdS$_3$ plane of the classical solution are 
\begin{equation}
    \rho_0\to \rho_0+\lambda^{-1/4}\tilde{\rho}(\tau,\sigma)\,,\qquad \varphi\to (w\tau+p\sigma)+\lambda^{-1/4}\tilde{\varphi}(\tau,\sigma)\,.
\end{equation}
The characteristic frequencies are then found as the four algebraic roots $\omega_{I,r}^{\rm{AdS}_3}$ ($I=1,2,3,4$) of the quartic polynomial
\begin{equation}\label{eq:ads3quartic}
   \mathcal{P}(r,\omega)\equiv  (\omega_r^2-r^2)^2+4\kappa^2\sinh^2\rho_0\,\omega_r^2-4\cosh^2\rho_0\left(\sqrt{\kappa^2+p^2}\,\omega_r-pr\right)^2=0\,.
\end{equation}
Since \eqref{eq:ads3quartic} is invariant under $\mathcal{P}(r,\omega)\to\mathcal{P}(-r,-\omega)$, we have that $\omega^{\rm{AdS}_3}_{I,r}=-\omega^{\rm{AdS}_3}_{I,-r}$. Furthermore, due to the absence of an $\omega_r^3$ term in \eqref{eq:ads3quartic}, the naive sum of the four roots vanishes and one cannot simply take the two positive roots at fixed $r$ in the one-loop sum. The prescription in \cite{Park:2005ji} is to include the four roots in the one-loop sum \eqref{eq:SJE1oneloopsum} as\footnote{The sign prescription originates from a large $\mathcal{J}$ expansion of the roots - see Section 4.1 of \cite{Park:2005ji} for more details}
\begin{equation}
    \frac{1}{2}\sum_{I=1}^4\mathrm{sign}(I)\,\omega_{I,r}^{\rm{AdS}_3}\,,\qquad \mathrm{sign}(1)=\mathrm{sign}(3)=+1\,,\qquad \mathrm{sign}(2)=\mathrm{sign}(4)=-1\,.
\end{equation}
One can only solve the above quartic explicitly in special regimes, an example of which is the $r=0$ sector, where one finds
\begin{equation}
    \omega_{(1,2)}^{\rm{AdS}_3}=0\,,\qquad \omega_{(3,4)}^{\rm{AdS}_3}=\pm 2\sqrt{\kappa^2+p^2\cosh^2\rho_0}\,.
\end{equation}
\subsection{Fermionic frequencies}
For the fermionic frequencies, one starts with the quadratic fermionic part of the $\kappa$-symmetry gauge-fixed Green-Schwarz action \eqref{kfixedfermionicaction}. We now use the orthonormal frame
\begin{equation}
    E^0_t=\cosh\rho_0\,\dd t\,,\qquad E_\rho^1=\dd\rho\,,\qquad E_\varphi^3=\sinh\rho_0\dd\varphi\,,\qquad E_{\varphi_1}^6=\dd\varphi_1\,,
\end{equation}
so that the two-dimensional projections of the 10d Gamma matrices are
\begin{equation}
\varrho_0=\kappa\cosh\rho_0\Gamma_0+w\sinh\rho_0\Gamma_3+\omega\Gamma_6\,,\; \varrho_1=p\sinh\rho_0\Gamma_3+\mu\Gamma_6\,,\; \varrho_{(a}\varrho_{b)}=\left(\mu^2+p^2\sinh^2\rho_0\right)\eta_{ab}\,.
\end{equation}
The non-trivial components of the spin-connection relevant to the AdS$_3$ motion are $\omega^{01}=\sinh\rho_0\,\dd t$ and $
\omega^{31}=\cosh\rho_0\,\dd\varphi$, with the non-zero components of their pull-backs being $\omega^{01}_{\tau}=\kappa\sinh\rho_0$, $\omega^{13}_{\tau}=-w\cosh\rho_0$ and $\omega^{13}_{\sigma}=-p\cosh\rho_0$. The fermionic kinetic operator is then
\begin{equation}\begin{split}
D_F=&\left(\kappa\cosh\rho_0\Gamma_0+w\sinh\rho_0\Gamma_3+\omega\Gamma_6\right)\left(\partial_0-\frac{\kappa\sinh\rho_0}{2}\Gamma_{10}-\frac{w\cosh\rho_0}{2}\Gamma_{13}\right)\\
&-\left(p\sinh\rho_0\,\Gamma_3+\mu\Gamma_6\right)\left(\partial_1-\frac{p\cosh\rho_0}{2}\Gamma_{13}\right)+\frac{p\kappa}{2}\sinh 2\rho_0\,\Gamma_{124}\,.
\end{split}\end{equation}
The above operator may be simplified by performing a constant spinor rotation in the $(36)$ plane and a boost in the $(06)$ plane (i.e. $\theta\to S_{06}S_{36}\Psi$), taking the form
\begin{equation}\begin{split}
&S_{36}=e^{-\frac{l}{2}\Gamma_{36}}\,,\quad \cos l=\frac{p\sinh\rho_0}{N}\,,\quad \sin l=\frac{\mu}{N}\,, \quad N^2\equiv p^2\sinh^2\rho_0+\mu^2\,,\\
&S_{06}=e^{-\frac{b}{2}\Gamma_{06}}\,,\quad \cosh b=\frac{\kappa\cosh\rho_0}{N}\,,\quad \sinh b=-\frac{M}{N}\,,\quad M\equiv\frac{\sinh\rho_0(p\omega-\mu w)}{N}\,.
\end{split}\end{equation}
Note that $\cos l<0$ since $p<0$ is required by the classical solution. Therefore, after the constant rotation and boost, one finds that $\varrho_{1}\to N\,\Gamma_3$ and $\varrho_0\to N\,\Gamma_0$. The Dirac operator can be rewritten (after removing the overall factor of $N$) as 
\begin{equation}
    D_F=\Gamma_0\partial_0-\Gamma_3\partial_1+ i\,\mathrm{f}_1\Gamma_1+\mathrm{f}_2\Gamma_{016}+\mathrm{f}_3\Gamma_{136}\,,
\end{equation}
where we have defined the coefficients
\begin{equation}\label{eq:fdefinition}
    \mathrm{f}_1\equiv \frac{\vert p\vert\kappa\sinh 2\rho_0}{2N}\,,\qquad \mathrm{f}_2\equiv \mu\frac{\kappa}{w}\frac{w^2-\omega^2}{\kappa^2-\nu^2}\,,\qquad \mathrm{f}_3\equiv \mu\frac{p\kappa\cosh^2\rho_0}{\kappa^2-\nu^2}\,.
\end{equation}
Note, in particular that $\mathrm{f}_3<0$ as a consequence of $p$ being a negative integer. Looking for plane-wave solutions of the form \eqref{eq:fermionplanewaves}, the mode equation (after multiplying by $\Gamma_0$) is
\begin{equation}
\bigg(-i\Omega_r^s\mathbb{1}-ir\Gamma_{03}\pm i\mathrm{f}_1\Gamma_{01}-\mathrm{f}_2\Gamma_{16}+\mathrm{f}_3\Gamma_{0136}\bigg)\psi_r^s=0\,.
\end{equation}
Similarly to before, we project the fermions onto eigenspaces defined by $i\Gamma_{0136}\,\psi_r^s=\pm\psi_r^s$. Then, $\Gamma_{0136}\psi_r^s=\mp i\psi_r^s$ and $\Gamma_{16}\psi_r^s=\pm i\Gamma_{03}\psi_r^s$. On each eigenspace, choosing a two-dimensional basis where $\Gamma_{03}=\sigma_3$ and $\Gamma_{01}=\sigma_1$,\footnote{This is valid since $(\Gamma_{03})^2=\mathbb{1}=(\Gamma_{01})^2$ and $\{\Gamma_{03},\Gamma_{01}\}=0$.} leads to the characteristic frequencies being found as the zeros of the $2\cross 2$ reduced blocks
\begin{equation}
    \det\begin{pmatrix}
    \Omega_r^s\pm(\mathrm{f}_3+\mathrm{f}_2)+r & -\mathrm{f}_1\\
    -\mathrm{f}_1 & \Omega_r^s\pm (\mathrm{f}_3-\mathrm{f}_2)-r
    \end{pmatrix}=(\Omega_r^s\pm\mathrm{f}_3)^2-(r\pm\mathrm{f}_2)^2-\mathrm{f}_1^2=0\,,
\end{equation}
which differ simply by the sign of the projection. The characteristic equation is the product of the determinants on each block, yielding the quartic
\begin{equation}
  (\Omega_r^s)^4  -2(r^2+\mathrm{f}_1^2+\mathrm{f}_2^2+\mathrm{f}_3^2)(\Omega_r^s)^2+8r\mathrm{f}_2\mathrm{f}_3\,\Omega_r^s+(r^2+\mathrm{f}_1^2+\mathrm{f}_2^2-\mathrm{f}_3^2)^2-4r^2\mathrm{f}_2^2=0\,,
\end{equation}
whose (positive) solutions are given by the four-fold degenerate set of two frequencies
\begin{equation}\label{eq:SJfermionicfreqs}
    \Omega_r^{+}=\left\vert \sqrt{(r+\mathrm{f}_2)^2+\mathrm{f}_1^2}-\mathrm{f}_3\right\vert\,,\qquad \Omega_r^{-}=\left\vert \sqrt{(r-\mathrm{f}_2)^2+\mathrm{f}_1^2}+\mathrm{f}_3\right\vert\,,
\end{equation}
which result in the expected eight physical fermionic modes.
\subsection{Different regimes of frequencies}\label{sect:SJfrequencies}
The bosonic frequencies in \eqref{eq:SJS5freqs}, \eqref{eq:SJAdSfreqs} and the ones found from \eqref{eq:ads3quartic}, admit large $r$ expansions of the form
\begin{equation}
4w_r\to 4\vert r\vert+\frac{2(\omega^2-\mu^2)}{\vert r\vert} \,,\quad 2w^{\rm{AdS}\perp}_r\to 2\vert r\vert+\frac{\kappa^2}{\vert r\vert}\,,\quad \frac{1}{2}\sum_{I=1}^4\mathrm{sign}(I)\omega_{r,I}^{\rm{AdS}_3}\to 2\vert r\vert+\frac{\kappa^2}{\vert r\vert}\,,
\end{equation}
up to and including $\mathcal{O}(\vert r\vert^{-1})$. Their overall sum in this regime gives
\begin{equation}\label{eq:SJoverallbosoniclarger}
    8\vert r\vert+\frac{2(\kappa^2+\omega^2-\mu^2)}{\vert r\vert}+\order{\vert r\vert^{-2}}\,.
\end{equation}
Expanding now at large $\kappa$ (or large $\mathcal{J}$) with fixed $r$ (and fixed $u=\frac{\mathcal{S}}{\mathcal{J}}$), one has that $\sinh^2\rho_0\simeq u$, $\mu=-pu$. Then, one finds
\begin{equation}
\begin{split}
   & w_r\to\mathcal{J}+\frac{(r^2-\mu^2)}{2\mathcal{J}}\,,\qquad w_r^{\rm{AdS}_{\perp}}\to \mathcal{J}+\frac{r^2+\mu^2+2p^2u}{2\mathcal{J}}\,,\\
   &\omega_{(1,2),r}^{\rm{AdS}_3}\to\frac{r}{2\mathcal{J}}\left(2p(1+u)\pm\sqrt{r^2+4p^2u(1+u)}\right)\,,\\
   &\omega_{(3,4),r}^{\rm{AdS}_3}\to\pm2\mathcal{J}\pm\frac{r^2\mp2pr(1+u)+2p^2(1+3u+u^2)}{2\mathcal{J}}\,,
\end{split}
\end{equation}
with the orbifold dependence entering through the classical relation $\mu=-pu$. The lighter AdS$_3$ branches are real since $p^2u(1+u)=\mu(\mu-p)>0$ from \eqref{eq:orbifoldSJvirasoro}, so the solution is completely stable at large $\mathcal{J}$. The novelty is that, as opposed to the analog solution on AdS$_5\cross S^5$, one can now choose $u\in\big[0,\frac{1}{\vert p\vert}\big)$ by considering purely fractional windings $\mu\in[0,1)$. This probes a novel stable subsector at large $\mathcal{J}$ with arbitrarily small, but non-zero $u=\frac{n}{\vert p\vert L}$ that is not accessible through the standard AdS$_5\cross S^5$ background and may be continuously connected to a BPS sector as $\mu\to 0$. 

A large $\kappa$ expansion with $x=\frac{r}{\kappa}$ held fixed gives a total bosonic contribution of the form
\begin{equation}
4w_r+2w_r^{\rm{AdS}}+\frac{1}{2}\sum_{I}\omega_{r,I}^{\rm{AdS}_3}\to 8\kappa\sqrt{1+x^2}+\frac{p^2(1+u)}{\kappa}\left(\frac{1+u}{(1+x^2)^{3/2}}-\frac{4u}{\sqrt{1+x^2}}\right)\,.
\end{equation}
The fermionic frequencies in \eqref{eq:SJfermionicfreqs} admit a large $r$ expansion at fixed $\kappa$, giving the total contribution
\begin{equation}
4(\Omega_{r}^{+}+\Omega_{r}^{-})\to 8\vert r\vert+\frac{4\mathrm{f}_1^2}{\vert r\vert}+\dots\,.
\end{equation}
Comparing to \eqref{eq:SJoverallbosoniclarger}, we see that the bosonic and fermionic divergent terms in the $r$ sum cancel out (cf. \eqref{eq:orbifoldSJvirasoro}). Similarly, an expansion in large $\kappa$ (or large $\mathcal{J}$) with $r$ held fixed of the form
\begin{equation}
 4(\Omega_{r}^{+}+\Omega_{r}^{-})\to 8\kappa+\frac{4r^2+(\mu-p)(p-3\mu)}{\kappa} +\order{\kappa^{-3}}\,.
\end{equation}
Finally, for completeness, the large $\kappa$ expansion with fixed $x=\frac{r}{\kappa}$ gives
\begin{equation}
    4(\Omega_r^++\Omega_r^-)\to8\kappa\sqrt{1+x^2}+\frac{(\mu-p)^2-4\mu(\mu-p)(1+x^2)}{\kappa(1+x^2)^{3/2}}+\order{\kappa^{-2}}\,.
\end{equation}
\section{Twisted Bethe equations in $\mathrm{SU}(3)$ sector}\label{app:2}
For the orbifolded SU$(3)$ Bethe ansatz, one first chooses an ordered basis of fields $\vec{\Phi}=(\Phi_1,\Phi_2,\Phi_3)$, where $\Phi_1$ specifies the vacuum $\mathrm{Tr}[\gamma^n\Phi_1^{J_{\rm{tot}}}]$ on top of which excitations are built. The charge vector $\vec{s}=(s_1,s_2,s_3)$ encodes the $\mathbb{Z}_L$ charges of $\vec{\Phi}$. There are two Bethe root types $\{u_{1,k},u_{2,k}\}$, with degeneracies $\{K_1,K_2\}$. The number of $(\Phi_1,\Phi_2,\Phi_3)$ sites in the spin-chain is $(J_{\rm{tot}}-K_1,K_1-K_2,K_2)$ respectively. Closure of the gauge index (analogous to \eqref{eq:angmomentaprojection}) imposes that
\begin{equation}
    s_1J_{\rm{tot}}+(s_2-s_1)K_1+(s_3-s_2)K_2=0\mod L\,,
\end{equation}
for the single-trace operator. The two nesting levels are associated with the simple-root transitions $\Phi_1\to\Phi_2$ and $\Phi_2\to\Phi_3$. The first-level roots $u_{1,k}$ are `momentum-carrying', whereas the second-level roots $u_{2,k}$ resolve the internal `type' of the first-level excitations. Upon orbifolding, the local magnon scattering is unchanged, with the twist appearing as an overall phase in the nested Bethe equations, which read
\begin{equation}\begin{split}\label{eq:SU3nestedbetheeqs}
&\xi^{n(s_2-s_1)}\left(\frac{u_{1,k}+i/2}{u_{1,k}-i/2}\right)^{J_{\rm{tot}}}=\prod_{\substack{j=1 \\ j\neq k}}^{K_1}\frac{u_{1,k}-u_{1,j}+i}{u_{1,k}-u_{1,j}-i}\prod_{j=1}^{K_2}\frac{u_{1,k}-u_{2,j}-i/2}{u_{1,k}-u_{2,j}+i/2}\,,\\
&\xi^{n(s_3-s_2)}=\prod_{\substack{j=1 \\ j\neq k}}^{K_2}\frac{u_{2,k}-u_{2,j}+i}{u_{2,k}-u_{2,j}-i}\prod_{j=1}^{K_1}\frac{u_{2,k}-u_{1,j}-i/2}{u_{2,k}-u_{1,j}+i/2}\,.
\end{split}\end{equation}
The phase associated to the momentum constraint is controlled by $\Phi_1$, resulting in
\begin{equation}\label{eq:SU3twistedpconstraint}
 \xi^{ns_1}=\prod_{k=1}^{K_1}\frac{u_{1,k}+i/2}{u_{1,k}-i/2}=e^{iP}\,,\qquad P=\sum_{k=1}^{K_1}p_k\,.
\end{equation}
In the continuum limit $J_{\rm{tot}}\to\infty$, one re-scales the rapidities as $x_{a,k}=\frac{u_{a,k}}{J_{\rm{tot}}}$, with $a\in\{1,2\}$. Taking logarithms, we choose common branches $r_{1,k}=-m_{1}\in\mathbb{Z}$, $r_{2,k}=-m_2\in\mathbb{Z}$ at each nesting level. Expanding the LHS of \eqref{eq:SU3nestedbetheeqs} at large $J_{\rm{tot}}$, one obtains
\begin{equation}\label{eq:SU3lognestedbetheeqns}
\begin{split}
&2\pi q_{1}+\frac{1}{x_{1,k}}=\frac{1}{i}\sum_{\substack{j=1 \\ j\neq k}}^{K_1}\log\frac{x_{1,k}-x_{1,j}+i/J_{\rm{tot}}}{x_{1,k}-x_{1,j}-i/J_{\rm{tot}}}+\frac{1}{i}\sum_{j=1}^{K_2}\log\frac{x_{1,k}-x_{2,j}-i/2J_{\rm{tot}}}{x_{1,k}-x_{2,j}+i/2J_{\rm{tot}}}\,,\\
&2\pi q_2=\frac{1}{i}\sum_{\substack{j=1 \\ j\neq k}}^{K_2}\log\frac{x_{2,k}-x_{2,j}+i/J_{\rm{tot}}}{x_{2,k}-x_{2,j}-i/J_{\rm{tot}}}+\frac{1}{i}\sum_{j=1}^{K_1}\log\frac{x_{2,k}-x_{1,j}-i/2J_{\rm{tot}}}{x_{2,k}-x_{1,j}+i/2J_{\rm{tot}}}\,,
\end{split}
\end{equation}
where 
\begin{equation}
 q_1\equiv m_1+\mu(s_2-s_1)\,,\qquad   q_2\equiv m_2+\mu(s_3-s_2)\,, 
\end{equation}
are analogous to $q_{\rm{SU}(2)}$ in \eqref{eq:SU2thermodynamic}, encoding the twist dependence. Note that the logarithms on the RHS of \eqref{eq:SU3lognestedbetheeqns} are not expanded in order to keep track of the `anomalous' nearby contributions $x_{a,k}-x_{b,j}=\order{J_{\rm{tot}}^{-1}}$. Defining the two resolvents
\begin{equation}\begin{split}
&G_1(x)=\frac{1}{J_{\rm{tot}}}\sum_{k=1}^{K_1}\frac{1}{x-x_{1,k}}\,,\qquad G_1(x\to\infty)\sim\frac{K_1}{J_{\rm{tot}}x}\,,\\
&G_2(x)=\frac{1}{J_{\rm{tot}}}\sum_{k=1}^{K_2}\frac{1}{x-x_{2,k}}\,,\qquad G_2(x\to\infty)\sim\frac{K_2}{J_{\rm{tot}}x}\,,
\end{split}\end{equation}
allows to rewrite the one-loop anomalous dimension as 
\begin{equation}
    \Delta-J_{\rm{tot}}=\frac{i\lambda}{8\pi^2}\left[G_1\left(\frac{i}{2J_{\rm{tot}}}\right)-G_1\left(-\frac{i}{2J_{\rm{tot}}}\right)\right]+\order{\lambda^2}\,,
\end{equation}
which, in the $J_{\rm{tot}}\to\infty$ limit becomes $\Delta-J_{\rm{tot}}=-\frac{\lambda}{8\pi^2J_{\rm{tot}}}G_1'(0)$. The total momentum may be written as $P=-G_1(0)=2\pi(k+\mu s_1)$ with $k\in\mathbb{Z}$. It is now more complicated to deal with the nearby roots than in \cite{Beisert:2005mq}, due to the nested structure of the Bethe equations.

Specialising to the cases of interest, when $J/2\geq J_3$ one chooses the basis $\vec{\Phi}=(X,Y,Z)$ with $\vec{s}=(1,-1,0)$. This fixes the `vacuum' $\mathrm{Tr}[\gamma^nX^{J_{\rm{tot}}}]$, which exists only when $J_{\rm{tot}}=0\,\mathrm{mod}\,L$ and $n=0$. In this case, the Bethe root type degeneracies are $K_1=J_3+J/2$ and $K_2=J_3$. For the SU$(3)$ state with purely fractional windings, as in \cite{Astolfi:2008yw}, we choose $m_1=m_2=k=0$ so that $q_1=-2\mu$, $q_2=\mu$ and $P=2\pi\mu$, in agreement with the twisted momentum constraint \eqref{eq:SU3twistedpconstraint}. For $J_3\geq J/2$, one instead chooses $\vec{\Phi}=(Z,X,Y)$ with $\vec{s}=(0,1,-1)$, and the vacuum is $\mathrm{Tr}[\gamma^nZ^{J_{\rm{tot}}}]$, which exists for all spin-chain lengths $J_{\rm{tot}}$. Here $K_1=J$ and $K_2=J/2$, so that for the fractional winding case we find $q_1=\mu$, $q_2=-2\mu$ and vanishing total momentum (i.e. $P=0$, which is consistent with $e^{iP}=1$). Note that although the twisted phases have been redistributed between the nested equations due to the change of ordered basis $\vec{\Phi}$, they describe the same physical spin-chain state, and so share the same anomalous dimension, which is representation-independent.

\bibliographystyle{JHEP}
\bibliography{semiclassical}
\end{document}